\documentclass[]{emulateapj}
\usepackage{natbib,tabularx}
\usepackage[english]{babel}
\usepackage{amsmath}
\usepackage[colorlinks=true,linkcolor=blue,citecolor=blue]{hyperref}%

\usepackage[caption=false]{subfig}
\usepackage{ amssymb }
\shorttitle{Optical Transients}
\shortauthors{Villar et al.}

\newcommand{\appropto}{\mathrel{\vcenter{
  \offinterlineskip\halign{\hfil$##$\cr
    \propto\cr\noalign{\kern2pt}\sim\cr\noalign{\kern-2pt}}}}}

\begin{document}


\title{Theoretical Models of Optical Transients. I. A Broad Exploration of the Duration-Luminosity Phase Space}

\author{V. Ashley Villar\altaffilmark{1}, Edo Berger\altaffilmark{1}, Brian D. Metzger\altaffilmark{2}, James Guillochon\altaffilmark{1} }
\altaffiltext{1}{Harvard-Smithsonian Center for Astrophysics, 60 Garden Street,
Cambridge, MA 02138, USA}
\altaffiltext{2}{Columbia University, New York, NY 10027, USA}
\email{vvillar@cfa.harvard.edu}

\begin{abstract}
The duration-luminosity phase space of optical transients is used, mostly heuristically, to compare various classes of transient events, to explore the origin of new transients, and to influence optical survey observing strategies. For example, several observational searches have been guided by intriguing voids and gaps in this phase space. However we should ask: Do we expect to find transients in these voids given our understanding of the various heating sources operating in astrophysical transients? In this work, we explore a broad range of theoretical models and empirical relations to generate optical light curves and to populate the duration-luminosity phase space (DLPS). We explore transients powered by adiabatic expansion, radioactive decay, magnetar spin-down, and circumstellar interaction. For each heating source, we provide a concise summary of the basic physical processes, a physically motivated choice of model parameter ranges, an overall summary of the resulting light curves and their the occupied range in the DLPS, and how the various model input parameters affect the light curves. We specifically explore the key voids discussed in the literature: the intermediate luminosity gap between classical novae and supernovae, and short-duration transients ($\lesssim 10$ days). We find that few physical models lead to transients that occupy these voids.  Moreover, we find that only relativistic expansion can produce fast and luminous transients, while for all other heating sources, events with durations $\lesssim 10$ days are dim ($M_\mathrm{R}\gtrsim -15$ mag). Finally, we explore the detection potential of optical surveys (e.g., LSST) in the DLPS and quantify the notion that short-duration and dim transients are exponentially more difficult to discover in untargeted surveys.
\end{abstract}


\section{Introduction}

The initial classification of astronomical transient sources is phenomenological by necessity. Focusing on optical light curves as an example, one can extract a number of salient features including durations, colors, peak luminosities and rise/decline times (see e.g., \citealt{richards2011machine}). The hope is that unique physical classes will ultimately become distinguishable in this multidimensional feature space without extensive photometric and spectroscopic datasets, leading in turn to physical insight about the underlying heating sources. Furthermore, the underlying distribution of objects within this multidimensional feature space can guide the design and optimization of future optical surveys. As larger optical surveys (such as the Zwicky Transient Facility, ZTF, and the Large Synoptic Survey Telescope, LSST) come online, understanding the distribution of transients within this space is essential for classification of optical light curves. There has been a number of works devoted to the computational and algorithmic problems of discovering and classifying transients from such surveys \citep{bailey2007find,karpenka2012simple,2016ApJS..225...31L,2017ApJ...837L..28C}. However, little work has been done on the {\it expected} distribution of astrophysical transients within this feature space utilizing well-motivated physical models.

This work focuses on a useful subspace of the full feature space of optical transients: the duration-luminosity phase space (DLPS). The DLPS is valuable to astronomers for a number of reasons. Both duration and peak magnitude are easily measured from the light curve, and optical transients span a wide range in both properties. Additionally, when coupled with survey parameters and progenitor properties, the DLPS can be used to measure expected observational rates of each transient class or to tune survey parameters to optimize the detectability of specific classes. 

To date, the DLPS has been used mainly to collate transients after they have been observed and to illuminate ``voids" in the observed DLPS. Of particular interest are two known voids: objects with short duration ($\lesssim10$ days) durations and objects which lie in the luminosity gap between classical novae and supernovae ($M_\mathrm{R}\sim -10$ to $-14$ mag). The latter has been noted in the literature as early as the 1930s \citep{baade1938absolute} and has been the focus of some observational searches (e.g.,\citealt{kasliwal2012bridging}). However, it also essential to explore the DLPS using theoretical models that couple various heating sources with expected ranges of physical parameters relevant for each model in order to to understand if the voids in the DLPS can, in principle, be heavily occupied (e.g., \citealt{berger2013search}). 

Here we aim to take the first step of exploring the full extent of the DLPS using physically-motivated models and input parameter ranges, as well as a uniform framework for generating the models and populating the DLPS. We review a broad range of heating sources for optical transients and generate $R$-band light curves for each physical class. We then use these light curves to produce the DLPS, and we explore the overlap of classes with observed transients.  We address the question of whether these voids appear to be occupied by theoretical models with reasonable ranges of physical parameters.

The layout of the paper is as follows.  In Section \ref{sec:mathematics}, we introduce the mathematical framework for our one-zone models. In Section \ref{sec:specific_engines}, we systematically explore a wide range of heating sources, their resulting light curves, and the ranges and trends that they follow in the DLPS. We discuss the resulting distributions and the broad discovery potential of untargeted surveys like LSST throughout the DLPS in Section \ref{sec:discussion}. We draw our primary conclusions about the DLPS, the observed voids, and the design of optical transient surveys in Section \ref{sec:conclusion}. 

In a series of follow-up papers, we will combine the insight from this work with estimates of volumetric event rates and luminosity functions to explore how different surveys (e.g., LSST) will sample the DLPS, and how these surveys can be optimized to produce scientifically valuable light curves from interesting regions of the DLPS (Villar et al.~in prep.).

\begin{figure*}[!htp]
\centering
\includegraphics[clip,width=2\columnwidth]{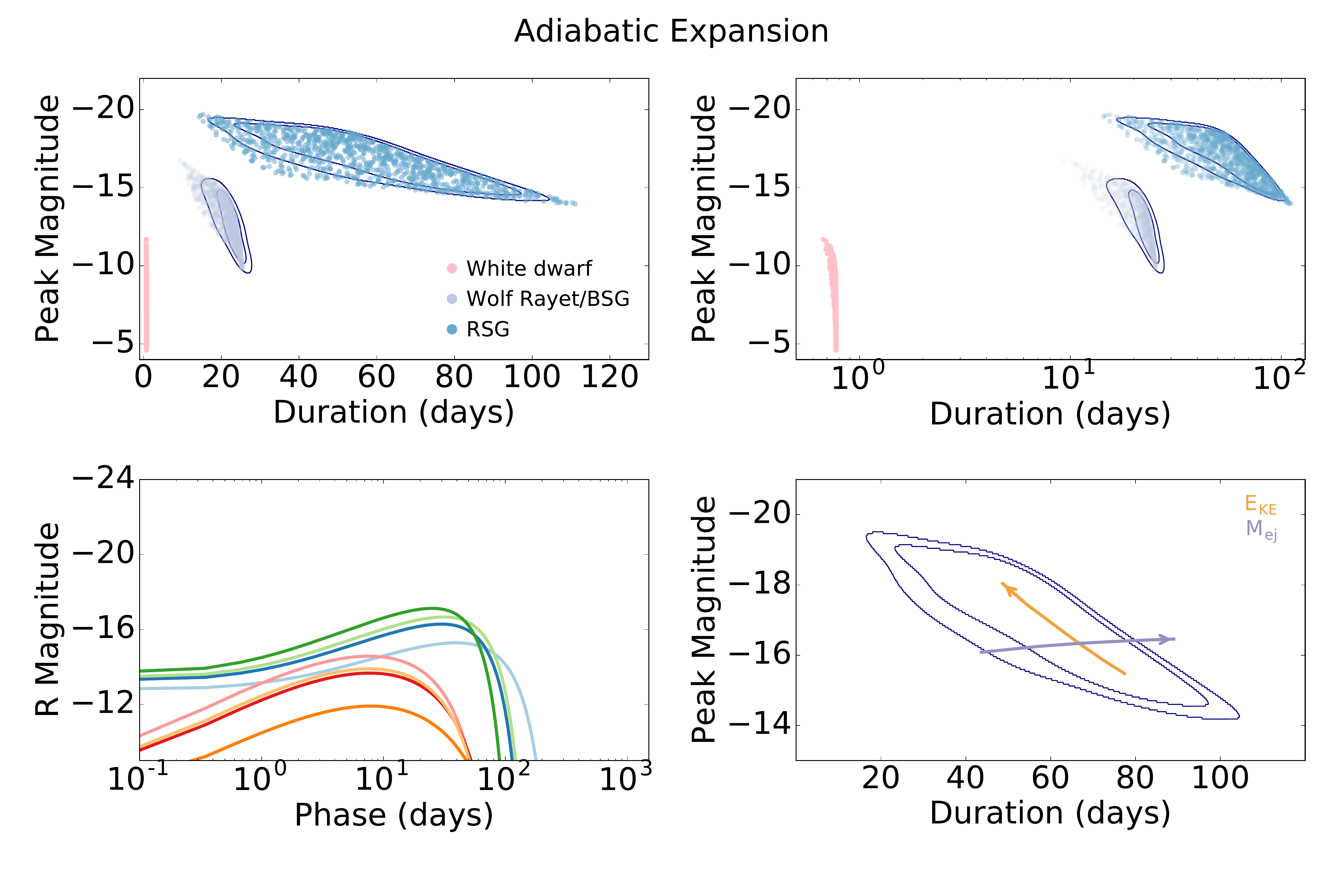}%
\caption{\textit{Top Row}: DLPS for adiabatically expanding explosions for white dwarf progenitors (pink), Wolf Rayet/BSG progenitors (purple) and RSG progenitors (blue). Also shown are 68th and 90th percentile contours for the realizations, estimated using a kernel density estimate (KDE). The WD contours are omitted for clarity. \textit{Bottom left}: Representative simulated light curves. \textit{Bottom right}: Effect of $E_\mathrm{KE}$ (orange) and $M_\mathrm{ej}$ (purple) on adiabatically expanding light curves. Arrows point towards increasing values of each parameter, with all other parameters held constant. Also shown are the contours of the simulated  light curve realizations for the RSG case.}
\label{fig:adiabatic}
\end{figure*}

\section{One-Zone Models and Mathematical Framework}\label{sec:mathematics}

Throughout this paper we use simple one-zone models of transients within the framework laid out by \cite{arnett1980analytic} for Type I SN light curves. For heating source we assume the following: 

\begin{enumerate}
    \item The ejecta are spherically symmetric and undergo homologous expansion, unless otherwise stated.
    \item Radiation pressure dominates over electron and gas pressure in the equation of state.
    \item The heating source is located in the center of the ejecta, unless otherwise stated.
    \item The optical opacity is a constant $\kappa=0.1$ cm$^2$ g$^{-1}$, unless otherwise stated. This is a typical value for stripped SNe (see e.g., \citealt{wheeler2015analysis}).
    \item The initial radius is small, unless otherwise stated.
\end{enumerate}

We can write the first law of thermodynamics as

\begin{equation}\label{eqn:thermo}
    \dot{E} = -P\dot{V} -\frac{\partial L}{\partial m} + \epsilon
\end{equation}
where $E = aT^4V$ is the specific internal energy, $P=aT^4V/3$ is the pressure, $V=1/\rho$ is the specific volume, $L$ is the radiated bolometric luminosity, $\epsilon$ is the energy generation rate of the heating source, $T$ is the temperature and $\rho$ is the density (e.g., \citealt{arnett1980analytic,chatzopoulos2012generalized}). In this framework, all available energy from the heating source supplies either the expansion of the ejecta or observable radiation, and we ignore neutrino losses. Following our assumption of homologous expansion, the radius grows as $R(t)= vt$ (assuming a negligible initial radius), where we approximate $v$ as the photospheric velocity, $v=v_\mathrm{ph}$. This assumption also means that no significant additional kinetic energy is added to the ejecta during the duration of the transient (i.e., the ejecta does not accelerate).

The sink terms in Equation 1 control the diffusion timescale of the system, which acts as a smoothing filter to the input luminosity function. During the photospheric phase at early times (i.e., the phase explored in this study), the luminosity and duration of the resulting transient depend heavily on the diffusion properties of the system. During the nebular phase at later times, the light curve of the transient should converge on the input luminosity of the heating source, assuming that the heating source is a smooth function.

The solution for generic input heat sources has been outlined in a number of works (e.g., \citealt{arnett1980analytic,kasen2010supernova,chatzopoulos2012generalized}). We cite the solutions derived by \cite{chatzopoulos2012generalized}. For the case of a homologously expanding photosphere, the output luminosity is given by:

\begin{multline}
    L_\mathrm{obs}(t) = \frac{2E_0}{t_\mathrm{d}}e^{-\left(\frac{t^2}{t_\mathrm{d}^2}+\frac{2R_0t}{vt_\mathrm{d}^2}\right)}\times \\
    \int_0^t e^{-\left(\frac{t'^2}{t_\mathrm{d}^2}+\frac{2R_0t'}{vt_\mathrm{d}^2}\right)}\left(\frac{R_0}{vt_\mathrm{d}}+\frac{t'}{t_\mathrm{d}}\right)L_\mathrm{in}(t')dt' + HS
\end{multline}
where $L_\mathrm{in}$ is the input luminosity from the central heating source, $E_0\sim M_\mathrm{ej}v^2_\mathrm{ph}/4$ is the initial energy of the transient, $R_0$ is the initial radius of the source, $t_\mathrm{d}=\sqrt{2\kappa M_\mathrm{ej}/\beta cv}$ is the diffusion timescale, and $HS$ is the homogeneous solution to Equation \ref{eqn:thermo} (the solution with no source term), which will only be considered in the case of no internal sources of heating (Section \ref{sec:adiabatic}). In most cases we consider $R_0=0$ (i.e., a small initial radius) and $\beta=\frac{4\pi^3}{9}\approx13.7$, a geometric correction factor \citep{arnett1982type}. 

In the case of transients powered by interaction of the shock wave and circumstellar material (CSM), we consider diffusion through a fixed photospheric radius. The luminosity is then described by:

\begin{equation}
    L_\mathrm{obs}(t) = \frac{2R_0}{vt_\mathrm{d}^2}e^{-\frac{2R_0}{vt_\mathrm{d}^2}t}\int_0^te^{-\frac{2R_0}{vt_\mathrm{d}^2}t'}L_\mathrm{in}(t')dt'+HS'
\end{equation}
where $HS'$ is the homogeneous solution to Equation 1 for the fixed photosphere conditions. Again, we will neglect this term for the case of transients powered by ejecta-CSM interaction.

To generate light curves from these models, we use the open-source program {\tt MOSFiT}\footnote{https://github.com/guillochon/mosfit} (Modular Open-Source Fitter for Transients v0.7.1). {\tt MOSFiT} is a Python-based package that generates and fits semi-analytical models of various transients using modular scripts for different input heating sources, diffusion methods, template spectral energy distributions (SEDs), fitting routines, etc. (Guillochon et al.~in prep.). We generate thousands of model light curves by sampling uniformly over reasonable parameter spaces for the various models (see Table \ref{tab:models} and Section 3). The \texttt{MOSFiT} modules used to generate the models are listed in Table \ref{tab:mosfit}. In all cases we assume a blackbody SED, and we report the properties of the $R$-band light curves at redshift $z=0$ with no reddening. Our blackbody assumption is a reasonable approximation when broadly exploring the DLPS rather than detailed modeling for individual sources. We are specifically interested in the ``first-order" properties of the optical light curves: the peak absolute magnitude and the duration. We define the duration as the timescale for the light curve to rise and decline by one magnitude relative to the peak. If the light curve is multi-peaked, with secondary peaks within 1 mag of the maximum luminosity, we include the secondary peak in the duration. We select $R$-band because it samples the mid-range of the optical wavelength regime. We stress that the choice of filter does not have a significant effect on our results, although in general the durations may be slightly shorter in bluer filters, and slightly longer in redder filters due to cooling of the blackbody SED as a function of time.  We also note that at substantial redshifts the duration will be stretched by a factor of $1+z$. However, given the resulting peak luminosities, most transients are expected to be detected at modest redshifts, and therefore time dilation will not be a significant factor compared to the ranges of physical parameters we consider in this work.

\section{Specific Engine Sources}\label{sec:specific_engines}

In this section we investigate various heating sources for optical transients and systematically explore the resulting light curves and the regions they occupy in the DLPS. In most cases, the free parameters of each class can be divided into two categories: those which contribute to the sink terms of Equation 1 (e.g., $v$, $M_\mathrm{ej}$) and those which contribute to the source term (e.g., $M_\mathrm{Ni}$, $P_\mathrm{spin}$, $B$). The parameters explored for each class, their ranges, and their sampling method (linearly or logarithmically spaced) are listed in Table \ref{tab:models}. 

In each subsection, we introduce the basic physics and free parameters of each heating source. We describe our choice of parameter ranges and the effect of each parameter on the light curves. We then present the simulated DLPS, specifically highlighting boundaries or interesting features. Finally, we compare our simulated DLPS with observed objects when possible.

\subsection{Adiabatic Expansion (No Central Heating Source)}\label{sec:adiabatic}
Without a heating source, the light curves are entirely defined by the homogeneous solution to Equation 2 for an expanding photosphere:

\begin{equation}
    L(t) = L_0 e^{-\left(\frac{t^2}{t_\mathrm{d}^2} + \frac{2R_0t}{v_\mathrm{ej}t_\mathrm{d}}\right)}
\end{equation}
where $R_0$ is the progenitor radius and $L_0\approx E_\mathrm{KE}/2t_\mathrm{d}$ is the initial luminosity from the explosion. Although the bolometric luminosity is monotonically decreasing, the $R$-band light curve rises and then declines as the photosphere expands, and the peak of the blackbody SED evolves from shorter to longer wavelengths through the optical regime. 

Rather than varying the progenitor radius across several orders of magnitude, we focus on three specific regimes that sample the full range of reasonable scenarios: white dwarfs ($R_0\sim0.01$ R$_\odot$, $M_\mathrm{ej}\sim0.1-1$ M$_\odot$), Wolf-Rayet/blue supergiant (BSG)-like stars ($R_0\sim10$ R$_\odot$, $M_\mathrm{ej}\sim1-10$ M$_\odot$) and red supergiant (RSG)-like stars ($R_0\sim500$ R$_\odot$, $M_\mathrm{ej}\sim1-10$ M$_\odot$). Luminous blue variables (LBVs), known for their eruptive mass loss events, have radii intermediate between the BSG and RSG progenitors, $R_0\sim10-100$ R$_\odot$.

In Figure \ref{fig:adiabatic}, we plot a sample of simulated light curves and the DLPS of each progenitor type, randomly sampling from uniform distributions of ejecta mass and logarithmically in kinetic energy ($E_\mathrm{KE}\sim10^{49}-10^{51}$ erg). As expected, we find the general trend that larger progenitors produce longer-duration and more luminous transients. The upper and lower bounds to the quadrilateral-like areas each of these models occupy are defined by our chosen energy limits, while the vertical (duration) boundaries are set by our chosen ejecta mass limits. Only compact (WD) progenitors produce transients with durations $t_\mathrm{dur}\lesssim1$ day all of which have a low luminosity ($M_\mathrm{R}\gtrsim-12$ mag). In contrast, the larger progenitors (BSG/RSG) only produce transients with longer durations ($t_\mathrm{dur}\gtrsim10$ days) which are brighter ($M_\mathrm{R}\gtrsim-10$ mag). Thus, there is an overall clear positive correlation between luminosity and duration for this type of explosions.

The effects of the main parameters ($E_{\rm KE}$ and $M_{\rm ej}$) are explicitly shown in Figure \ref{fig:adiabatic}. For a constant $E_\mathrm{KE}$, the transients become longer and somewhat brighter with increasing $M_\mathrm{ej}$. On the other hand, transients become shorter and brighter for increasing values of $E_\mathrm{KE}$ given a constant value of $M_\mathrm{ej}$. Therefore, the brightest (dimmest) transients with no central heating have large (small) kinetic energies and small (large) ejecta masses for a fixed value of $R_0$. We find that these trends are generally true in all other heating sources as well.

\begin{figure}[!htp]
\centering
  \includegraphics[clip,width=\columnwidth]{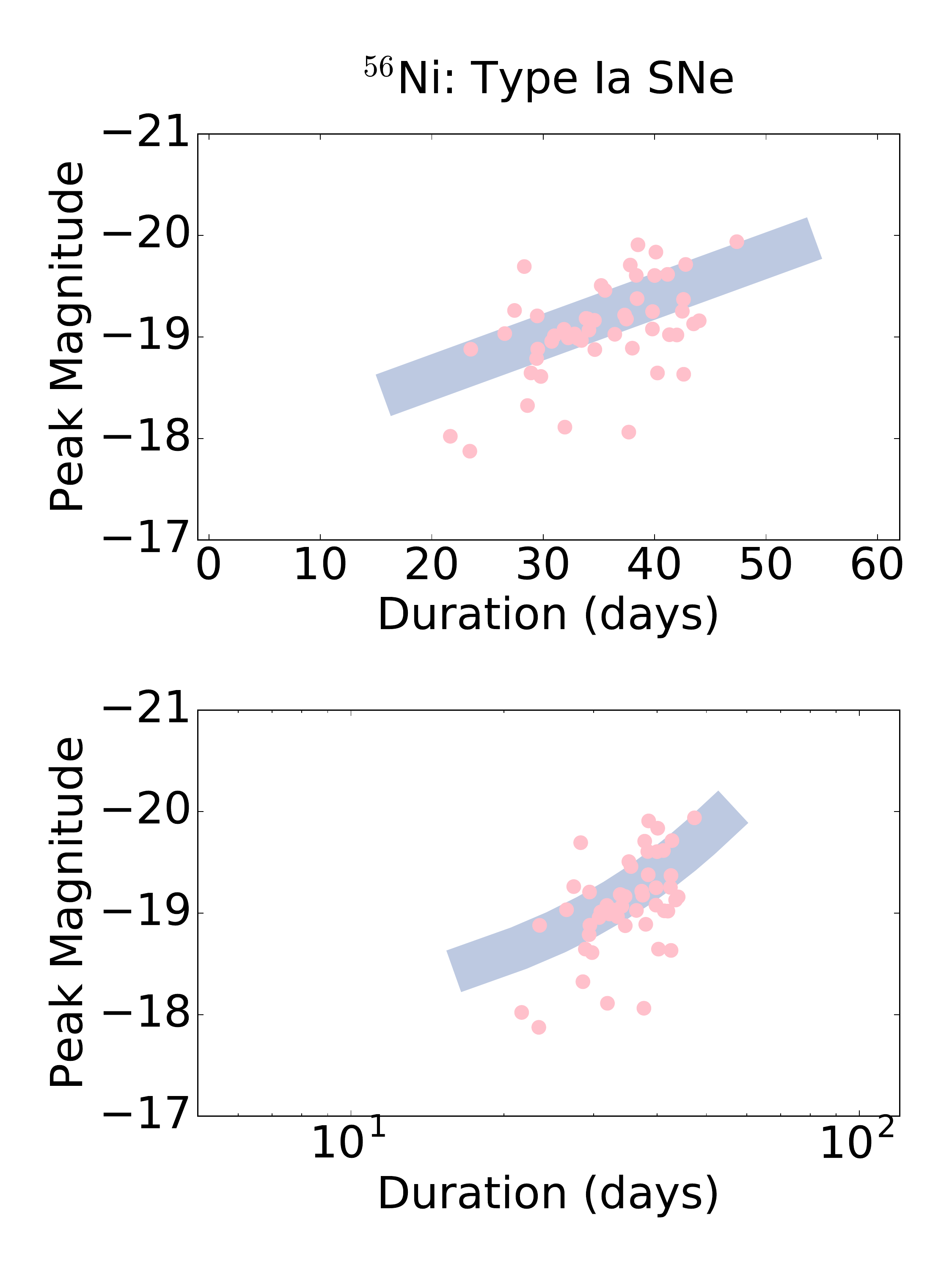}%
\caption{Derived duration-luminosity relation for Type Ia SNe using Equations 6 and 7, assuming a luminosity scatter of $\approx0.2$ mag (purple) with a sample of 50 well-sampled Type Ia SNe from the OSC after correcting for reddening and time dilation. This sample contains the 50 objects with redshifts $z\geq0.1$ (to avoid distance-dominated errors in nearby SNe) with the greatest number of data points on the OSC at the time of writing. Objects from \cite{1999AJ....117..707R,2002AJ....124.2100S,2003A&A...397..115V,2006AJ....131..527J,2007MNRAS.377.1531P,2009ApJ...700..331H,2010AJ....139..519C,2010ApJS..190..418G,silverman2011fourteen,2012ApJS..200...12H,2012MNRAS.425.1789S,2015ApJS..219...13W,firth2015rising,2016yCat..35920040F}.}
\label{fig:ia_objects}
\end{figure}

\subsection{Radioactive Heating from Decay of $^{56}$Ni}\label{sec:radidecay}
One of the most important and well-studied heating sources, responsible for the bulk of thermonuclear and stripped-envelope core-collapse SNe, is the radioactive decay of $^{56}$Ni (and $^{56}$Co) synthesized in the explosion (see \citealt{arnett1979theory,arnett1980analytic,arnett1982type,chatzopoulos2012generalized}, etc). The input luminosity is given by

\begin{equation}
    L_\mathrm{in}(t) = M_\mathrm{Ni}\left[\epsilon_\mathrm{Co} e^{-t/\tau_\mathrm{Co}} + \left(\epsilon_\mathrm{Ni} - \epsilon_\mathrm{Co}\right)e^{-t/\tau_\mathrm{Ni}}\right]
\end{equation}
where $M_\mathrm{Ni}$ (the initial nickel mass) is the only free parameter of this heating source. The energy generation rates of $^{56}$Ni and $^{56}$Co ($\epsilon_\mathrm{Ni}=3.9\times10^{10}$ erg s$^{-1}$ g$^{-1}$ and $\epsilon_\mathrm{Co}=6.8\times10^9$ erg s$^{-1}$ g$^{-1}$) and the decay rates ($\tau_\mathrm{Ni} = 8.8$ days and $\tau_\mathrm{Co} = 111$ days) are known constants. 

The radioactive decay of $^{56}$Ni powers objects spanning a broad range of properties. We explore four regimes in this work: Type Ia SNe, Type Ib/c SNe, pair-instability SNe and Iax-like SNe.

\subsubsection{Type Ia SNe}\label{sec:type_ia} White dwarfs can explode as Type Ia SNe after thermonuclear ignition \citep{hoyle1960nucleosynthesis}, although there is ongoing debate about whether this ignition arises from pure deflagration, delayed detonation or other mechanisms (see e.g., \citealt{khokhlov1991delayed,arnett1994delayed,phillips2007peculiar}). Although it is unclear whether the progenitors of Type Ia SNe are single- or double-degenerate systems, Type Ia SNe occupy a narrow range of the DLPS due to their homogeneity. 

Type Ia SNe have low ejecta masses ($M_\mathrm{ej}\approx1.4$ M$_\odot$) with a relatively large fraction of this ejecta being radioactive $^{56}$Ni ($f_\mathrm{Ni}\sim0.3-0.5$). Rather than modelling these light curves using simple blackbody SEDs, we use an empirical relation described by \cite{tripp1999determination} and \cite{betoule2014improved}:

\begin{align}
    t_\mathrm{dur} &= 35s\; \mathrm{\, days} \\
    M_{\mathrm{peak}} &= -19.4 + 1.4(s-1)
\end{align}
where $s$ is the stretch of the light curve, which we range from 0.6 to 1.2 (roughly matching the range explored in \citealt{guy2005salt}). We use the canonical ($s=1$) template from \cite{nugent2002k} to extract a template $R$-band light curve to stretch.  As a consistency check, we find that this relation roughly agrees with that found using the $R$-band templates from \cite{2006ApJ...647..501P}. 

Using this relation, we find the well-known result that brighter (dimmer) Type Ia SNe have longer (shorter) durations. Type Ia SNe are constrained to a small subset of the DLPS, with durations of $t_\mathrm{dur}\sim25-50$ days and $M_\mathrm{R}\sim-18$ to $-20$ mag. In Figure \ref{fig:ia_objects}, we plot the duration-luminosity relation and several Type Ia SNe from the Open Supernova Catalog (OSC; \citealt{guillochon2017open}; see caption for details). To summarize, Type Ia SNe occupy a narrow portion of the DLPS, with no sources having durations of $t_{\rm dur}\lesssim 20$ days.

\subsubsection{Type Ib/c SNe} Type Ib/c SNe occur when stripped-envelope massive stars undergo core-collapse at the end of their lives and are identified by their hydrogen-free (and helium-free in the case of Type Ic SNe) spectra. Type Ib/c SNe contain a relatively low fraction of $^{56}$Ni  ($f_{\mathrm{Ni}}\sim 0.01 - 0.15$) in their ejecta ($M_\mathrm{ej} \sim 1-10$ M$_\odot$; \citealt{drout2011first}). In this class, we include the parameter ranges for both normal and broad-lined Type Ib/c SNe and sample uniformly across their typical kinetic energies ($E_\mathrm{KE}\sim10^{51}-10^{52}$ erg). 

\begin{figure}[htbp]
\centering
\includegraphics[clip,width=\columnwidth]{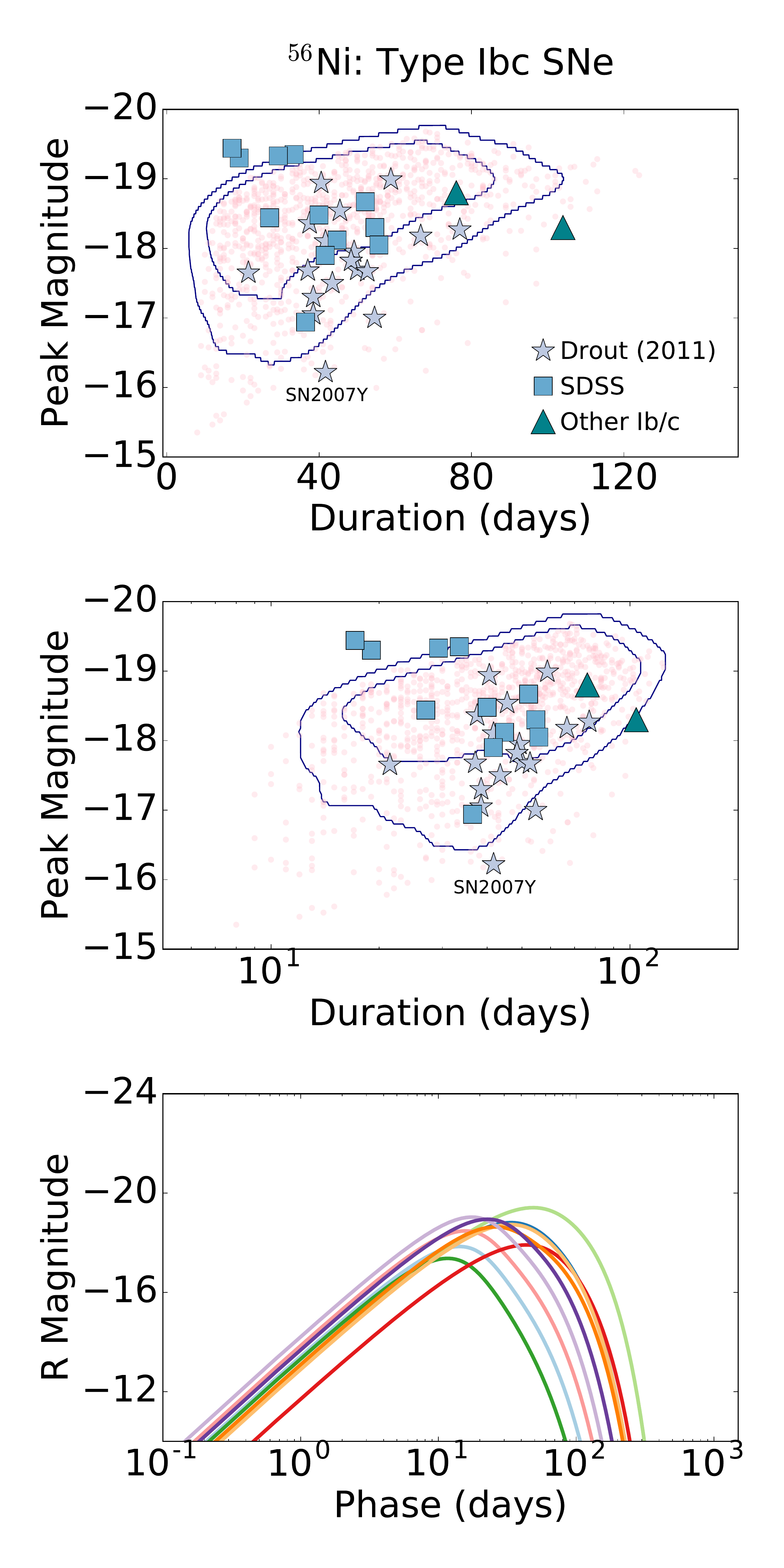}%
\caption{\textit{Top \& Middle}: DLPS of Type Ib/c SNe (pink) with a sample of observed objects from \cite{drout2011first} (purple stars) and \cite{taddia2015early} (blue squares). For both samples, we estimate the transient durations using the reported $\Delta m_{15}$ values, assuming symmetric light curves about the peak. Exceptional long-duration events: SN 2011bm \citep{valenti2012spectroscopically} and iPTF15dtg \citep{taddia2016iptf15dtg} are also plotted for comparison (green triangles). Also shown are 68th and 90th percentile contours for the realizations, estimated using a KDE. \textit{Bottom}: Representative simulated light curves for Type Ib/c SNe.}
\label{fig:ibc_objects}
\end{figure}

Our simulated light curves (Figure \ref{fig:ibc_objects}) span a wide range in both absolute magnitude ($M_R\sim -16\text{ to }-19$ mag) and duration ($t_\mathrm{dur}\sim 10-120$ days). The brightest (dimmest) transients have the largest (smallest) $^{56}$Ni masses ($f_\mathrm{Ni}M_\mathrm{ej}$), while the longest (shortest) durations are largely set by the ejecta velocity ($v\approx [E_\mathrm{KE}/M_\mathrm{ej}]^{1/2}$). Due to this positive correlation, the shortest duration transients ($t_\mathrm{dur}\sim10$) also have the lowest luminosities ($M_\mathrm{R}\sim-16$ mag). We find essentially no transients with $t_{\rm dur}\lesssim 10$ d. Such transients would require faster ejecta velocities than typically observed in these sources.

We compare our generated light curve properties to samples from the literature \citep{drout2011first,taddia2015early}. The observed sample generally overlaps our simulated DLPS, although our models extend to shorter durations of $t_\mathrm{dur}\sim10-20$ days which have not been observed.

\subsubsection{Pair-instability SNe (PISNe)} It is predicted that stars with $M\sim 140-260$ M$_\odot$ will reach sufficiently high core temperatures to produce electron-positron pairs leading to a loss of pressure and a resulting thermonuclear runaway and explosion that leaves no remnant \citep{barkat1967dynamics}. Due to the large ejecta masses and kinetic energies, the optical light curves are expected to be both bright and long-duration \citep{kasen2011pair,dessart2013radiative}. We expect PISNe to have similar (extending to slightly larger) $^{56}$Ni fractions as Type Ib/c SNe ($f_\mathrm{Ni}\sim1-30$\%) but to have much larger ejecta masses ($M_\mathrm{ej}\sim50 - 250$ M$_\odot$, with the lower masses representing stripped progenitors) and kinetic energies ($E_\mathrm{KE}\sim10^{51}-10^{53}$ erg); see \citealt{kasen2011pair}. 

\begin{figure}[htbp]
\centering
\includegraphics[clip,width=\columnwidth]{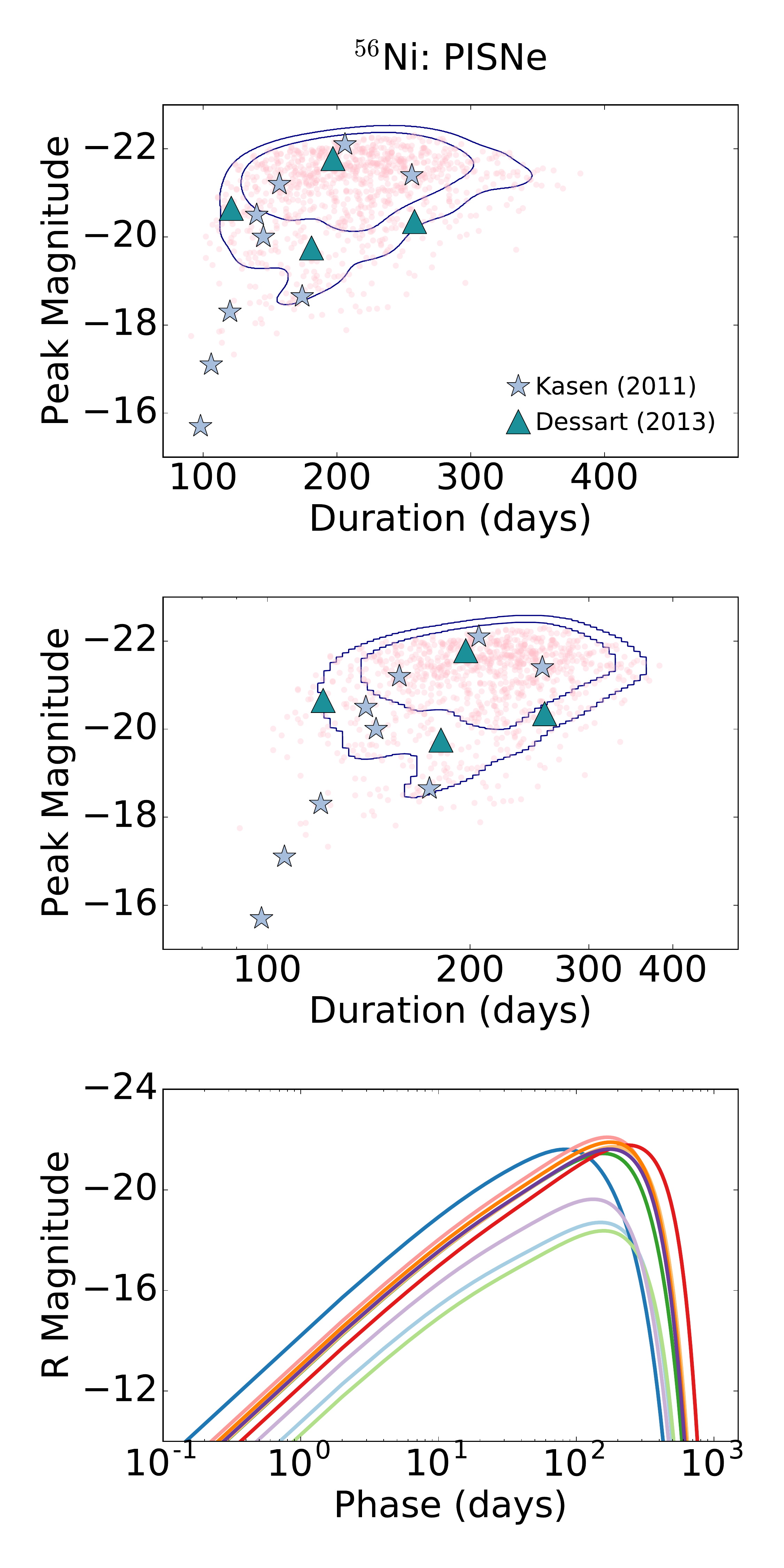}
\caption{\textit{Top \& Middle}: DLPS for PISNe (pink) with a sample of models from \cite{kasen2011pair} (purple stars) and \cite{dessart2013radiative} (green triangles). Also shown are 68th and 90th percentile contours for the realizations, estimated using a KDE. \textit{Bottom}: Representative simulated light curves for PISNe. }
\label{fig:pi_objects}
\end{figure}

In Figure \ref{fig:pi_objects}, we show a sample of simulated PISNe light curves and the associated DLPS. Compared to Type Ib/c SNe, the PISNe typically have longer durations ($t_\mathrm{dur}\sim100-400$ days) and higher luminosities ($M_\mathrm{R}\sim-18$ to $-22$ mag). Like the other $^{56}$Ni-decay powered models, the durations and luminosities are  positively correlated, with the shortest duration transients ($t_\mathrm{dur}\sim100$) being the least luminous ($M_\mathrm{R}\sim-18$ mag). 

We compare our results to more detailed calculations by \cite{kasen2011pair} and \cite{dessart2013radiative}. The first order properties of the light curves are in rough agreement (Figure \ref{fig:pi_objects}), although the \cite{kasen2011pair} models allow for less energetic/luminous explosions. By allowing energy and ejecta mass to vary independently, our model explores a larger parameter space than comprehensive PISNe models in which the progenitor masses and kinetic energies are linked.

\subsubsection{Ultra-stripped SNe/Iax-like SNe}\label{sec:iaxultra} Although ultra-stripped SNe and Iax SNe have some similar properties, they emerge from distinct physical scenarios and more detailed simulations predict unique spectral features. Ultra-stripped SNe are theorized to arise from helium star-neutron star binary systems which undergo significant stripping of the helium envelope \citep{tauris2015ultra,moriya2016light}. Iax SNe define a loose observational class which are spectroscopically similar to Type Ia SNe, although they are dimmer in optical bands \citep{foley2013type}. 

\begin{figure}[htbp]
\centering
\includegraphics[clip,width=\columnwidth]{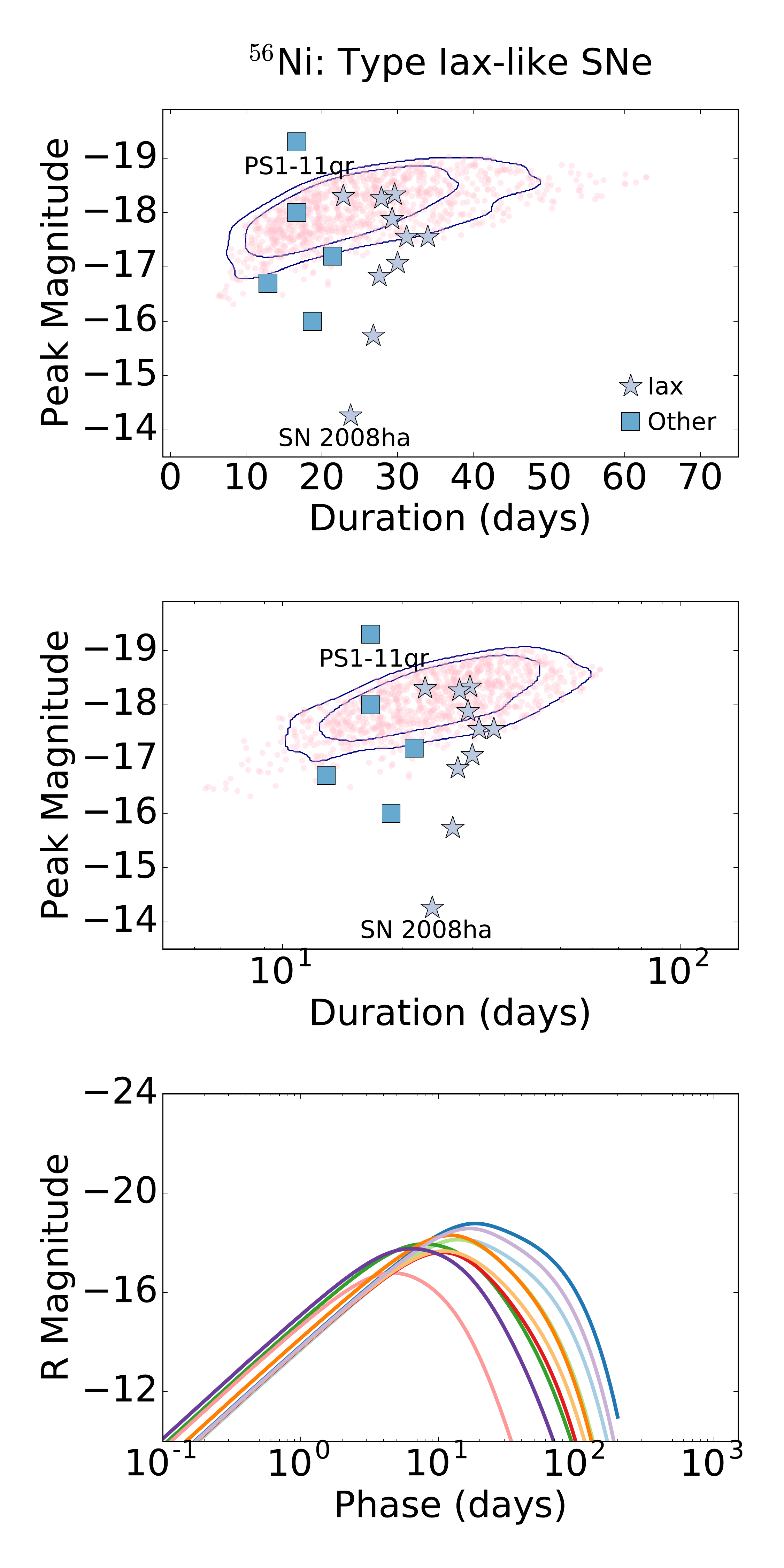}%
\caption{ \textit{Top \& Middle}: DLPS for Iax-like SNe (pink) with a sample of Iax SNe from \cite{foley2013type} (purple stars) and rapidly evolving transients from \cite{drout2014rapidly} (blue squares). Also shown are 68th and 90th percentile contours for the realizations, estimated using a KDE. Note that the \cite{drout2014rapidly} objects are not necessarily powered by $^{56}$Ni decay (see Section \ref{sec:discussion}). For both samples, we estimate the transient durations using the reported $\Delta m_{15}$ values, assuming symmetric light curves about the peak. We remove SN2008ge from the \cite{foley2013type} sample due to its highly uncertain duration. \textit{Bottom}: Representative simulated light curves for Iax-like SNe.}
\label{fig:iax_objects}
\end{figure}

Ultra-stripped SNe and Iax-like SNe have a high nickel content ($f_\mathrm{Ni}\sim0.1-0.5$) similar to Type Ia SNe but have lower ejecta masses ($M_\mathrm{ej}\sim0.01-1$ M$_{\odot}$) and kinetic energy ($E_{\rm KE}\sim10^{49}-10^{51}$ erg) compared to Type Ib/c SNe. In Figure \ref{fig:iax_objects} we present a sample of Iax-like SNe light curves and the associated DLPS. We find a tighter positive correlation between duration ($t_\mathrm{dur}\sim10-50$ days) and peak magnitudes ($M_\mathrm{R}\sim-16$ to $-19$ mag) compared to the Type Ib/c SNe due to the narrower ranges of $f_\mathrm{Ni}$ and $M_\mathrm{ej}$. Unlike the other $^{56}$Ni-decay powered models, the Iax-like model can produce a small fraction of transients with durations $t_\mathrm{dur}\lesssim10$ days (with a shortest duration of $t_\mathrm{dur}\sim1$ week), but these transients are also the dimmest ($M_\mathrm{R}\sim-16.5$ mag). 

We compare our models to the Iax SNe sample from \cite{foley2013type} and a sample of short-duration transients from \cite{drout2014rapidly} in Figure \ref{fig:iax_objects}. Our models do not account for the lowest luminosity observed objects, which likely have lower $^{56}$Ni and ejecta masses than our model ranges. Furthermore, the realizations extend to longer durations ($t_\mathrm{dur}\sim30-60$ days) than seen in current observations.

\subsubsection{General Trends}

\begin{figure}[htbp]
\centering
\subfloat[]{%
  \includegraphics[clip,width=\columnwidth]{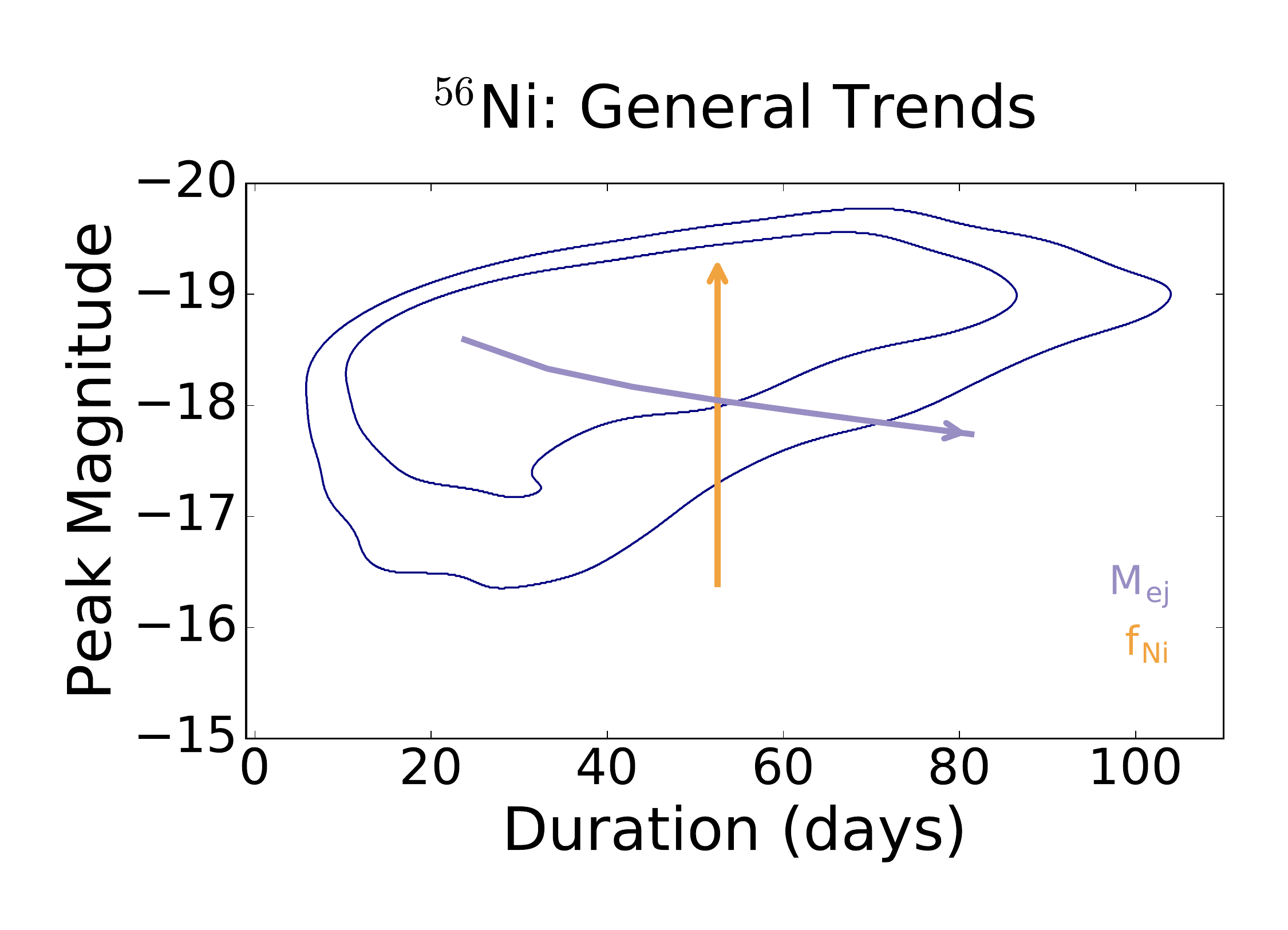}%
}
\vspace{-1cm}
\subfloat[]{%
  \includegraphics[clip,width=\columnwidth]{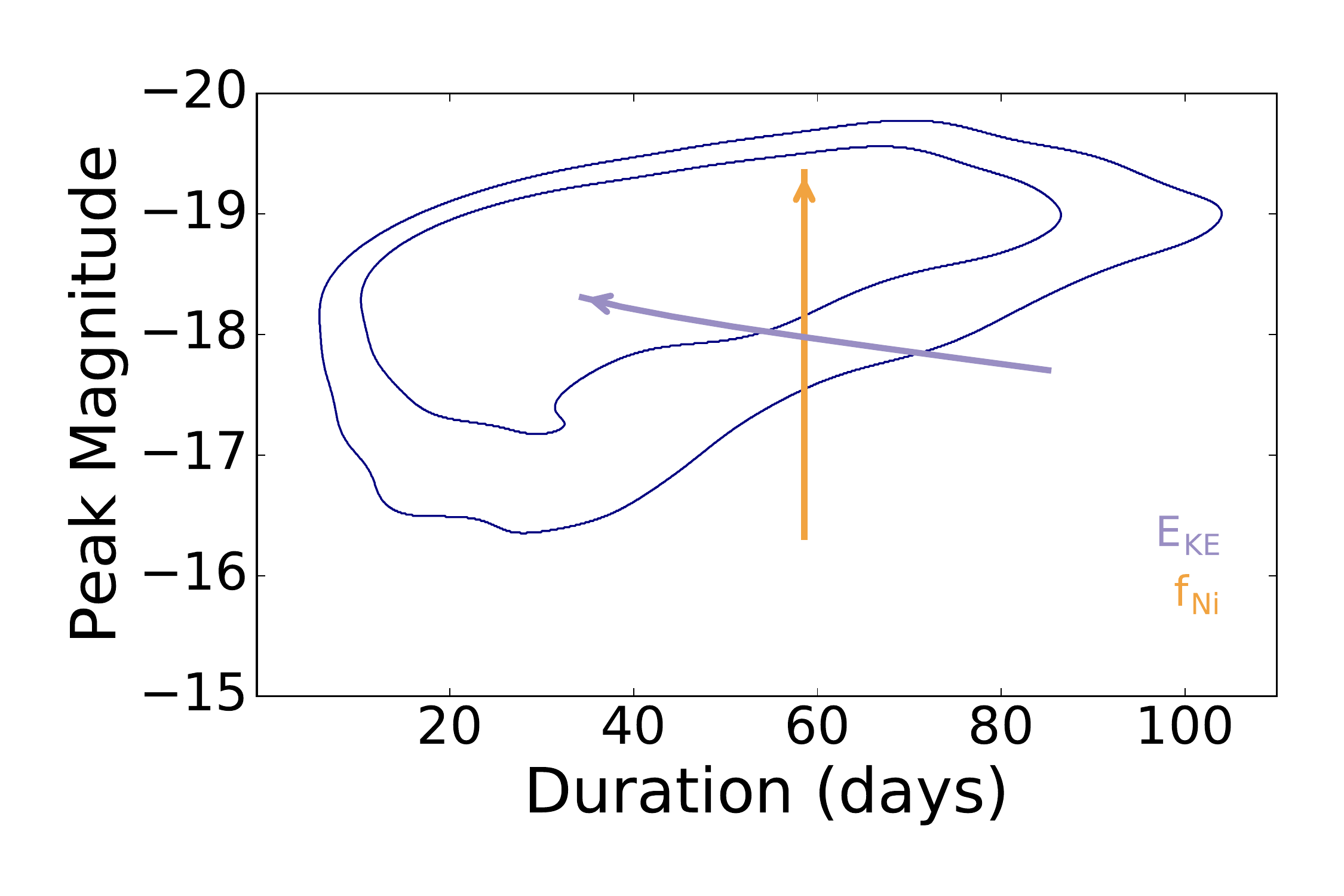}%
}
\caption{\textit{Top}: The effects of $f_\mathrm{Ni}$ (orange) and $M_\mathrm{ej}$ (purple) on light curves powered by $^{56}$Ni decay given a constant kinetic energy. Arrows points towards increasing values of each parameter. Also shown are contours of our simulated DLPS. \textit{Bottom}: Same, but for of $f_\mathrm{Ni}$ (orange) and $E_\mathrm{KE}$ (purple). }
\label{fig:sne_ibc_mni_mej}
\end{figure}

The effects of the kinetic energy, ejecta mass and nickel fraction on all $^{56}$Ni-powered models are explored in Figure \ref{fig:sne_ibc_mni_mej}. Unsurprisingly, the unique free parameter of this engine, $M_\mathrm{Ni}$,  exclusively impacts the brightness of the transient with no impact on its duration. $M_\mathrm{ej}$ and $E_\mathrm{KE}$ have degenerate and opposing effects on the light curve parameters (the same effect as in the adiabatic case, Figure \ref{fig:adiabatic}.). For a given kinetic energy, larger ejecta masses lead to longer and dimmer transients as the diffusion process becomes less efficient. For a given ejecta mass, larger kinetic energies lead to shorter and brighter transients due to the resulting larger velocities. Thus, the shortest duration transients have small ejecta masses and large kinetic energies and vice versa for the longest-duration transients. The brightest (dimmest) transients have large (small) nickel masses, corresponding to either large (small) nickel fractions or ejecta masses. We specifically find that $^{56}$Ni heating cannot power transients with durations $t_\mathrm{dur}\lesssim1$ week, unless their peak luminosities are also small, $M_\mathrm{R}\gtrsim-16.5$ mag.

\begin{figure*}[!htbp]
\centering
\includegraphics[clip,width=2\columnwidth]{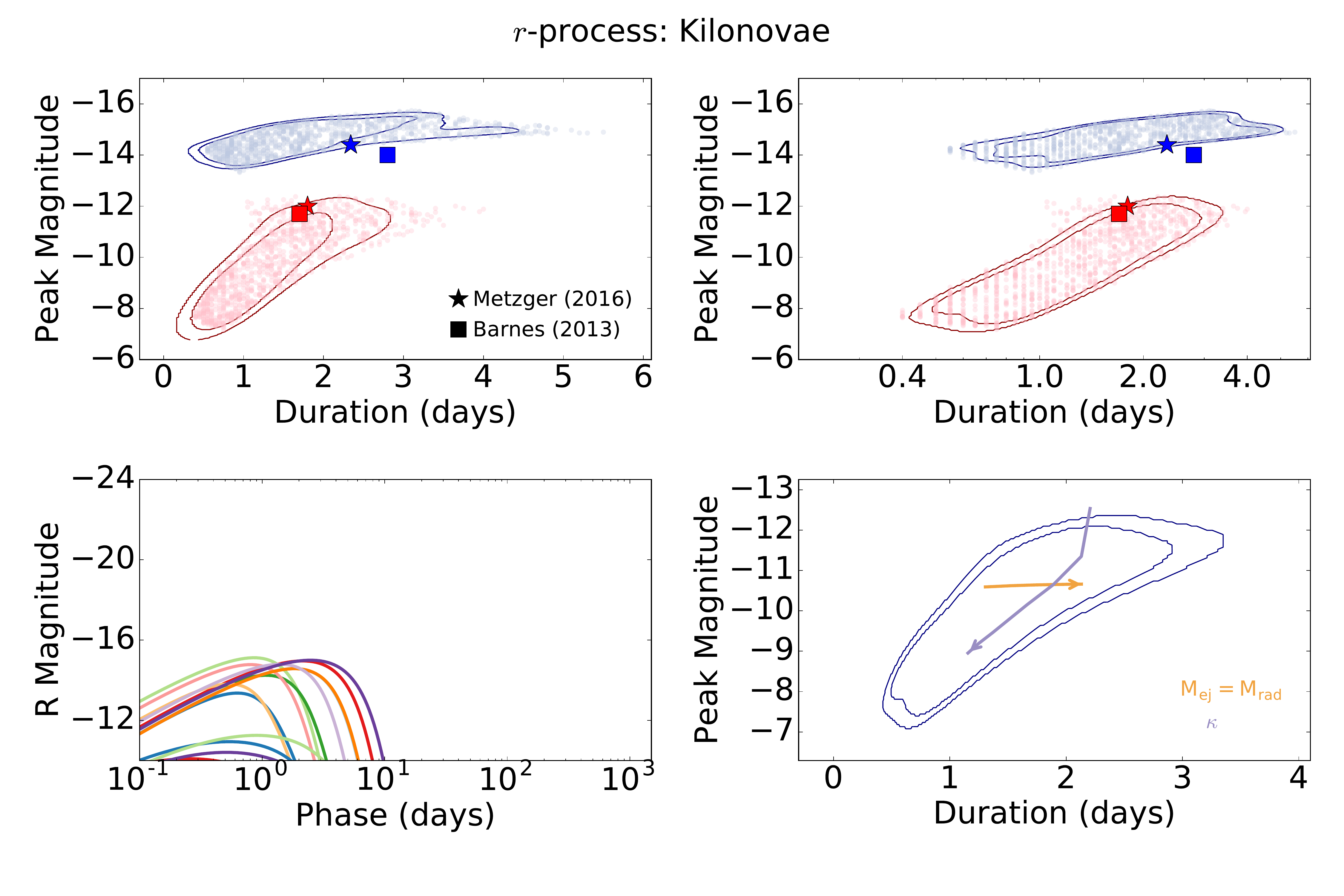}%
\caption{\textit{Top Row}: Kilonova DLPS assuming Lanthanide-rich (red) and Lanthanide-free (blue) ejecta. Also shown is a sample of detailed models from \cite{metzger2016kilonova} (stars) and \cite{barnes2013effect} (squares), and 68th and 90th percentile contours for the realizations, estimated using a KDE. \textit{Bottom Left}: Representative simulated light curves for $r$-process explosions. \textit{Bottom Right}: Effect of $M_\mathrm{ej}$ (orange) and $\kappa$ (purple) on light curves powered by $r$-process decay assuming a constant mass fraction of $r$-process material. Arrows point towards increasing values of each parameter. Also shown are contours of our simulated DLPS.}
\label{fig:kilonova}
\end{figure*}

\subsection{$r$-Process Radioactive Heating (Kilonovae)}
Neutron-rich ejecta from binary neutron star or black hole-neutron star mergers are expected to undergo $r$-process nucleosynthesis due to the neutron-rich ejecta from either the initial merger event or a remnant disk outflow \citep{li1998transient,metzger2010electromagnetic}. The radioactive decay of $r$-process products provides a heating source, while the synthesis of Lanthanides provides a high opacity \citep{barnes2013effect}. The ejecta masses are expected to be small, $M_{\rm ej}\sim 10^{-3}-0.1$ M$_\odot$ \citep{li1998transient,metzger2016kilonova}. The input luminosity can be parameterized by \citep{korobkin2012astrophysical,metzger2016kilonova}:

\begin{multline}
    L_\mathrm{in}(t) = 4\times10^{18}\epsilon_{th}(t)M_\mathrm{rp}\times\\
    \left[0.5 - \pi^{-1}\arctan\left(\frac{t-t_0}{\sigma}\right)\right]^{1.3} \text{erg s}^{-1}
\end{multline}
where $M_\mathrm{rp}$ is the mass of the $r$-process material, $t_0=1.3$ s and $\sigma=0.11$ s are constants, and $\epsilon_\mathrm{th}(t)$ is the thermalization efficiency \citep{barnes2016radioactivity,metzger2016kilonova} parameterized as:

\begin{equation}
    \epsilon_{\rm th}(t) = 0.36\left[e^{-0.56t}+\frac{\ln(1+0.34t^{0.74})}{0.34t^{0.74}}\right]
\end{equation}

Because kilonovae have not yet been conclusively observed (with the potential exception of GRB\,130603B; \citealt{berger2013r,tanvir2013kilonova}), there are a number of uncertainties in the light curve properties. Notably, the optical opacity of the Lanthanide-rich ejecta is unknown due to the complex structure of their valence f-shells. Early work assumed that Lanthanide-rich material had opacities similar to that of iron-peak elements, leading to bluer transients \citep{li1998transient}. Recent work suggests that Lanthanide-rich material will have an optical opacity $\kappa\sim 10^2-10^3$ times larger than that of iron-peak elements \citep{barnes2013effect,barnes2016radioactivity}. However, it is possible that both cases exist, if binary neutron star mergers leave a NS remnant with a survival timescale of $\gtrsim 0.1$ s \citep{kasen2015kilonova}. We consider these two possibilities in our models by generating two sets of light curves, following parameter ranges from \cite{metzger2016kilonova}: one with a fixed $\kappa=0.2$ cm$^2$ g$^{-1}$ (a ``blue" kilonova, similar to that originally explored by \citealt{li1998transient}) and one with a variable $\kappa$ sampled  logarithmically in the range $\kappa\sim10-200$ cm$^2$ g$^{-1}$ (a ``red" kilonova). For each group, we logarithmically sample from ejecta masses of $M_\mathrm{ej}\sim10^{-3}-10^{-1}$ M$_\odot$, uniformly sample from ejecta velocities of $v_\mathrm{ej}\sim0.1\text{c}-0.3$c and fix the $r$-process mass fraction $f_r=1$ \citep{metzger2016kilonova}. We additionally choose the geometric factor $\beta=3$ to calculate the diffusion timescale, following \citet{metzger2016kilonova}.

A sample of these models and their associated DLPS are shown in Figure \ref{fig:kilonova}. Both classes are dim ($M_\mathrm{R}\gtrsim-15$ mag) and short-duration ($t_\mathrm{dur}\lesssim5$ days). The red kilonovae are dimmer ($M_\mathrm{R}\sim-7$ to $-13$ mag) than the blue kilonovae ($M_\mathrm{R}\sim-13$ to $-15$ mag). Both subclasses have similar average durations of $t_\mathrm{dur}\sim2$ days, although a large fraction of the red kilonovae have even shorter durations of $t_\mathrm{dur}\lesssim1$ day. We note that although red kilonovae are expected to last $\sim 1$ week in the near-infrared, the transients are short-lived and dim in the $R$-band even if the ejecta is Lanthanide-poor. As with the $^{56}$Ni-powered models, the duration and peak luminosities are positively correlated. 

In Figure \ref{fig:kilonova} we explicitly show the effects of $\kappa$ and $M_\mathrm{ej}$ on the light curve properties. Increasing opacity makes the transient shorter and dimmer in the optical, while larger ejecta masses increase the duration and luminosity. The shortest (longest) duration transients have large (small) values of $\kappa$ and small (large) values of $M_\mathrm{ej}$. The brightest kilonovae have small opacity values, and vice versa for the dimmest kilonovae. We find that kinetic energy has a similar effect on the light curves as seen in transients lacking a central heating source (see Section \ref{sec:adiabatic} and Figure \ref{fig:adiabatic}); larger (smaller) kinetic energy leads to more (less) luminous transients with shorter (longer) durations. 

Finally, we compare our simple model with more detailed calculations from \cite{metzger2016kilonova} and \cite{barnes2013effect}. Our models are in rough agreement with the detailed calculation, both in duration and luminosity. However, our models include even dimmer kilonovae ($M_\mathrm{R}\gtrsim-11$ mag) which have low ejecta masses. We conclude that $r$-process heating in the context of compact object mergers can lead to short duration transients ($\lesssim {\rm few}$ days), but that these transients are invariably dim ($\gtrsim -15$ mag).

\begin{figure*}[htbp]
\centering
  \includegraphics[clip,width=2\columnwidth]{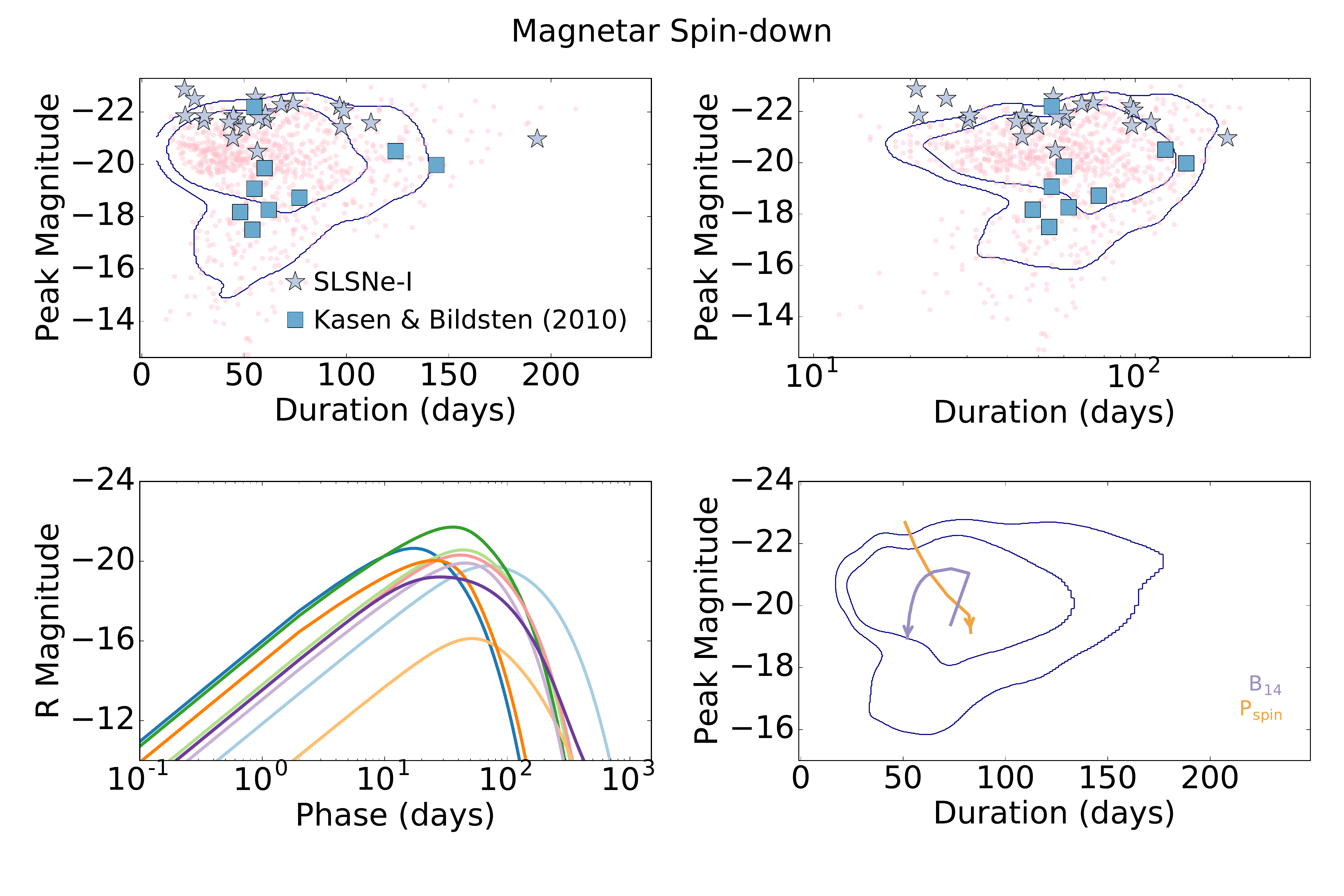}%
\caption{\textit{Top row}: DLPS of magnetar powered transients (pink) along with a sample of observed Type I SLSNe from the literature (\citealt{2010A&A...512A..70Y,2011Natur.474..487Q,2013ApJ...779...98H,2013ApJ...770..128I,2013ApJ...771...97L,2013Natur.502..346N,2014MNRAS.437..656M,2014MNRAS.444.2096N,2014ApJ...797...24V,2015ApJ...807L..18N,2015MNRAS.449.1215P,2015A&A...579A..40S,2016ApJ...831..144L,2016ApJ...826...39N,2017ApJ...835L...8N}; purple stars) and theoretical models from \citealt{kasen2010supernova} (blue squares). Also shown are 68th and 90th percentile contours for the realizations, estimated using a KDE. \textit{Bottom left}: Representative simulated light curves. \textit{Bottom right}:  Effect of $B_\mathrm{14}$ (purple) and $P_\mathrm{spin}$ (orange). Arrows point towards increasing values of each parameter, with all other parameters held constant. Also shown are contours of our simulated DLPS. }
\label{fig:magnetar}
\end{figure*}

\subsection{Magnetar Spin-down}\label{sec:magnetar}
Young magnetars, or highly magnetized neutron stars, can power optical transients as they spin down and deposit energy into the expanding ejecta \citep{woosley2010bright,kasen2010supernova,metzger2015diversity}. For a dipole field configuration, the input luminosity is given by

\begin{equation}
    L_\mathrm{in}(t) = \frac{E_\mathrm{p}}{t_\mathrm{p}} \frac{1}{(1+\frac{t}{t_\mathrm{p}})^2}
\end{equation}
where $E_\mathrm{p}=I_\mathrm{NS}\Omega^2/2$ is the initial magnetar rotational energy, described by the moment of inertia ($I_\mathrm{NS}$) and angular velocity of the neutron star ($\Omega$), and $t_p$ is the spin-down characteristic timescale:
\begin{equation}
    t_{p} = (1.3\times10^5)\frac{P_\mathrm{spin}^{-2}}{B_{14}^2} \text{ s}
\end{equation}
where $P_\mathrm{spin} =  5\left(\frac{E_\mathrm{p}}{10^{51}\mathrm{erg/s}}\right)^{-0.5}$ ms is the spin period, and $B_{14}$ is the magnetic field in units of $10^{14}$ G. Recently, the magnetar model has been used to explain Type I superluminous supernovae (SLSNe) \citep{quimby2011hydrogen,gal2012luminous,dessart2012superluminous,nicholl2013slowly,2017arXiv170600825N}. 

In this work we explore magnetar-powered transients with spin periods $P_\mathrm{spin}\sim1-10$ ms, magnetic fields $B\sim10^{13}-10^{15}$ G, ejecta masses $M_\mathrm{ej}\sim1-10$ M$_\odot$ and kinetic energies $E_\mathrm{KE}\sim10^{51}-10^{52}$ erg. These parameter ranges are designed to span realistic values where magnetar spin-down can be the dominant power source. Large spin periods of $\gtrsim 10$ ms, and low magnetic fields of $B\lesssim 10^{13}$ will result in low input power, and the transients will likely be dominated by $^{56}$Ni-decay (see Section \ref{sec:magni}). We additionally eliminate unphysical models with $E_\mathrm{KE}-E_\mathrm{SN,min}>E_\mathrm{p}$, where $E_\mathrm{SN,min}=10^{51}$ erg is the minimum energy required to leave a NS remnant. This condition removes models in which most of the rotational energy feeds into ejecta expansion rather than radiation. The magnetar-powered models and the associated DLPS are shown in Figure \ref{fig:magnetar}. While increasing spin periods lead to dimmer transients, the transient luminosity is actually optimized at intermediate values of $B_{14}$ which depend on $P_\mathrm{spin}$ when the spin-down timescale roughly matches the diffusion timescale.

There are several notable features caused by these dependencies in the magnetar DLPS. First, the paucity of long duration and low luminosity transients reflects the lower bound of our magnetic field range. In contrast, the luminosity upper limit is set by the lower bound on the spin period, which we set at the maximal NS spin (1 ms). There is also an absence of shorter duration transients with $M_\mathrm{R}\sim-18$ to $-20$ mag. The upper boundary of this void is set by our magnetic field lower limit, while the lower boundary is set by the lower ejecta mass limit; all of the transients below this void have low magnetic field strengths. The effects of $P_\mathrm{spin}$ and $B$ on the magnetar light curves are shown explicitly in Figure \ref{fig:magnetar}. We conclude that magnetar-powered transients are typically luminous ($M_\mathrm{R}\lesssim-19$ mag) with long durations ($t_\mathrm{dur}\gtrsim30$ days).

Finally, we compare our DLPS with a sample of Type I SLSNe from the literature and with detailed models by \cite{kasen2010supernova}. We find that the majority of our realizations agree with the observed population ($M_\mathrm{R}\sim-19$ to $-23$ mag and $t_\mathrm{dur}\sim20-200$ days). Additionally, we also reproduce the lower luminosity models ($M_\mathrm{R}\sim-17$ to $-19$ mag) explored by \cite{kasen2010supernova}.

\subsection{Ejecta-CSM Interaction}\label{sec:csm}

\begin{figure*}
\centering
\subfloat[]{%
  \includegraphics[clip,width=2\columnwidth]{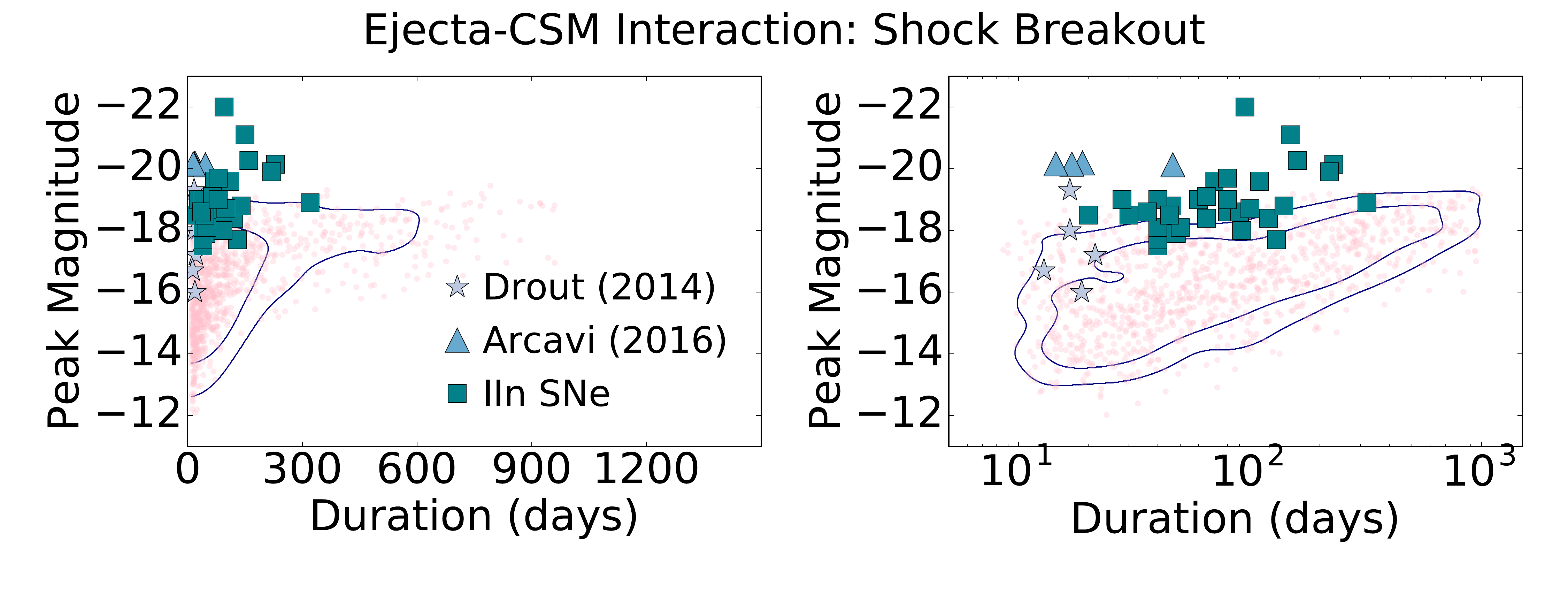}%
}
\vspace{-1cm}
\newline
\subfloat[]{%
  \includegraphics[clip,width=\columnwidth]{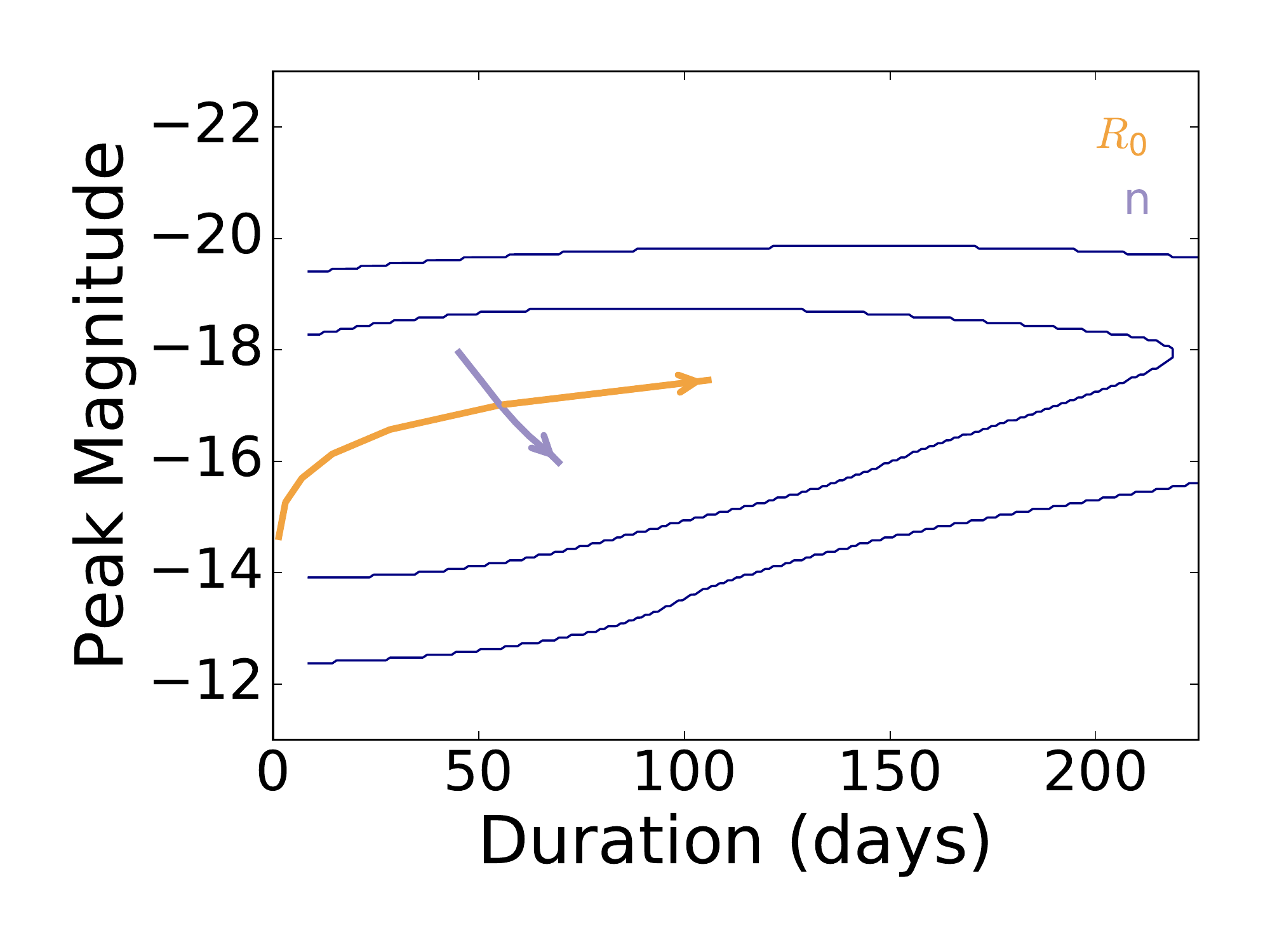}%
}
\vspace{-1cm}
\subfloat[]{%
  \includegraphics[clip,width=\columnwidth]{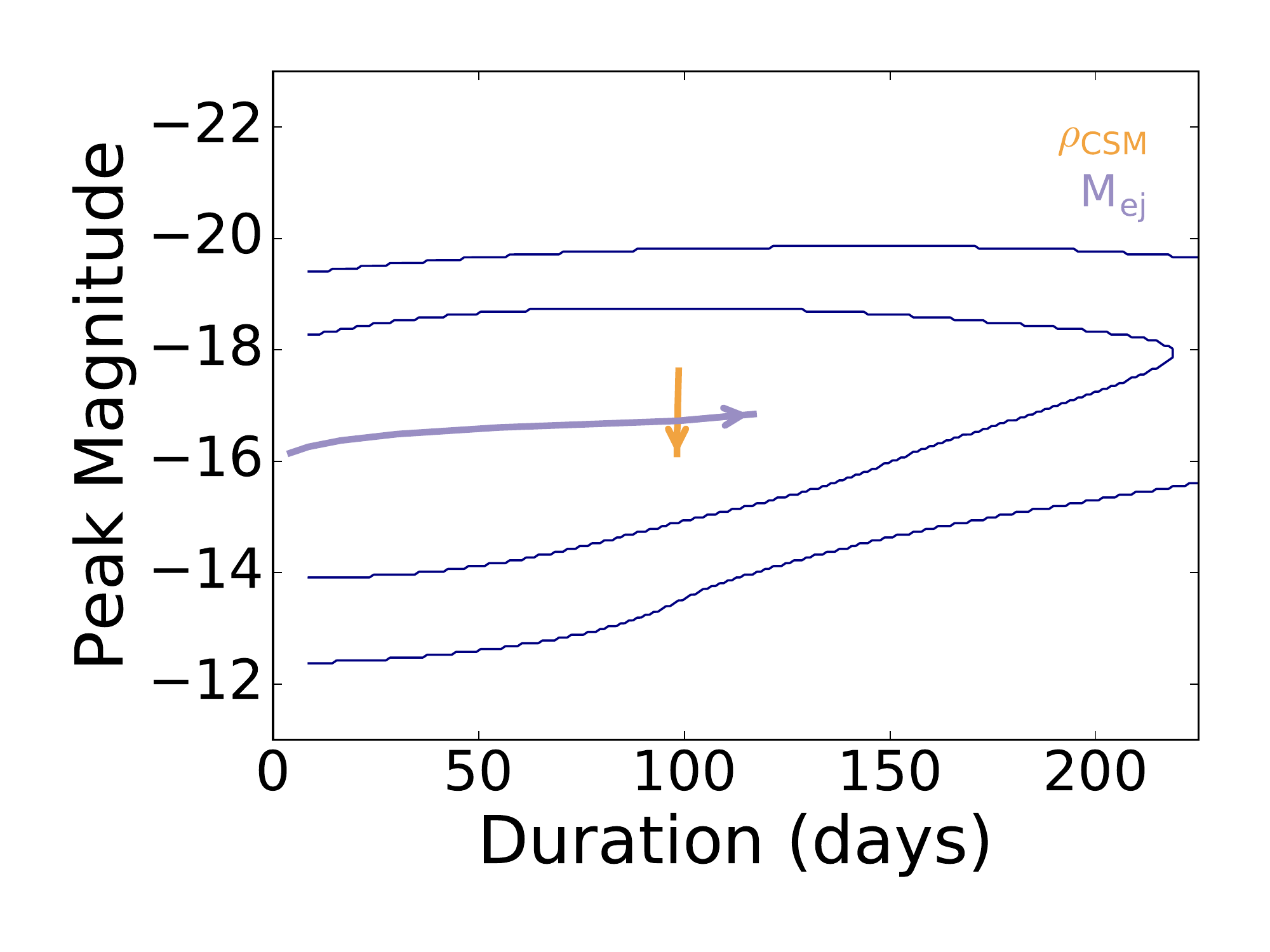}%
}
\newline
\subfloat[]{%
  \includegraphics[clip,width=\columnwidth]{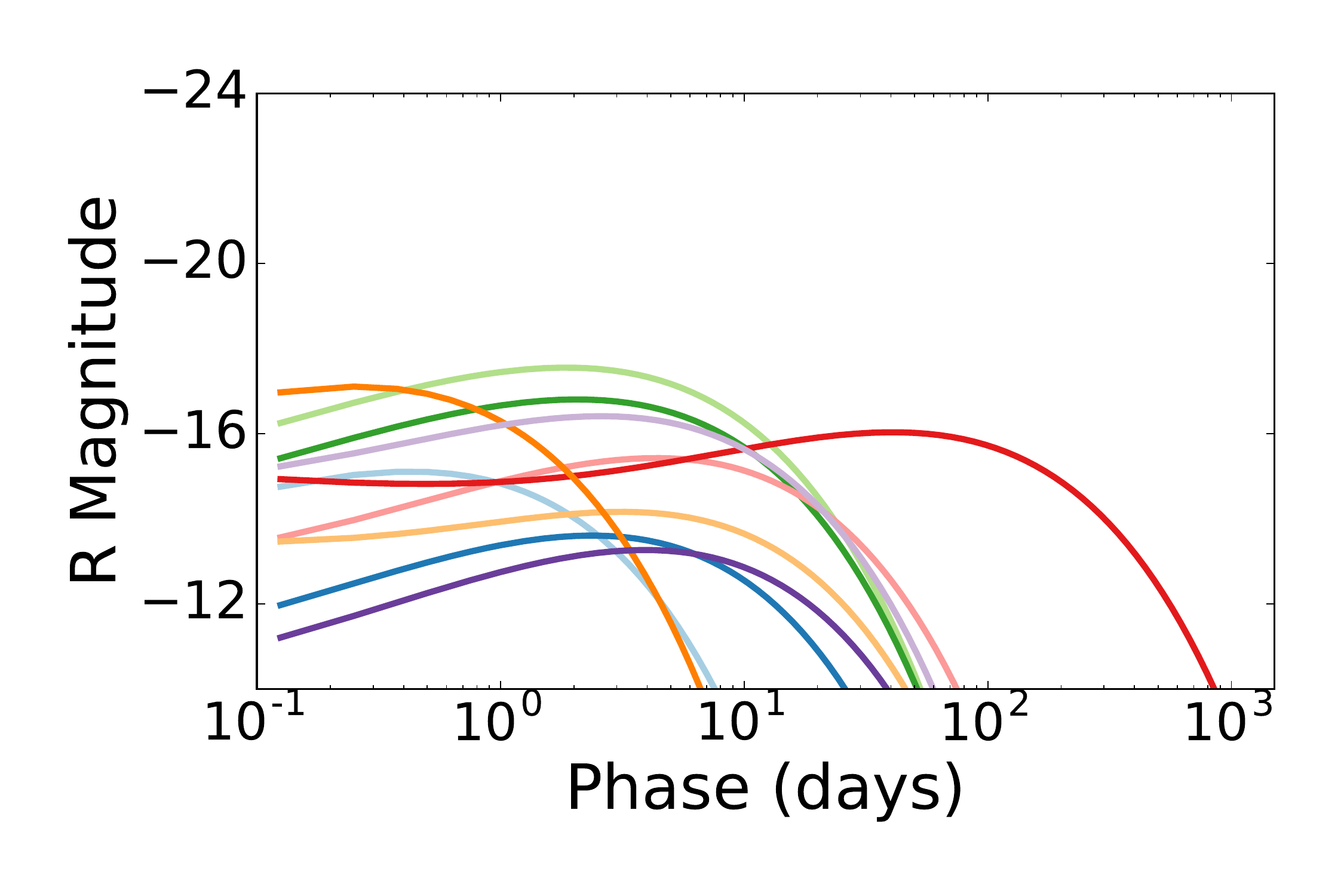}%
}
\caption{\textit{Top row}: DLPS of CSM shock breakout transients with a complete sample of V,R or I Type IIn SNe and SLSNe-II with well-sampled light curves ($>20$ data points) from the OSC, a sample of short-duration transients from \cite{drout2014rapidly} and a sample of short-duration and bright transients from \cite{arcavi2016rapidly}. Also shown are 68th and 90th percentile contours for the realizations, estimated using a KDE. Type IIn SNe light curves from \cite{taddia2013vizier,brown2014sousa,2015ATel.7207....1W,tsvetkov1993sternberg,inserra2013vizier,ofek2014precursors,smith2012sn,tartaglia2016interacting,kiewe2011caltech,smartt2015pessto,2015ATel.8325....1K,2014ATel.6700....1W,2011ApJ...729...88R,2014ATel.5875....1W,2007ApJ...666.1116S,2014AcA....64..197W,2012ATel.4689....1W,2014ATel.5975....1W,2013ATel.5663....1W,2014ATel.6494....1W,2014ATel.6861....1W}. \textit{Middle left}: Effect of $R_0$ (orange) and $n$ (purple) on light curves. \textit{Middle right}: Effect of $\rho_\mathrm{CSM}$ (orange) and $M_\mathrm{ej}$ (purple) on light curves. Arrows points towards increasing values of each parameter. Also shown are contours of our simulated DLPS. \textit{Bottom}: Representative simulated light curves.}
\label{fig:sbo}
\end{figure*}

Several types of optical transients, including Type IIn SNe and Luminous Blue Variable (LBV) outbursts, display clear signs of interaction between their ejecta and dense surrounding circumstellar material (CSM). Properties such as narrow hydrogen and metal emission lines, bright H$\alpha$ luminosities, and considerable X-ray/radio luminosities can be explained by a shock propagating through a CSM \citep{chevalier1994emission,matzner1999expulsion}. Similarly, bright, blue and short-duration transients have been linked to shock breakout from dense CSM ``cocoons" \citep{chevalier2011shock,drout2014rapidly,arcavi2016rapidly}. Because CSM interaction can describe an expansive range of transients, we consider two primary regimes: SN-like with ejecta masses ($M_\mathrm{ej}\sim1-10$ M$_\odot$) and kinetic energies ($E_\mathrm{KE}\sim10^{51}-10^{52}$ erg) typical of Type IIn SNe; and outburst-like, with low ejecta masses ($M_\mathrm{ej}\sim10^{-3}-1$ M$_\odot$) and wind-like velocities ($v\sim10^2-10^3$ km s$^{-1}$), typical of intermediate luminosity optical transients (ILOTs; including LBV outbursts and Type IIn precursors).

Many semi-analytical models have been created to describe optical light curves powered by shock heating \citep{chatzopoulos2012generalized,smith2013model,moriya2013analytic,ofek2014sn}. Most of these models follow the same formalism presented by \cite{chevalier1982self} and \cite{chevalier1994emission} and track a shock through the CSM as it thermalizes the large kinetic energy reservoir \citep{chevalier1982self,chevalier2011shock,dessart2015numerical}. Due to the current uncertainty in the analytical models available, we explore two interaction models described by \cite{chatzopoulos2012generalized} and \cite{ofek2014sn} and discuss their key differences. We specifically use the \cite{ofek2014sn} model for CSM shock breakout transients, and an altered \cite{chatzopoulos2012generalized} model for both SN-like and outburst-like transients. The details of these models are presented in the Appendix. In the subsections below, we discuss the input luminosities and model parameters.

\subsubsection{Shock Breakout from a Dense CSM}\label{sec:sbo}
Shock breakouts (SBO) from dense CSM winds surrounding massive stars have been used to describe Type IIn SNe and other bright, blue transients (e.g. see \citealt{ofek2014sn,margutti2013panchromatic}). This model assumes that the forward shock from the ejecta-CSM interaction radiates efficiently ($t_\mathrm{d}=0$) such that:

\begin{equation}
    L_\mathrm{in}=L_\mathrm{obs}=2\pi\epsilon \rho_\mathrm{CSM}(r_\mathrm{sh})r_\mathrm{sh}^2v_\mathrm{sh}^3
\end{equation}
where $\epsilon=0.5$ is an efficiency factor, $\rho_\mathrm{CSM}(r_\mathrm{sh})$ is the density of the CSM as a function of the shock radius $r_\mathrm{sh}$, and $v_\mathrm{sh}=dr_\mathrm{sh}/dt$ is the shock velocity. The shock radius and velocity depend on the geometry of the explosion ejecta and CSM. Here we assume that the CSM is distributed as a wind-like profile, $\rho_\mathrm{CSM}(r)\propto r^{-2}$. The ejecta density profile is described as a broken-power law, with an outer profile of $\rho_\mathrm{ej}(r)\propto r^{-n}$ where $n$ is a free parameter. We find that the inner profile has little effect on the light curves, and thus we assume a flat inner profile. (See the Appendix for details.) 

Thus, the free parameters are the ejecta density index ($n$), the kinetic energy of the explosion ($E_\mathrm{KE}$), the ejecta mass ($M_\mathrm{ej}$), the inner radius of the CSM ($R_0$), and the CSM density at $R_0$ ($\rho_\mathrm{CSM}$). We sample over the following ranges: $n\sim7-12$, $E_\mathrm{KE}\sim10^{51}-10^{52}$ erg,  $M_\mathrm{ej}\sim1-10$ M$_\odot$, $R_0\sim1-10^2$ AU and $\rho_\mathrm{CSM}\sim10^{-17}-10^{-14}$ g cm$^{-3}$. We then eliminate realizations with mass-loss rates $\dot{M}\equiv4\pi R_0^2\rho_\mathrm{CSM}v_\mathrm{w}<10^{-6}$ M$_\odot$ yr$^{-1}$, assuming a wind velocity of $v_\mathrm{w}\sim 10^2$ km s$^{-1}$. This cut corresponds to the lower end of expected RSG mass-loss rates \citep{smith2014mass}. Our parameters therefore correspond to mass-loss rates of $\dot{M}\sim10^{-6}-10^{-2}$ M$_\odot$ yr$^{-1}$, roughly matching the range of mass-loss rates of RSGs, YSGs and LBVs \citep{smith2014mass}.

We note that our range of $R_0$ values extends beyond the radii of most progenitor stars. However, it is possible that $R_0$ is the location of a so-called cool dense shell formed by an earlier eruption and not always representative of the progenitor radius \citep{smith2016interacting}. 

Finally, we note that our range of $n$ is representative of ejecta density profiles inferred for degenerate progenitors ($n\sim7$; \citealt{colgate1969early}) and RSG progenitors ($n\sim12$; \citealt{matzner1999expulsion}). Although previous work has typically set $n\sim12$, \cite{chevalier2011shock} suggest that the shallower portions of density profiles may play larger roles in the ejecta-CSM interaction, so we leave $n$ as a free parameter. Finally, we remove events with $v>15000$ km s$^{-1}$.

We show sample light curves and the associated DLPS in Figure \ref{fig:sbo}. Our models span a wide range in both luminosity ($M_\mathrm{R}\sim-13$ to $-19$ mag) and duration ($t_\mathrm{dur}\sim10-10^3$ days). Like most of our models, luminosity and duration are positively correlated. The shortest duration transients have $t_{\rm dur}\approx 10$ days, and peak brightness of $M_\mathrm{R}\gtrsim-17$ mag. The duration is largely determined by the mass-loss rate, with higher (lower) mass-loss rates leading to the longest (shortest) duration transients. The luminosities of the brightest transients are set by our minimum value of $n$ and maximum velocities, while the luminosities of the dimmest transients are set by the minimum velocities ($\sim4\times10^3$ km s$^{-1}$) of our parameter ranges. We explicitly show the effects of each free parameter in Figure \ref{fig:sbo}. As shown in the figure, larger values of $\rho_\mathrm{CSM}$ actually lead to \textit{less} luminous transients in the optical. This is due to the fact that large values of $\rho_\mathrm{CSM}$ lead to hotter effective temperatures that actually decrease the visible luminosity assuming a blackbody SED.

In Figure \ref{fig:sbo} we also compare our simulated distribution to the distribution of all well-sampled Type IIn SN listed on the OSC at the time of writing, a sample of short-duration transients from Pan-STARRs \citep{drout2014rapidly} and a sample of rapidly-rising, bright transients from \cite{arcavi2016rapidly}. We note that neither of the latter two samples are claimed to be from CSM SBO models; however, the CSM model is able to roughly reproduce the peak luminosities and durations of the rapid transients from \cite{drout2014rapidly}. The simulated shock breakout models generally produce dimmer transients than observed and allow for longer duration transients ($t_\mathrm{dur}\gtrsim400$ days).

\begin{figure}[htbp]
\centering
\subfloat[]{%
  \includegraphics[clip,width=\columnwidth]{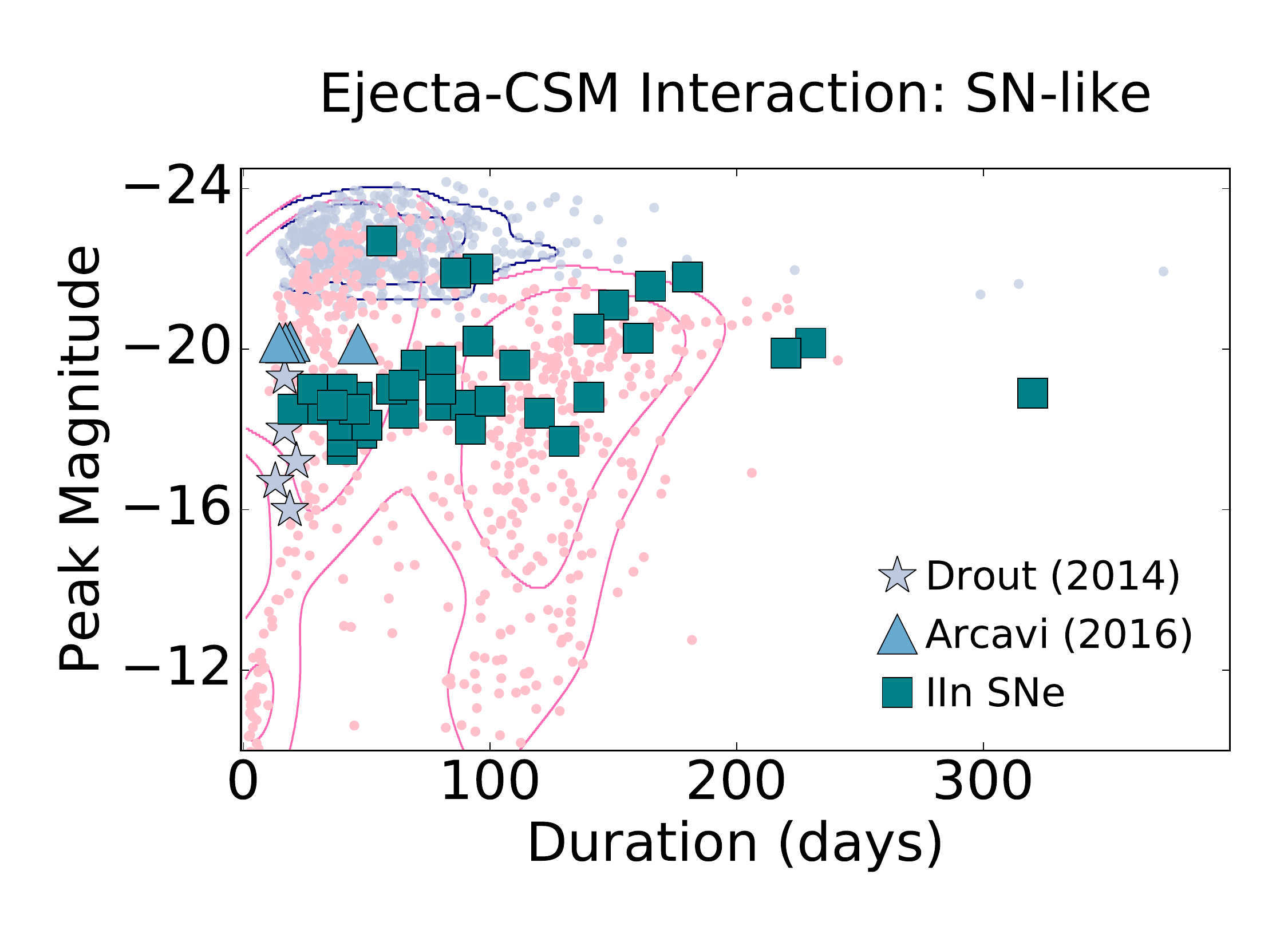}%
}
\vspace{-1.cm}
\subfloat[]{%
  \includegraphics[clip,width=\columnwidth]{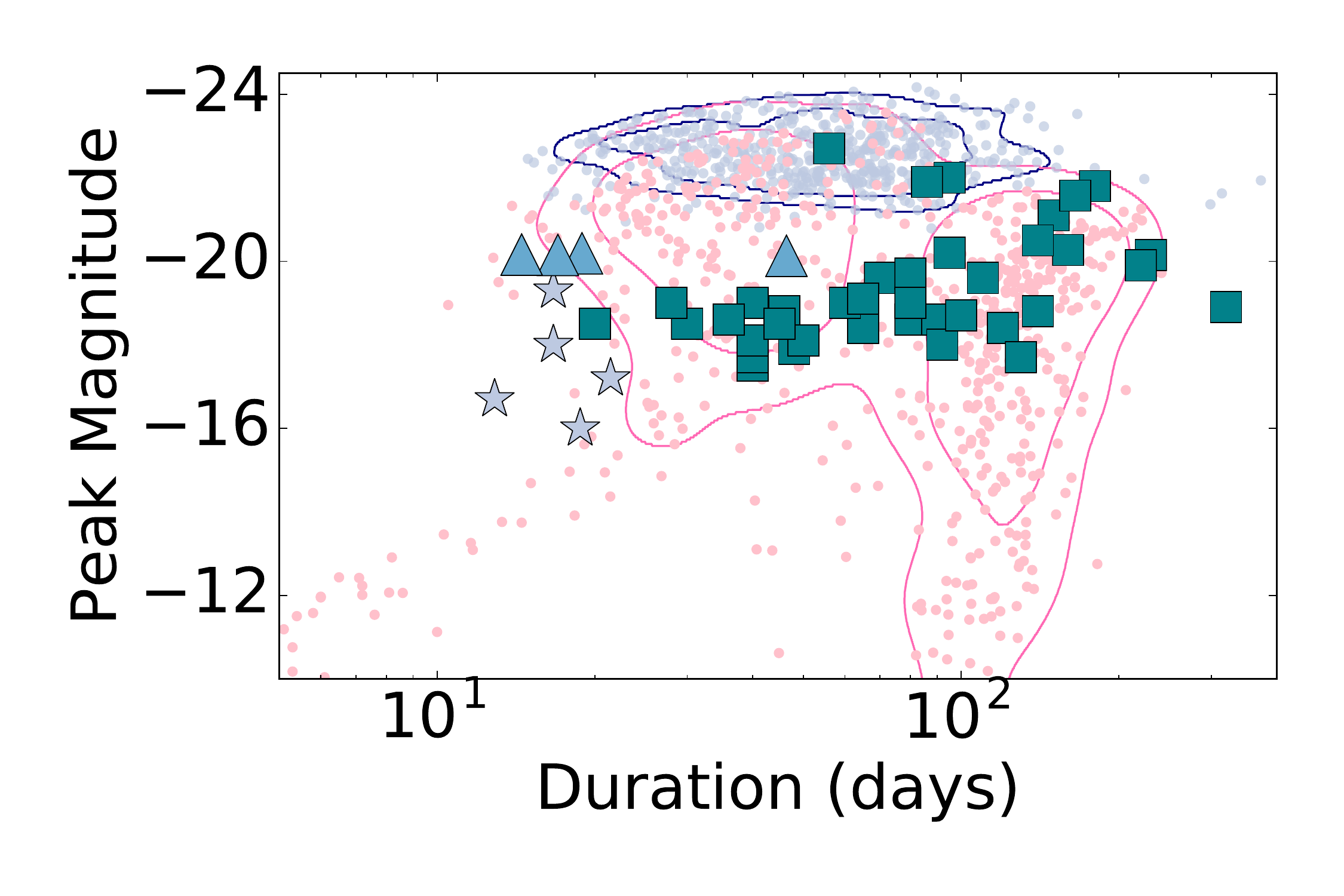}%
}
\vspace{-1.cm}
\subfloat[]{%
  \includegraphics[clip,width=\columnwidth]{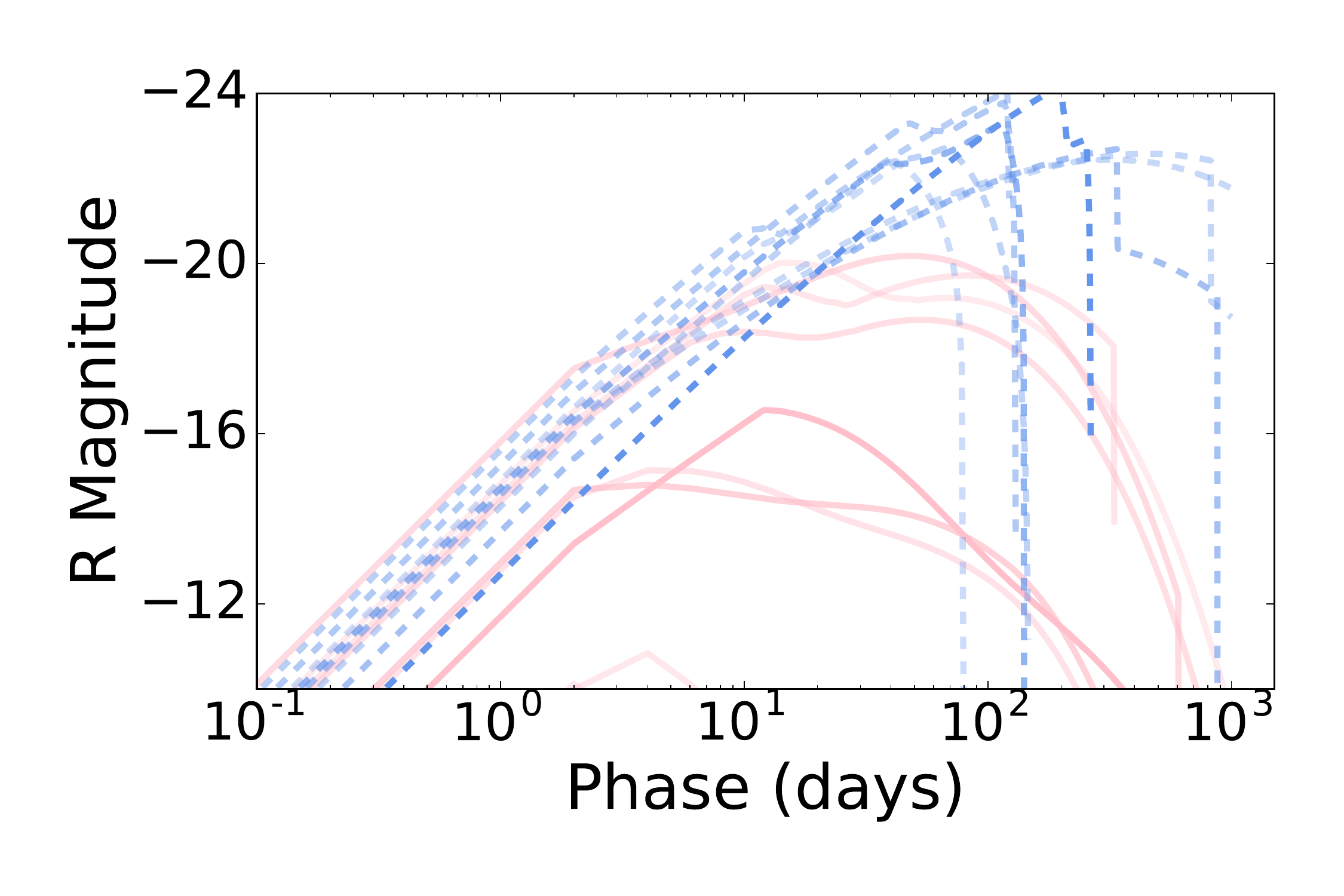}%
}
\caption{ \textit{Top \& Middle}: Ejecta-CSM interaction DLPS for SN-like transients assuming $s=0$ (blue) and $s=2$ (pink) and the 68th and 90th percentile contours for the realizations, estimated using a KDE. Also shown is the same sample of Type IIn SNe and SLSNe-II light curves from Figure \ref{fig:sbo}. \textit{Bottom}: Simulated light curves for shell-like $s=0$ (blue, dashed) and wind-like $s=2$ (pink) mass-loss.}
\label{fig:csm}
\end{figure}

\subsubsection{Ejecta-CSM Interaction with Diffusion}\label{sec:csm_diffusion}
We now explore the generalized problem of ejecta-CSM interaction with diffusion assuming a stationary photosphere \citep{chatzopoulos2012generalized}. This model is similar to that in Section \ref{sec:sbo}, but in this case we consider the reverse shock contribution and a stationary photosphere (see Appendix). The input luminosity is given by:

\begin{multline}
    L_\mathrm{in} = \epsilon(\rho_\mathrm{CSM}R_0^s)^\frac{n-5}{n-s}t^\frac{2n+6s-ns-15}{n-s}\times\\
    \Big[C_1\theta\left(t_\mathrm{FS}-t\right)+C_2\theta\left(t_\mathrm{RS}-t\right)\Big]
\end{multline}
where $s$, $C_1$ and $C_2$ are geometric parameters of the CSM, and $\theta(t)$ is the heaviside function which controls the input times for the forward ($t_\mathrm{FS}$) and reverse ($t_\mathrm{RS}$) shocks. There are seven free parameters of the model: $s$, $n$, $R_0$, $E_\mathrm{KE}$, $M_\mathrm{ej}$, $\rho_\mathrm{CSM}$, and the total CSM mass ($M_\mathrm{CSM}$). We set $s=0$ for ``shell-like" CSM models and $s=2$ for ``wind-like" CSM models.

We place a number of additional physical constraints on these models:
\begin{enumerate}
    \item We require the photospheric radius to be within the CSM shell: $R_0\leq R_\mathrm{ph}\leq R_\mathrm{CSM}$ 
    \item We require the CSM mass to be less than the ejecta mass: $M_\mathrm{CSM}\leq M_\mathrm{ej}$
    \item We require the velocity of the ejecta $v_{\mathrm{min}}\leq v_\mathrm{ph}\equiv\sqrt{10E_\mathrm{KE}/3M_\mathrm{ej}}\leq v_{\mathrm{max}}$, where $v_{\mathrm{min}}= 5000$ km s$^{-1}$ and $v_{\mathrm{max}}=15000$ km s$^{-1}$ for SN-like sources and $v_{\mathrm{min}}=100$ km s$^{-1}$ and $v_{\mathrm{max}}=1000$ km s$^{-1}$ for outburst-like sources.
    \item We require the diffusion time ($t_\mathrm{d}$) through the CSM to be less than the shock crossing time through the CSM ($t_\mathrm{FS}$). If this were \textit{not} the case, the light curve would exponentially decline as in the case of adiabatic expansion (Section \ref{sec:adiabatic}; see the shell-shocked model described by \citealt{smith2007shell}). \cite{moriya2013synthetic} and \cite{dessart2015numerical} argue that the optical depths in typical CSMs are significantly lower than the regime of a shell-shocked model, implying that $t_\mathrm{d}<t_\mathrm{FS}$. 
\end{enumerate}

Finally, we choose reasonable parameter ranges for SN- and outburst-like sources. For both subclasses, we sample logarithmically from $R_0\sim 1-100$ AU and $\rho_{\mathrm{CSM}}\sim10^{-17}-10^{-14}$ g cm$^{-3}$, typical ranges in Type IIn SNe studies \citep{moriya2013synthetic,dessart2015numerical}. For the SN-like models, we explore both shell-like ($s=0$) and wind-like ($s=2$) CSM profiles. For the outburst-like models, we only explore wind-like CSM profiles.

For SN-like transients, we sample logarithmically from $E_{\mathrm{KE}}\sim10^{51}-10^{52}$ erg and $M_{\mathrm{CSM}}\sim0.1-10$ M$_\odot$, and uniformly from $M_{\mathrm{ej}}\sim1-10$ M$_\odot$. Simulated light curves and the DLPS of our models are shown in Figure \ref{fig:csm} for both shell-like and wind-like CSM profiles. In the shell-like case, the light curves decline rapidly following peak brightness due to our use of the heaviside function to abruptly discontinue the input luminosity once the forward and reverse shocks have traversed the CSM. In the wind-like case, these light curves are smoother due to the continuous $\rho(r)\propto r^{-2}$ CSM profile.

One notable difference between the shell-like and wind-like models is the range of peak magnitudes, with shell-like models ($M_{\mathrm{peak}}\sim -21$ to $-24$ mag) spreading a narrower range than the wind models ($M_{\mathrm{peak}}\gtrsim-23$ mag) for the same range of physical parameters. This is likely due to the fact that $L_\mathrm{in}\propto\frac{2n+6s-ns-15}{n-s}$, or $L_{\mathrm{in}}\appropto t^{0.7} (t^{-0.3})$ for $s=0$ ($s=2$) assuming $n=12$, a typical value for RSGs \citep{chevalier1982self}. In other words, the input luminosity is always \textit{decreasing} in the wind-like model, while it actually \textit{increases} in the shell-like model for $t<t_{\rm FS}$. This leads to brighter transients in the shell-like case.

\begin{figure*}[htbp]
\includegraphics[width=16cm]{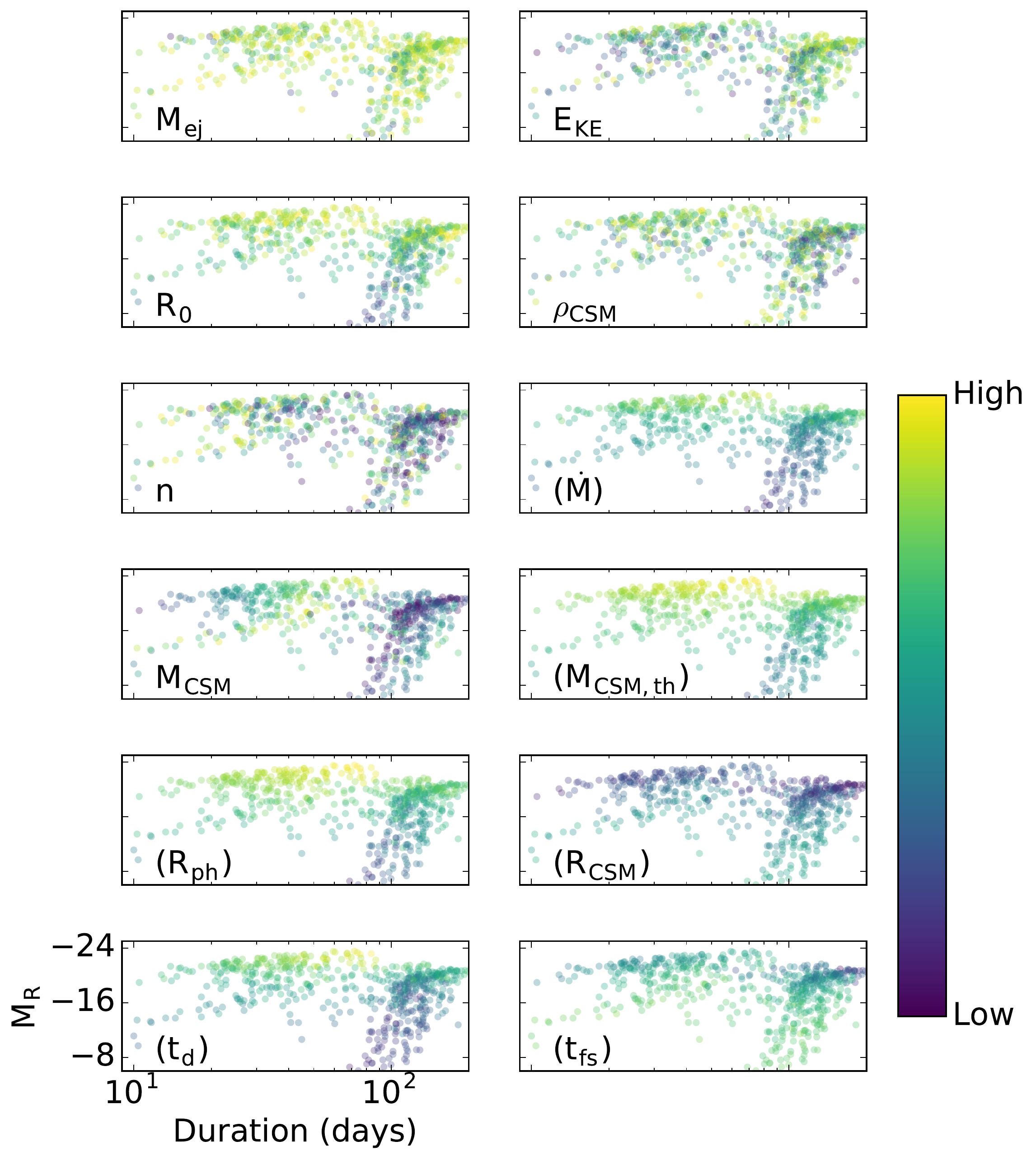}
\centering
\caption{Ejecta-CSM interaction DLPS for the Type IIn SNe transients assuming $s=2$. In each panel, the models are color coded based on the values of the inset parameters (e.g., the top left panel is color coded based on $M_\mathrm{ej}$). The ``high" and ``low" values are based on the parameter ranges listed in Table 1. The mass-loss rate ($\dot{M}$), optically thick CSM mass ($M_\mathrm{CSM,th}$), photospheric radius ($R_\mathrm{ph}$), CSM radius ($R_\mathrm{CSM}$) diffusion time ($t_\mathrm{d}$) and forward shock-crossing time ($t_\mathrm{fs}$) are all derived parameters. We note that many of the short-duration events are dominated by the forward shock (with the reverse shock contributing a less-luminous peak not included in the duration), while the long-duration events have durations that typically include both the forward- and reverse-shock peaks.}
\label{fig:csm_all_param}
\end{figure*}

There is little correlation between duration and luminosity for both shell-like and wind-like models due to the complicated effects of the multiple parameters. In Figure  \ref{fig:csm_all_param} we show how the free and derived parameters affect the wind-like CSM models. We highlight several global trends. The brightest transients typically have the largest mass-loss rates, optically thick CSM masses ($M_\mathrm{CSM,th}$) and photospheric radii ($R_\mathrm{ph}$), and vice versa for the dimmest transients. The shortest duration ($t_\mathrm{dur}\sim10$ days) transients with relatively high luminosities ($M_\mathrm{R} \lesssim -16$ mag) have small CSM masses although most of this CSM is optically thick. 

There are two low luminosity ($M_\mathrm{R}\gtrsim-15$ mag) ``branches" in the wind-like DLPS: one extending to shorter durations ($t_\mathrm{dur}<20$ days) and the other at $t_\mathrm{dur}\sim100$ days. The dearth of models between these branches is due to the fact that models within this area of phase space have optically thin CSM masses which are eliminated by our physical constraints. Realizations in the shorter-duration branch have larger CSM masses, mass-loss rates and inner CSM radii compared to the branch at $\sim 100$ days. Realizations in the shorter-duration branch have more peaked light curves due to thinner shells at larger radii, while those in the longer-duration branch have flatter light curves. We note that no transients with SN-like properties have been observed to date in either branch.

As with previous classes, we find that the transients with shortest durations and SN-like luminosities have $t_\mathrm{dur}\approx15$ days. Transients with shorter durations (down to $t_{\rm dur}\approx 15$ days) all have low luminosities of $\gtrsim-14$ mag.

Finally, we compare our DLPS to the sample of Type IIn SNe and other objects as in Section \ref{sec:sbo}. The wind-like DLPS largely overlaps with the sample, while the shell-like DLPS is only able to reproduce the brightest Type IIn SNe.

For outburst-like transients, we sample logarithmically from $v\sim 10^2-10^3$ km s$^{-1}$ and $M_{\mathrm{ej}}\sim 0.001-1$ M$_\odot$ and assume $s=2$. The corresponding kinetic energy limits are $E_{\mathrm{KE}}\sim 10^{44}-10^{49}$ erg. These limits were chosen to roughly match the velocities of LBV eruptions and explore a full range of the lowest luminosity transients \citep{humphreys1994luminous}.

\begin{figure}[htbp]
\centering
\subfloat[]{%
  \includegraphics[clip,width=\columnwidth]{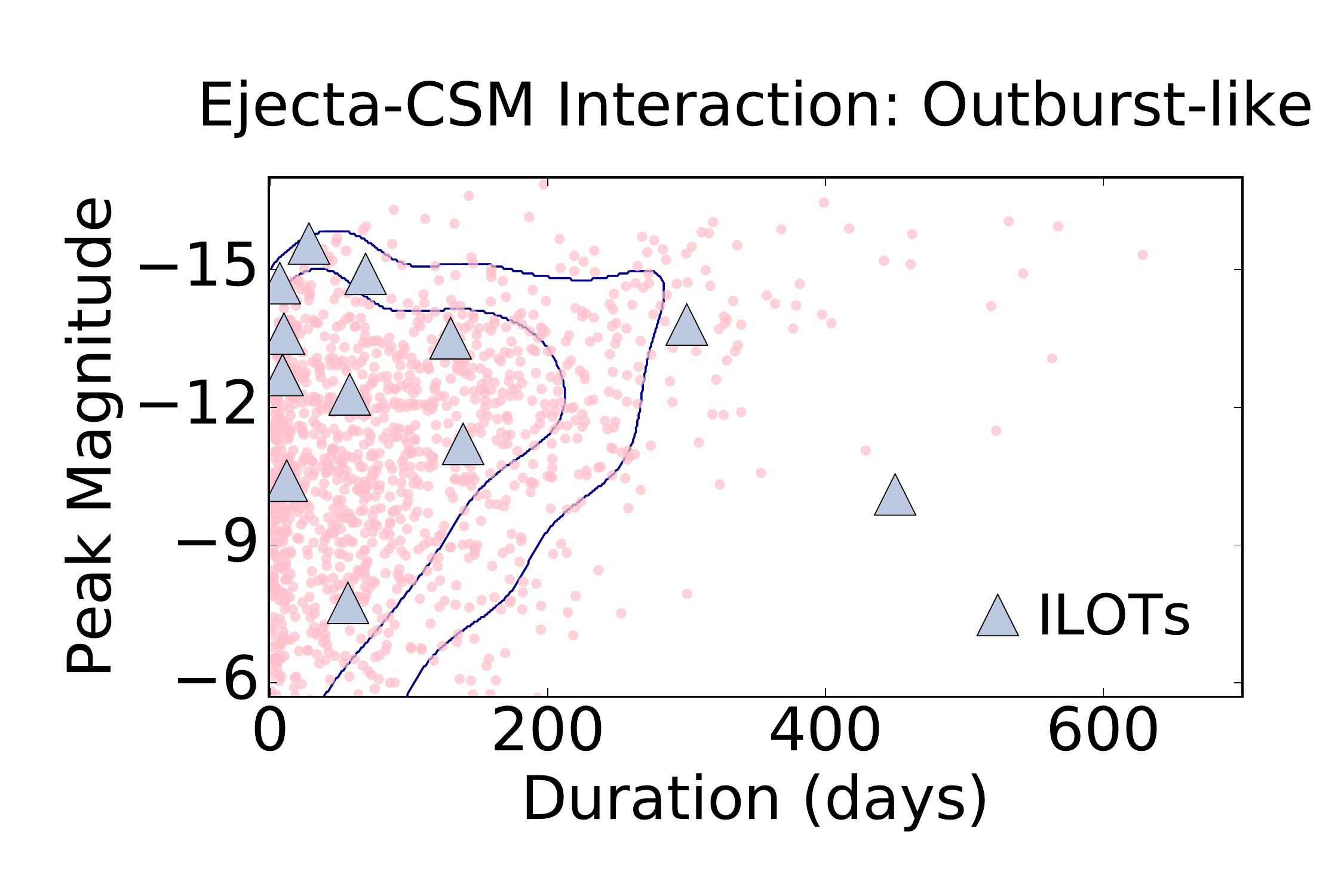}%
}
\vspace{-1.cm}
\subfloat[]{%
  \includegraphics[clip,width=\columnwidth]{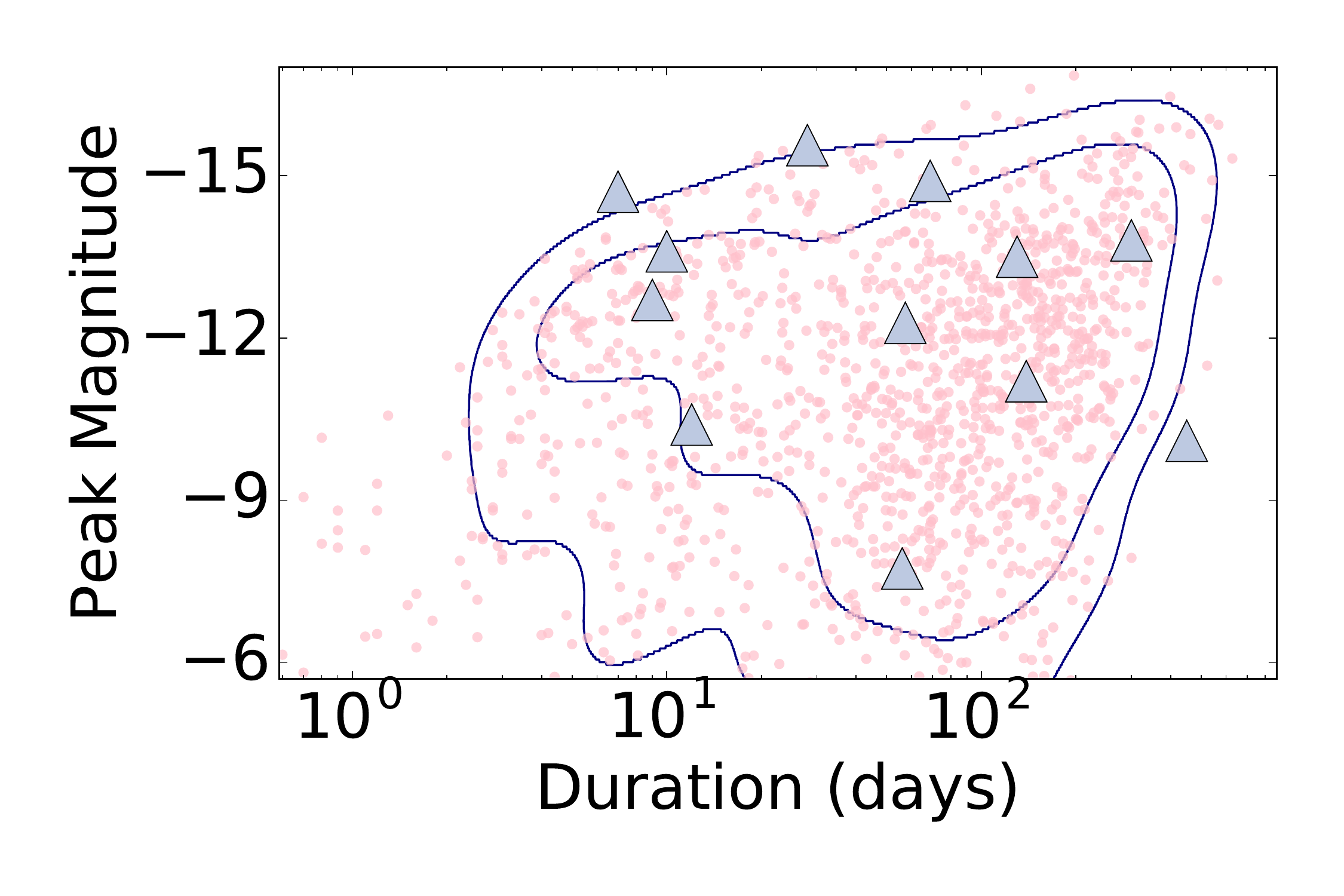}%
}
\vspace{-1.cm}
\subfloat[]{%
  \includegraphics[clip,width=\columnwidth]{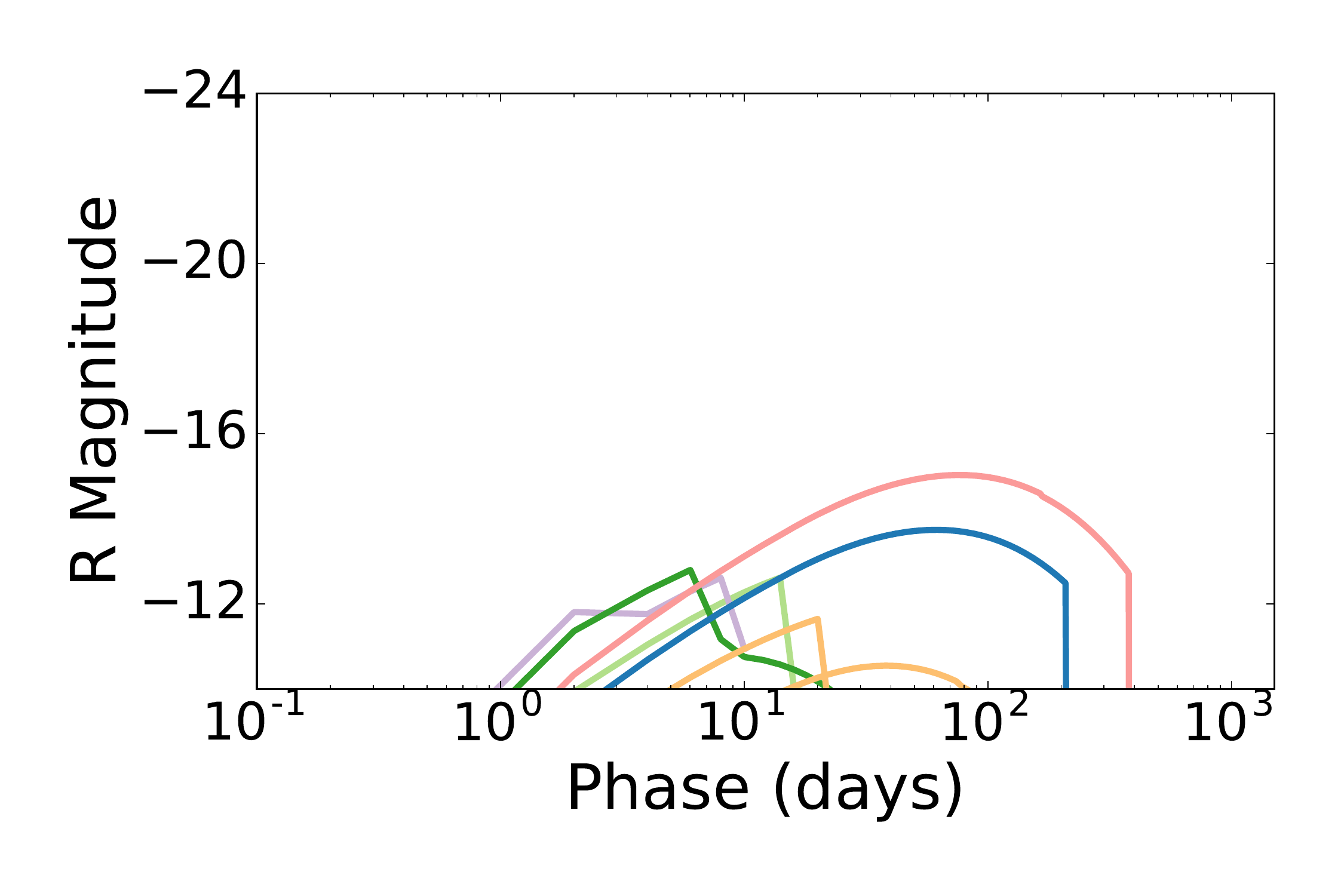}%
}
\caption{\textit{Top \& Middle}: Ejecta-CSM interaction DLPS for outburst-like transients and 68th and 90th percentile contours for the realizations, estimated using a KDE. Also shown is a sample of well-sampled ILOT light curves from the literature that are thought to be powered by CSM interaction: SN 1954J, SN 1961V, SN 2000ch, SN 2002bu, SN 2008S, SN 2009ip \citep{smith2011luminous}; SN 2002kg \citep{2006astro.ph..3025V}; SN 2010da \citep{villar2016intermediate}; PTF10fqs \citep{kasliwal2011ptf}; V838 Mon \citep{munari2002mysterious}; Eta Car \citep{smith2011revised}. \textit{Bottom}: Representative simulated light curves. }
\label{fig:csm_ilot}
\end{figure}

Sample light curves and the DLPS for outburst-like transients are shown in Figure \ref{fig:csm_ilot}. These models span a large range in both duration ($t_\mathrm{dur}\sim\mathrm{few}-100$ days) and luminosity ($M_\mathrm{R}\sim0$ to $-16$ mag). The light curve properties generally follow the same trends as the Type IIn SNe models. Short-duration transients ($t_\mathrm{dur}\lesssim10$ days) are less luminous ($M_\mathrm{R}\gtrsim-14$ mag). The parameter trends shown in Figure \ref{fig:csm_all_param} also hold for ILOT-like models.

Finally, we compare the simulated DLPS to a number of events from the literature, including LBV outbursts and Type IIn precursor events (see caption for details). In general, our models cover plausible timescales and magnitudes for ILOTs with signs of CSM interaction and overlap with many known objects.

\subsection{Hydrogen Recombination (Type IIP SNe)}

Type IIP SNe are explosions of red supergiants with masses of $\approx 8-17$ M$_\odot$ that have retained their hydrogen envelopes \citep{smartt2009death}. Following the explosion, a shock wave ionizes the hydrogen envelope. The characteristic flat, ``plateau" phase of their optical light curves is powered by hydrogen recombination as the expanding ejecta cools to $\sim 5000$ K at an approximately constant radius (i.e., the photosphere recedes in Lagrangian coordinates). The duration of the plateau phase is determined by the extent of the hydrogen envelope and kinematic properties of the blast wave. Following the plateau is a rapid decline in luminosity to a predominately $^{56}$Co-powered tail \citep{arnett1980analytic,weiler2003supernovae}. 

The bolometric and optical light curves of Type IIP SNe have been studied extensively (e.g., \citealt{patat1994light,hamuy2003observed,kasen2009type,sanders2015toward,rubin2016type}). Multi-zone semi-analytical and numerical models generally reproduce the observed light curves \citep{popov1993analytical,kasen2009type} and provide scaling relations for both the plateau durations and luminosities. Here, we use these theoretical scaling relations in conjunction with empirical trends found by \cite{sanders2015toward} to construct $R$-band light curves, neglecting contributions from both the shock breakout and $^{56}$Ni radioactive decay. While shock breakout should primarily affect the early light curve, significant amounts of $^{56}$Ni can extend the plateau duration. However, recent work has shown that $M_\mathrm{Ni}/M_\mathrm{ej}\lesssim0.01$ \citep{2017arXiv170200416M}, so we choose to ignore this contribution.

To construct light curves, we first assume instantaneous rise-times. In reality, Type IIP SNe have rise times which range from a few days to a week \citep{rubin2016type}. This is a minor effect given the long plateau durations. We then use the bolometric scaling relations derived in \cite{popov1993analytical} to estimate both the peak $R$-band luminosity ($L_\mathrm{p}$) and duration ($t_\mathrm{p}$) of the plateau phase:

\begin{equation}
    L_{p} = 1.64\times10^{42} \frac{R_{0,500}^{2/3}E_{51}^{5/6}}{M_{10}^{1/2}} \,\,{\rm erg}\,{\rm s}^{-1}
\end{equation}

\begin{equation}
    t_\mathrm{p} = 99 \frac{M_{10}^{1/2}R_{0,500}^{1/6}}{E_{51}^{1/6}}\mathrm{ days}
\end{equation}
where $R_{0,500}$ is the progenitor radius in 500 R$_\odot$, $E_{51}$ is the kinetic energy in $10^{51}$ erg and $M_{10}$ is the ejecta mass in 10 $M_\odot$. Here we have assumed that the $R$-band bolometic correction is negligible (approximately true during the plateau; \citealt{bersten2009bolometric}). Additionally, the blackbody SED has a temperature of 5054 K (the ionization temperature of neutral hydrogen) and the opacity is $\kappa=0.34 \,\mathrm{cm}^2$ g$^{-1}$. Following \citealt{sanders2015toward}, we assume that the light curve reaches a maximum and then monotonically declines during the plateau phase. The decline rate is strongly correlated to the peak luminosity and is parameterized by \citep{sanders2015toward}:

\begin{equation}
    L(t) = \frac{L_\mathrm{p}}{e^{(-13.1 - 0.47 M_\mathrm{R})}t}
\end{equation}
where $M_\mathrm{R}$ is the peak magnitude. Finally, we assume that for $t>t_\mathrm{p}$, the light curve drops off instantaneously. This assumption is justified by our definition of duration (within 1 mag of peak), which is minimally impacted by the late-time behaviour of the light curve. We generate light curves by sampling uniformly from ejecta mass ($M_\mathrm{ej}\sim5-15$ M$_\odot$) and progenitor radii ($R_0\sim100-1000$ R$_\odot$), and logarithmically in kinetic energy ($10^{50}-5\times10^{51}$ erg).

\begin{figure*}[htbp]
\centering
\includegraphics[clip,width=2\columnwidth]{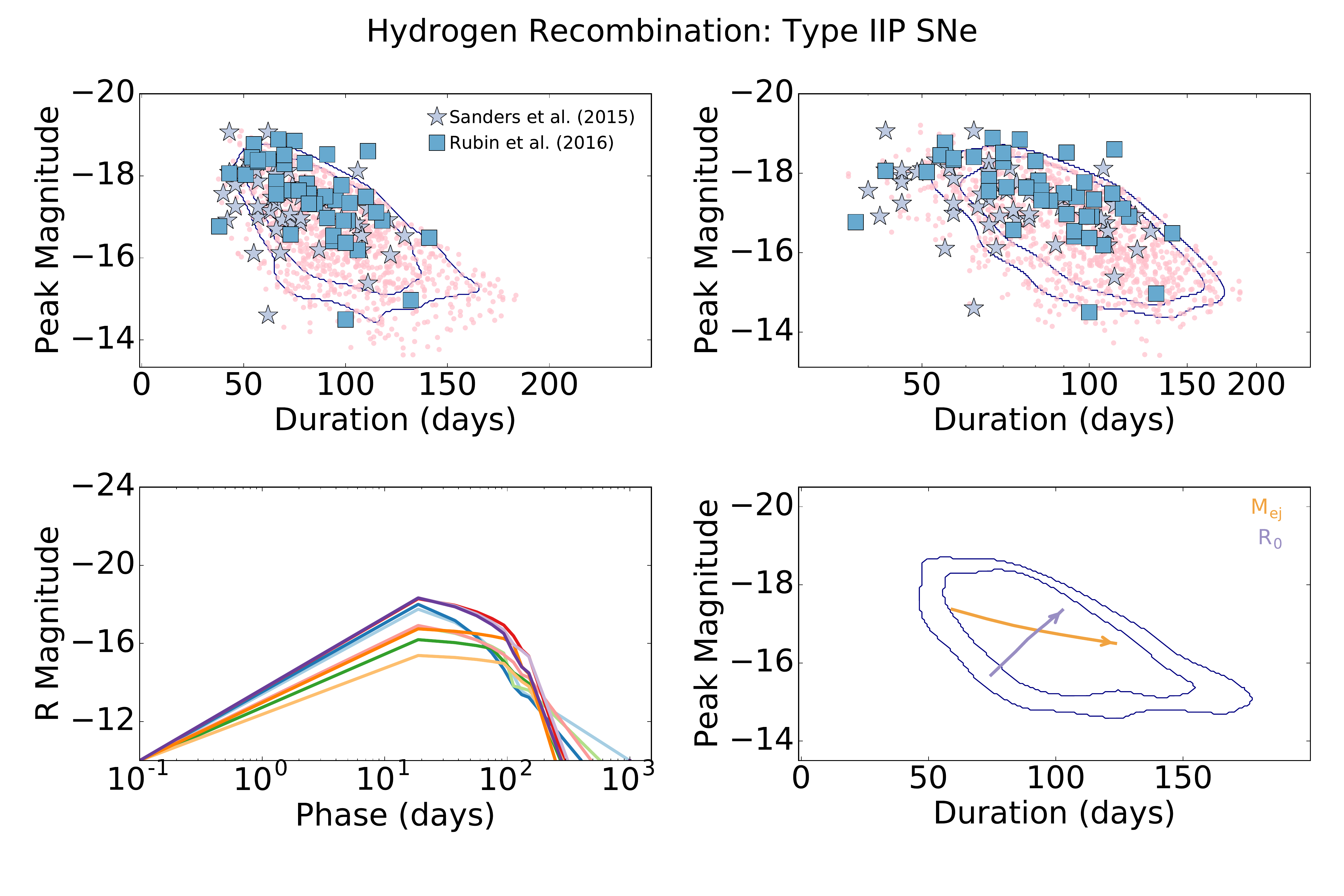}%
\caption{\textit{Top row}: Simulated Type IIP SNe DLPS (pink) with a sample of objects from the PS1/MDS (\citealt{sanders2015toward}; purple stars) and PTF (\citealt{rubin2016type}; blue squares). For the PS1/MDS sample, we construct light curves using the model and parameters described in \cite{sanders2015toward}, and we remove objects with less than 5 datapoints in the $R$-band. \textit{Bottom left}: Sample light curves. \textit{Bottom right}: Effect of $M_\mathrm{ej}$ (orange) and $R_0$ (purple) on the light curves of transients powered by hydrogen recombination given a constant kinetic energy. Arrows point towards increasing values of each parameter. Also shown are 68th and 90th percentile contours for the realizations, estimated using a KDE.}
\label{fig:iip}
\end{figure*}

The simulated DLPS and sample light curves are shown in Figure \ref{fig:iip}. The transient durations ($t_\mathrm{dur}\sim40-150$ days) and luminosities ($M_\mathrm{R}\sim-16$ to $-19$ mag) are negatively correlated. The upper luminosity boundary reflects our $M_\mathrm{ej}$ upper limit. In Figure \ref{fig:iip} we also explore the effects of the progenitor radius and the ejecta mass on the model light curves. Larger progenitor radii lead to brighter and longer duration transients as a result of the fixed photosphere. Increasing the ejecta mass produces less luminous and longer duration transients. Thus, the brightest (dimmest) Type IIP models have large (small) radii and small (large) ejecta masses. The longest (shortest) transients have large (small) radii and ejecta masses. The shortest duration events have $t_\mathrm{dur}\approx40$ days and high luminosities of $M_\mathrm{R}\approx-19$ mag.

In Figure \ref{fig:iip} we compare our generated light curve properties to samples from PanSTARRs \citep{sanders2015toward} and the Palomar Transient Factor (PTF; \citealt{rubin2016type}). It is worth noting that both samples contain so-called Type IIL SNe, which are spectroscopically similar to Type IIP SNe but decline linearly in magnitude more rapidly than most Type IIP SNe. Both \citealt{sanders2015toward} and \citealt{rubin2016type} find no evidence that Type IIP and Type IIL SNe arise from separate progenitor populations, so we also choose to keep Type IIL SNe in the observed sample. These samples largely overlap with our generated light curves. 


\subsection{GRB Afterglows}
Following a gamma-ray burst (GRB), the interaction of the relativistic jet with the CSM leads to a long-lived afterglow powered by synchrotron radiation \citep{1998ApJ...497L..17S}.  The afterglow emission can also be detected for off-axis sight lines (an ``orphan" afterglow; \citealt{rhoads1997tell,rossi2002afterglow,2010ApJ...722..235V}). There are two types of GRBs, long-duration (resulting from core-collapse of stripped massive stars; \citealt{woosley1993gamma}) and short-duration (likely produced by neutron star binary mergers; \citealt{berger2014short}). The energy scale of short GRBs (SGRBs) is about 20 times lower than for long GRBs (LGRBs), and their circumburst densities are at least an order of magnitude lower \citep{berger2014short,fong2015decade}. 

Here we explore both LGRB and SGRB afterglow models. Rather than generating analytical models, we use the publicly available\footnote{http://cosmo.nyu.edu/afterglowlibrary/offaxis2010broadband.html} broadband GRB afterglow model presented by \cite{2010ApJ...722..235V}. This model calculates broadband SEDs of both on- and off-axis GRB afterglows using a high-resolution two-dimensional relativistic hydrodynamics simulation. Typical LGRB values of isotropic energy $E_\mathrm{iso} = 10^{53}$ erg, jet half opening angle $\theta_\mathrm{jet}=11.5^o$, circumburst medium density $n_0 = 1$ cm$^{-3}$, accelerated particle slope $p = 2.5$, accelerated particle energy density fraction of thermal energy density $\epsilon_\mathrm{e} = 0.1$, and magnetic field energy density as fraction of thermal energy density $\epsilon_\mathrm{B} = 0.1$ are assumed. From this model, we can then generate the parameter space of long and short GRBs using scaling relation presented in \cite{van2012gamma}:

\begin{equation}
\begin{split}
    L_\mathrm{\nu,obs}&\appropto \epsilon_\mathrm{B}^{1/2}n_0^{-1/2}E_\mathrm{iso}\\
    t_\mathrm{dur}&\appropto \frac{E_\mathrm{iso}}{n_0}^{1/3}
\end{split}
\end{equation}

\begin{figure}[htbp]
\includegraphics[width=\columnwidth]{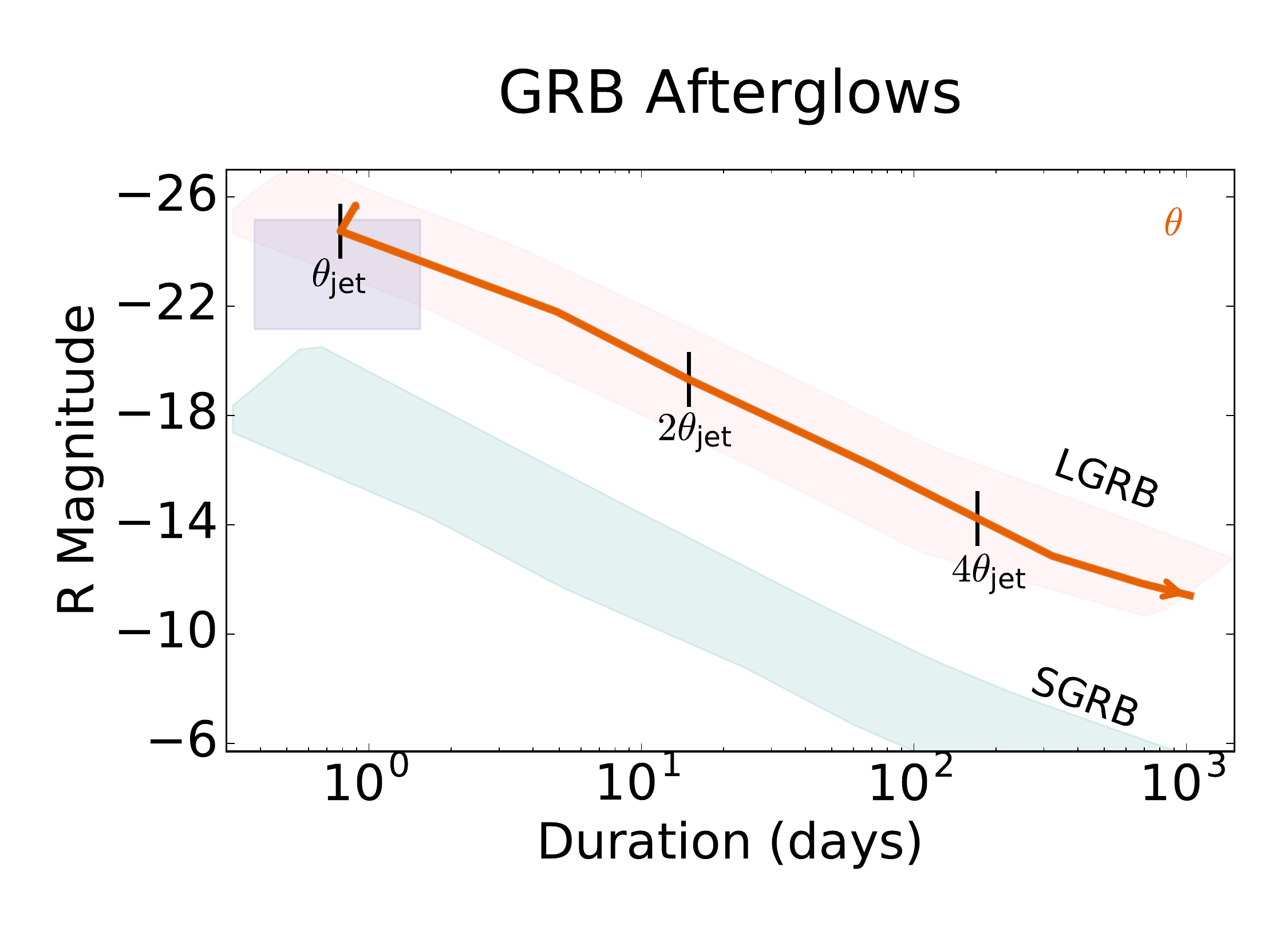}
\centering
\caption{GRB optical afterglow DLPS as a function of the observer viewing angle in units of the jet opening angle.  We show both LGRBs (red) and SGRBs (blue). The arrow points toward an increasing value of viewing angle $\theta$. Black vertical lines mark models where $\theta=\theta_\mathrm{jet}=0.2$ rad, $\theta=0.4$ rad and $\theta=0.8$ rad. Also shown is a $1\sigma$ region of on-axis LGRB afterglows observed by \cite{melandri2014optical} (purple rectangle).}
\label{fig:afterglow}
\end{figure}

Our simulated model, using the \cite{2010ApJ...722..235V} parameters and assuming that the afterglow is first observed $\approx 0.5$ days after the GRB, is shown in Figure \ref{fig:afterglow}. As the orientation becomes increasingly off-axis, the $R$-band transient becomes dimmer and longer duration.  We also plot scaled versions of this model, assuming $n_0\sim1-10$ cm$^{-1}$ and $E_\mathrm{iso}\sim(0.3-3)\times10^{53}$ erg for LGRBs, and $n_0\sim0.01-0.1$ cm$^{-1}$ and $E_\mathrm{iso} \sim(0.3-3)\times10^{51}$ erg for SGRBs. Both the SGRB and LGRB models span a wide range of durations ($t_\mathrm{dur}\sim1-1000$ days) and luminosity ($M_\mathrm{R}\sim-2$ to $-21$ mag for SGRBs and $M_\mathrm{R}\sim-12$ to $-16$ mag for LGRBs). The duration and luminosity are tightly negatively correlated, with the shortest duration events being the brightest. The only events with $t_\mathrm{dur}\lesssim10$ days are on-axis, which are known to be rare. For off-axis sight lines, the luminosity drops rapidly as the duration increases such that at $\approx2\theta_\mathrm{j}$, the afterglow is comparable to SNe in terms of timescale and luminosity. For larger angles, the events are much dimmer than SNe with longer durations.

We additionally plot a region corresponding to the $1\sigma$ observed properties of a sample of rest-frame $R$-band afterglows of on-axis LGRBs from the BAT6 sample (after removing LGRBs with early flares; \citealt{melandri2014optical}). The sample is in general agreement with the afterglow model, with short durations ($t_\mathrm{dur}\sim0.4-2$ days) and bright luminosities ($M_\mathrm{R}\sim-22$ to $-25$ mag), as expected for these on-axis events.

\subsection{Tidal Disruption Events}

Tidal disruption events (TDEs) occur when a star passes near a supermassive black hole (SMBH) and becomes tidally disrupted \citep{frank1976effects,hills1988hyper}. About half of the star's mass forms an accretion disk around the SMBH leading to an optically bright transient that lasts for weeks to months depending on the system's characteristics \citep{guillochon2009three}. A number of complications arise when modelling these transients, including the complex 3D geometry of the system, the hydrodynamics forming the accretion disk, the possibility of existing CSM surrounding the event and reprocessing of the disk emission by outflowing gas \citep{guillochon2014ps1}. We therefore present basic scaling relations for the durations and luminosities of TDEs. 

Assuming that the accretion rate onto the SMBH is less than the Eddington limit, the peak bolometric luminosity of the transient scales as \citep{stone2013tidal}:

\begin{equation}
    \label{eqn:tde:lum}
    L_{\text{peak}}\propto\dot{M}\propto M_\mathrm{BH}^{-1/2}M_*^2R_*^{-3/2}
\end{equation}
where $\dot{M}$ is the peak accretion rate of the disrupted star calculated at the tidal radius, $M_\mathrm{BH}$ is the black hole mass, and $M_*$ and $R_*$ are the star's mass and radius, respectively. The peak accretion rate is typically near the SMBH Eddington accretion rate, which leads to a plateau at the corresponding Eddington luminosity. Super-Eddington accretion will likely lead to an outflow of material (see e.g., \citealt{alexander2016discovery}). 

There are three timescales which potentially affect the transient duration: the diffusion time ($t_\mathrm{d}$), the viscous time ($t_\nu$) and the timescale of peak fallback accretion ($t_\mathrm{peak}$). In most cases, the diffusion timescale is small relative to at least one of the other two \citep{guillochon2009three,guillochon2013hydrodynamical}. Assuming a low disk viscosity, the duration of the transient will be proportional to \citep{lodato2012challenges}:

\begin{equation}
    \label{eqn:tde:dur_peak}
    t_\mathrm{dur}\propto t_{\text{peak}}\propto M_\mathrm{BH}^{1/2}M_*^{-1}R_*^{3/2}
\end{equation}
For canonical parameters (a sun-like star and $M_\mathrm{BH}=10^6$ M$_\odot$), this duration is about 40 days. 

If the accretion rate is near-Eddington, the light curve will plateau for a duration roughly corresponding to \citep{stone2013tidal}:

\begin{equation}
    \label{eqn:tde:dur_visc}
    t_\mathrm{dur}\propto t_{\text{edd}} \propto M_\mathrm{BH}^{-2/5}M_*^{1/5}R_*^{3/5}
\end{equation}
For canonical parameters, this corresponds to a duration of about 750 days.

Lost in these scaling relations is the fact that more massive black holes cannot disrupt less massive stars, because the tidal radius will be inside the horizon. The limiting SMBH mass (i.e., the Hills mass, $M_\mathrm{H}$) $M_\mathrm{H}=(1.1\times10^8 M_\odot) R_*^{3/2}M_*^{-1/2}$ \citep{hills1988hyper} is proportional to $M_*^{0.7}$, assuming $R_*\propto M_*^{0.8}$ for main sequence stars \citep{1991Ap&SS.181..313D}. 

\begin{figure}[htbp]
\includegraphics[width=\columnwidth]{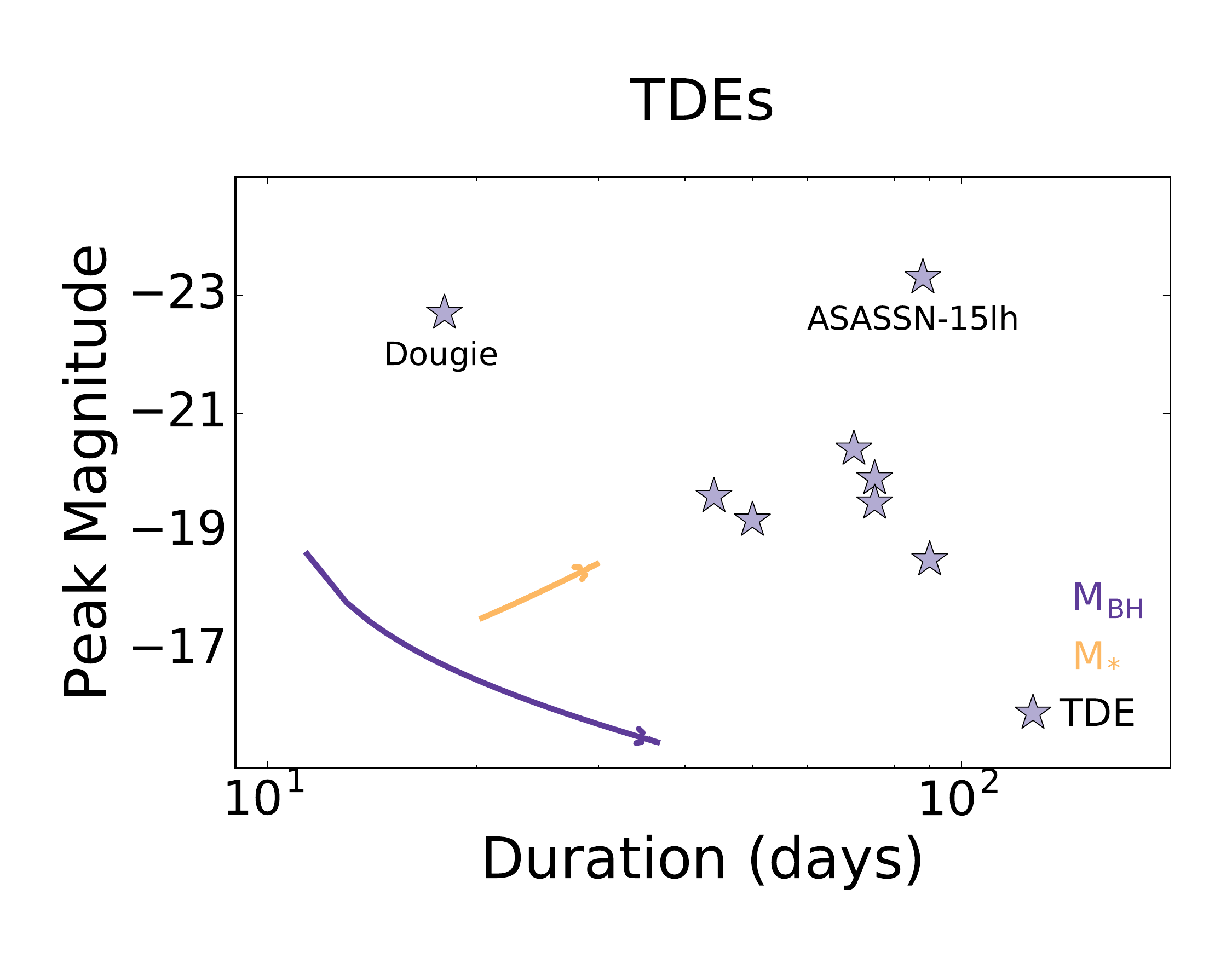}
\centering
\caption{Rough effects of $M_\mathrm{BH}$ (purple) and $M_*$ (orange) on TDE light curves assuming that the duration is proportional to $t_\mathrm{peak}$ (Equation \ref{eqn:tde:dur_peak}) and that $R_*\propto M_*^{0.8}$. Arrows point towards increasing values of each parameter. Also shown is a complete sample of well-sampled R/V/I TDE light curves from which we can measure a duration from the Open TDE Catalog at the time of writing. Light curve data gathered from \cite{dong2016asassn,holoien2014asassn,arcavi2014continuum,gezari2012ultraviolet,chornock2013ultraviolet,gezari2006ultraviolet,vinko2014luminous,wyrzykowski2014ogle}.}
\label{fig:tde}
\end{figure}

The scaling relations in Equations \ref{eqn:tde:lum} and \ref{eqn:tde:dur_peak}, along with a sample of TDEs from the literature, are shown in Figure \ref{fig:tde}. The majority of these transients follow the scaling relation with black hole mass, with the notable exception of extremely luminous ASASSN-15lh \citep{dong2016asassn}. A rapid spin rate and large black hole masses were necessary to explain the unique optical light curve of this claimed TDE \citep{margutti2016x,leloudas2016superluminous,2017arXiv170703458V}. From the sample of observed objects and the above scaling relations, it is clear that TDEs are not expected to produce short duration ($\lesssim20$ days) transients.

\subsection{Other Subclasses}\label{sec:other_subclasses}

In this section we enumerate additional types of transients which are either observed in small numbers or only hypothesized to exist, but whose physical models we do not explore in detail.

\subsubsection{Accretion Induced Collapse (AIC)} As an accreting white dwarf approaches the Chandrasekhar limit, it can collapse into a NS \citep{bailyn1990neutron,nomoto1991conditions,fryer1999can} with a rotationally-supported disk with mass $M_\mathrm{disk}\lesssim0.1$ M$_\odot$ \citep{dessart2006multidimensional}. The disk will then accrete onto the NS and eventually unbind as free nucleons recombine to form He. The radioactive heating of this ejecta is predicted to produce a fast ($t_\mathrm{dur}\sim1$ day) and dim ($M_\mathrm{R}\sim -13.5$ mag) optical transient \citep{metzger2009nickel,darbha2010nickel}. These are somewhat dimmer and shorter-duration compared to those of the $^{56}$Ni models explored in Section \ref{sec:radidecay}. We plot several models from \cite{darbha2010nickel} in Figure \ref{fig:other_transients}.

\textbf{Sub-Chandrasekhar Models/.Ia SNe}  If a WD is accreting hydrogen or helium from a companion, it can undergo unstable thermonuclear ignition, which may then lead to detonation given high enough densities of accreted material \citep{bildsten2007faint,shen2010thermonuclear,woosley2011sub}. The resulting transient is specifically referred to as a ``.Ia SN" if the binary companion is He-rich \citep{shen2010thermonuclear}. Theoretical models of .Ia SNe peak in the optical ($M_\mathrm{R}\sim-17$ to $-19$ mag) and have intermediate durations ($t_\mathrm{dur}\sim10-20$ days). More generally, these types of transients are described as sub-Chandrasekhar detonations and explosions (e.g., \citealt{sim2010detonations,woosley2011sub}) and are powered by radioactive decay. No convincing cases of such a model have been observed to date.  In Figure \ref{fig:other_transients} we show several models from the literature \citep{2010ApJ...715..767S,2012MNRAS.420.3003S,2011ApJ...734...38W}.

\subsubsection{Ca-rich Transients} Ca-rich transients are an observational class of dim transients ($M_\mathrm{R}\sim-15$ to $-16$ mag) with intermediate durations ($t_\mathrm{dur}\sim20$ days) whose nebular-phase spectra are rich in Ca and are primarily found in the outskirts of elliptical galaxies \citep{kasliwal2012calcium,lyman2014progenitors,2017ApJ...836...60L}. Like many low-luminosity classes, the exact origin of these transients is uncertain, although they are likely powered by radioactive decay. One suggested origin is a WD-NS merger \citep{metzger2012nuclear}.  A sample of these transients \citep{2017ApJ...836...60L} is shown in Figure \ref{fig:other_transients}. 

\subsubsection{Electron-Capture SNe} Electron-capture SNe (ECSNe) are explosions of super-asymptotic giant branch (SAGB) stars ($M_{\mathrm{MS}}\sim7-9.5$ M$_\odot$) with O+Ne+Mg cores rather than Fe cores. As the density of these cores increase, the electron capture onto Mg nucei leads to a decrease in the degeneracy pressure leading to collapse \citep{miyaji1980supernova, tominaga2013supernova}. Like Type IIP SNe, ECSNe are powered by hydrogen recombination and radioactive decay. The resulting optical transients are expected to be dim ($M_\mathrm{R}\sim-16$ to $-18$ mag), due to the small ejecta masses and kinetic energies and have intermediate durations ($t_\mathrm{dur}\sim40-100$ days). We show the theoretical light curves produced by \cite{tominaga2013supernova} in Figure \ref{fig:other_transients}.

\subsubsection{Luminous Red Novae (LRNe)} LRNe are an observational class of terminal transients which are characterized by their dim ($M_\mathrm{R}\sim -10$ to $-13$ mag) and red light curves ($g-r>1$) with durations ($t_\mathrm{dur}\sim50-100$ days) typically longer than those of classical novae at the same brightness \citep{martini1999nova,kulkarni2007unusually}. The class is heterogeneous, although many LRNe have double-peaked light curves, with the peaks separated by $\sim100$ days. The origin of these events is unclear and theoretical explanations range from planetary capture \citep{retter2003model} to stellar mergers entering the common envelope phase \citep{soker2006violent, rau2007spitzer,2013Sci...339..433I,2017ApJ...834..107B,2017arXiv170503895M}. We show a number of observed events which were identified as LRNe in the literature in Figure \ref{fig:other_transients}. 

\subsubsection{Classical Novae} \label{sec:other:novae}Novae have a rich observational history due to their high observed rate and utility as standardizable candles \citep{della1995calibration}. They occur when H-rich matter accretes onto a white dwarf from a binary companion, and the surface undergoes thermonuclear ignition \citep{gallagher1978theory}. We use the empirical maximum magnitude relation with decline time (MMRD) to place classical novae in the duration-luminosity phase space diagram. The MMRD relates the $V$-band peak magnitude with the decline time, $t_2$ ($t_3$), or the time to dim by two (three) magnitudes from peak. We approximate the duration as twice the time is takes to fall by one magnitude ($2t_1$). However, $t_1$ is not often reported in studies of the MMRD and can be much faster than the naive assumption of $t_2/2$ or $t_3/3$. We approximate $t_1$ by assuming that $t_3-t_2 = t_2-t_1$, or that the light curve decays linearly between $t_3$ and $t_2$. We use the relation from \cite{1990ApJ...360...63C} to transform between $t_2$ and $t_3$ and solve our above equation for $t_1$:

\begin{equation}
    t_1 = 0.31 t_2 - 1.9\,\,{\rm days}
\end{equation}

We then use the MMRD relation measured by \citet{della1995calibration} to estimate the peak magnitude:

\begin{equation}
    M_\mathrm{R} = -7.92 - 0.81 \arctan\frac{1.32-\log t_2}{0.23}
\end{equation}

\begin{figure}[!t]
\centering
\includegraphics[clip,width=\columnwidth]{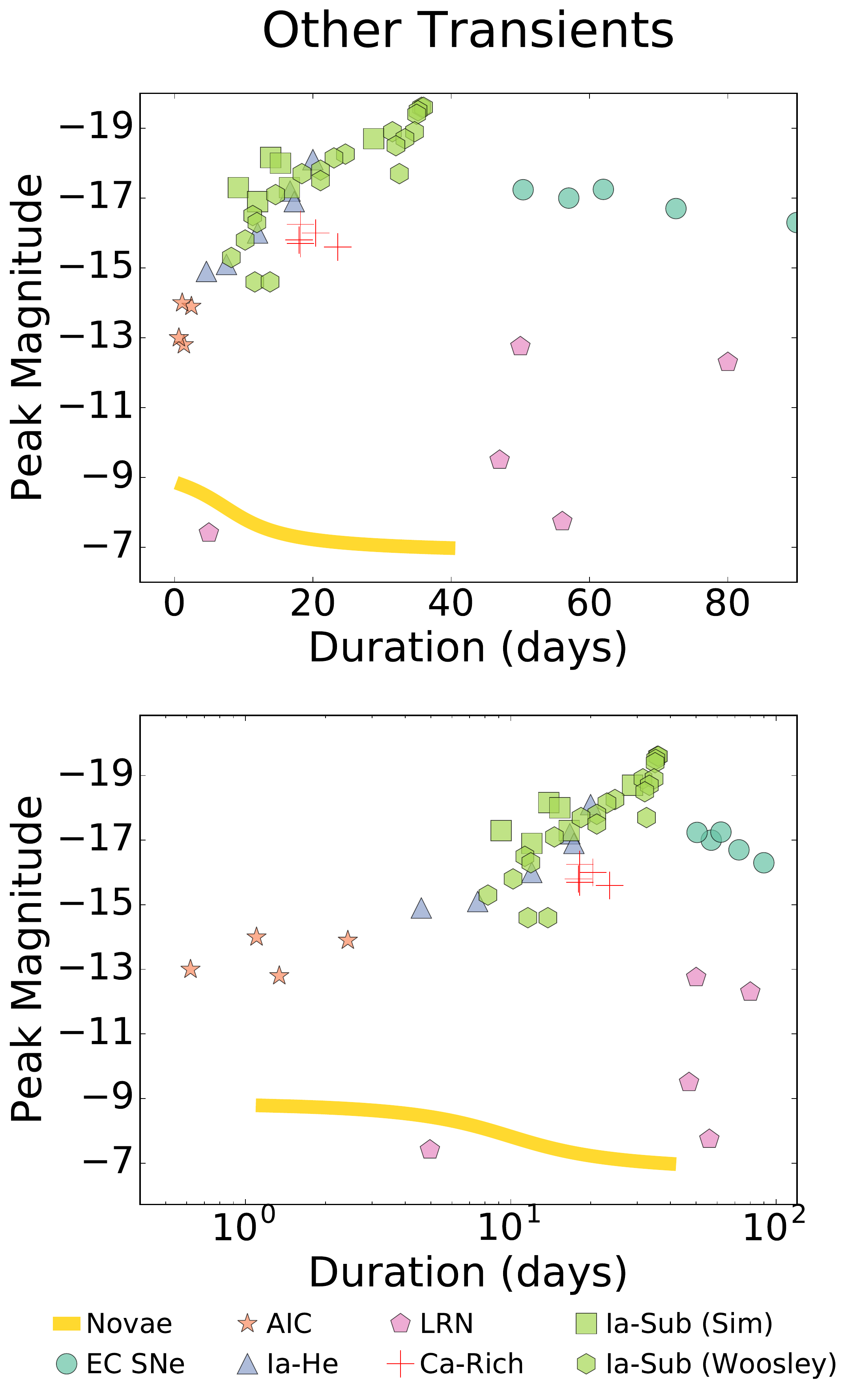}%
\caption{DLPS of transients described in Section \ref{sec:other_subclasses}. Specifically shown are electron-capture SNe models from \cite{2013ApJ...771L..12T}, accretion-induced collapse models from \cite{2010MNRAS.409..846D}, Ia-He models from \cite{2010ApJ...715..767S}, and sub-Chandrasekhar Ia models from \cite{2012MNRAS.420.3003S} and \cite{2011ApJ...734...38W}. We additionally show samples  of Ca-rich transients from \cite{kasliwal2012calcium} and \cite{2017ApJ...836...60L}, and LRNe from the literature \citep{williams2016progenitors,kashi2010common,kasliwal2011ptf,goranskij2016photometry}. }
\label{fig:other_transients}
\end{figure}

Novae $R$-band light curves tend to be brighter and longer-duration than $V$-band \citep{cao2012classical}, but we do not make an explicit correction for this. The modified MMRD relation described above is shown in Figure \ref{fig:other_transients}.

\subsubsection{Other Theoretical Merger Models} 
There are several other theorized merger models which might additionally occupy the short-duration regime of the DLPS, which we will not discuss in detail in this work. For example, following a WD-NS merger, both $^{56}$Ni and shocks powered by wind-ejecta interaction may produce a luminous ($L\sim10^{43}$ erg s$^{-1}$) and short-duration ($t_\mathrm{dur}\sim$ week) transient \citep{2016MNRAS.461.1154M}. Similarly, WD-WD mergers which do not produce Type Ia SNe might produce less-luminous ($L\sim10^{41}-10^{42}$ erg s$^{-1}$) and shorter duration ($t_\mathrm{dur}\sim1$ day) optical transients powered by the outflow of a differentially rotating merger product \citep{2014MNRAS.438..169B}.

\subsection{Combined Models: $^{56}$Ni Decay and Magnetar Spin-down}\label{sec:magni}
Until now we have assumed that each transient class is powered by a single energy source. In reality, we expect SN-like explosions to have multiple heating sources. We specifically expect newly synthesized $^{56}$Ni within SNe ejecta.

In this section, we consider light curves generated from a combination of two power sources: $^{56}$Ni decay and magnetar spin-down. We generate these light curves by adding the input luminosities from both contributions and diffuse the input luminosity through the expanding ejecta using {\tt MosFIT}. 

\begin{figure}[htbp]
\centering
\subfloat[]{%
  \includegraphics[clip,width=\columnwidth]{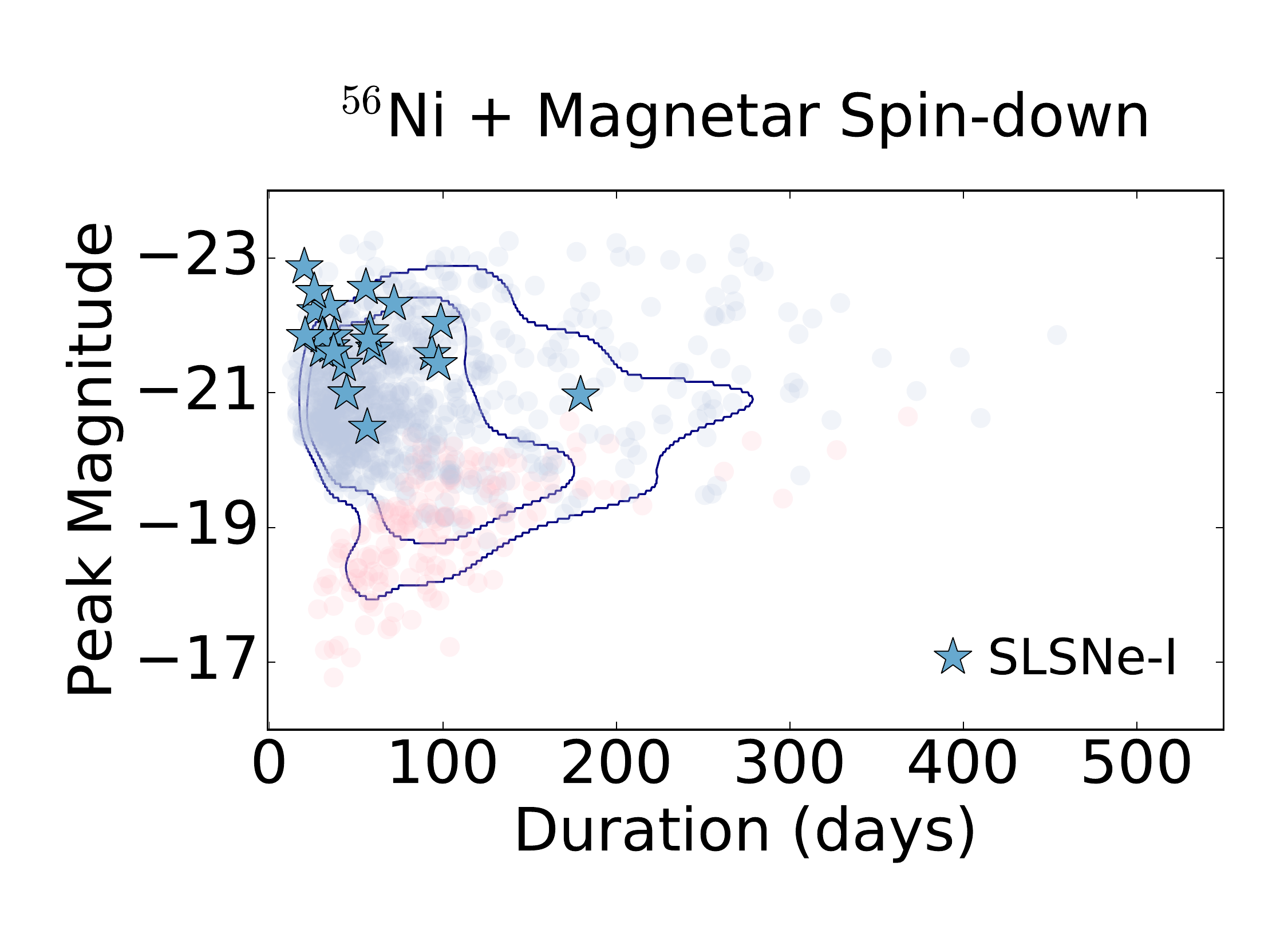}%
}
\vspace{-1.cm}
\subfloat[]{%
  \includegraphics[clip,width=\columnwidth]{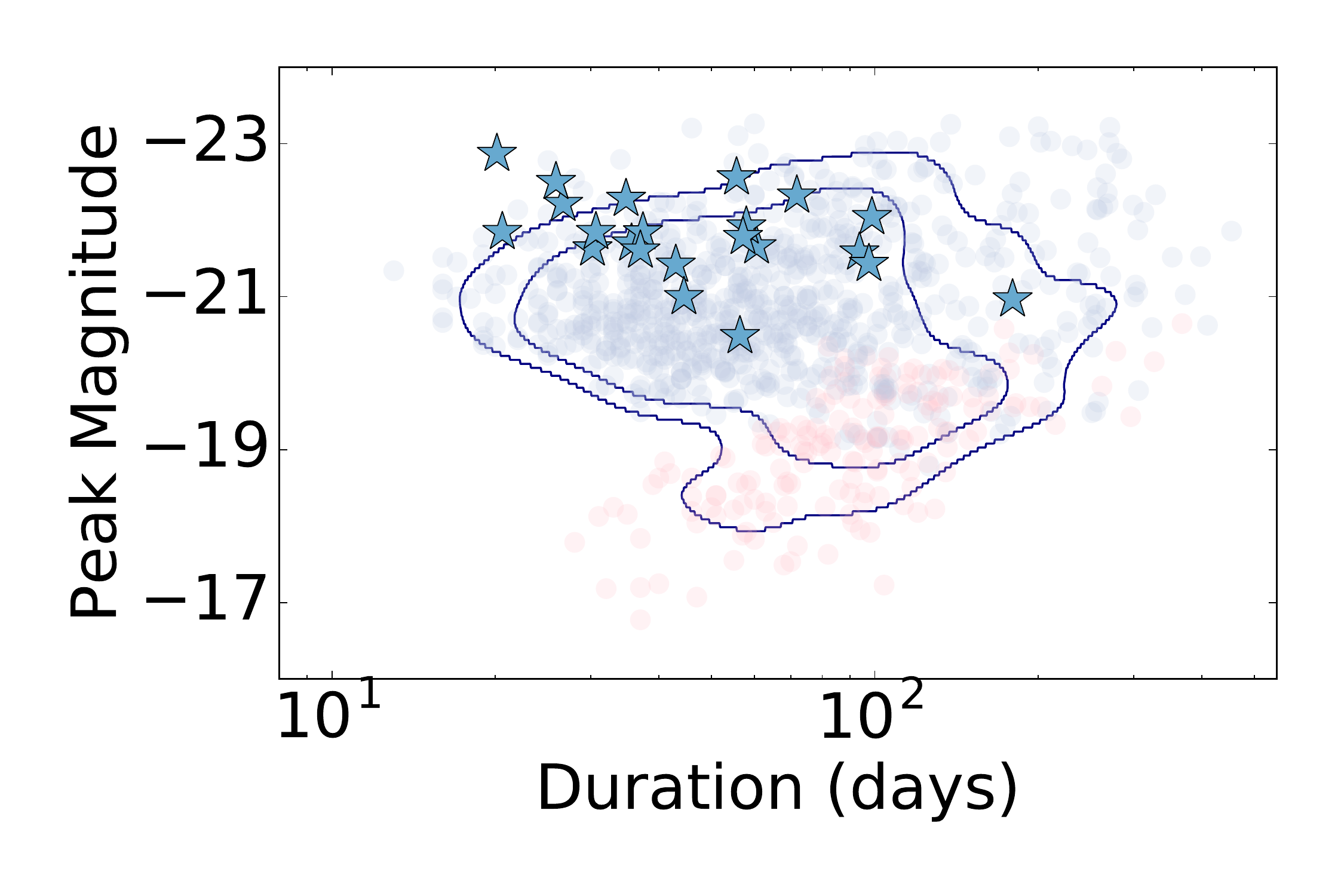}%
}
\vspace{-1.cm}
\subfloat[]{%
  \includegraphics[clip,width=\columnwidth]{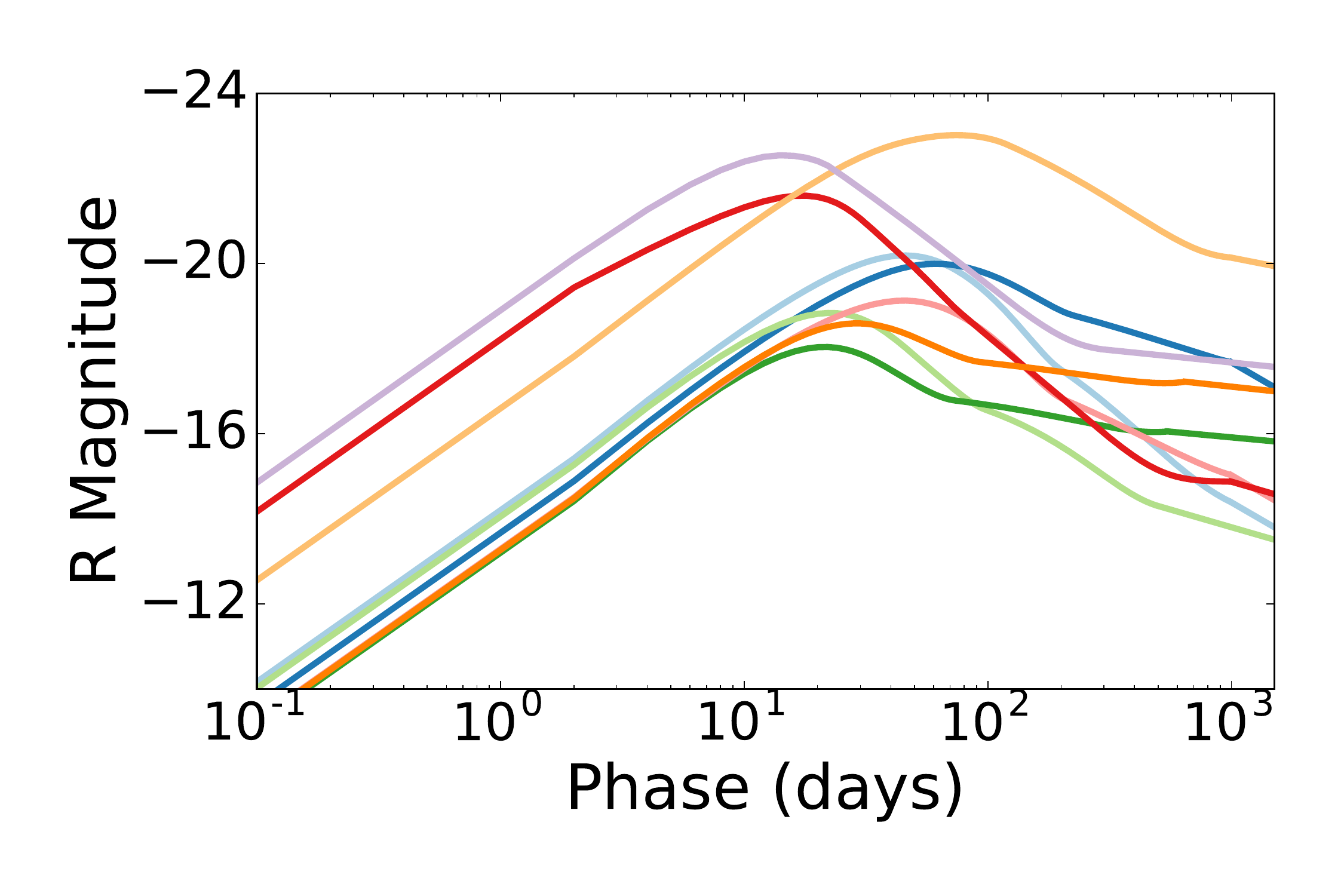}%
}
\caption{ \textit{Top \& Middle}: DLPS for explosions powered by magnetar spin-down and $^{56}$Ni radioactive decay. The color indicates the dominated heating source at peak luminosity (blue for magnetar spin-down and red for $^{56}$Ni). Also shown are 68th and 90th percentile contours for the realizations, estimated using a KDE. \textit{Bottom}: Representative simulated light curves.}
\label{fig:magni}
\end{figure}

Using the same parameter distributions as in Sections \ref{sec:radidecay} and \ref{sec:magnetar}, we generate $R$-band light curves for the combined power sources, and show these distributions in Figure \ref{fig:magni}. The joint DLPS generally overlaps with the distribution of solely magnetar-powered transients, although the low-luminosity ($M_\mathrm{R}\gtrsim-16$ mag) transients are missing, since in these cases the heat input from $^{56}$Ni decay dominates over the magnetar heating.

We separate the models based on the dominant (contributing $\geq50$\%) heat source at peak ($R$-band) luminosity and find that almost all transients brighter than $M_\mathrm{R}\sim-19.5$ mag are dominated by magnetar spin-down; conversely, all transients fainter than this value are dominated by $^{56}$Ni. The transition between dominating power sources is mainly controlled by the magnetic field of the magnetar. All models dominated by $^{56}$Ni have $B\lesssim5\times10^{13}$ G. At lower luminosities, the presence of a newly-formed magnetar will not be apparent photometrically.

\subsection{Combined Models: $^{56}$Ni Decay and Ejecta-CSM Interaction}\label{sec:csmni}
We next explore combined $^{56}$Ni decay and ejecta-CSM interaction. The diffusion processes for these two models are different: the input luminosity from the $^{56}$Ni decay diffuses through the ejecta and optically thick CSM ($M_\mathrm{ej}+M_\mathrm{CSM,th}$), while the ejecta-CSM input luminosity diffuses through $M_\mathrm{CSM,th}$. We assume that these two components evolve independently and add together their final luminosities. 

\begin{figure}[htbp]
\centering
\subfloat[]{%
  \includegraphics[clip,width=\columnwidth]{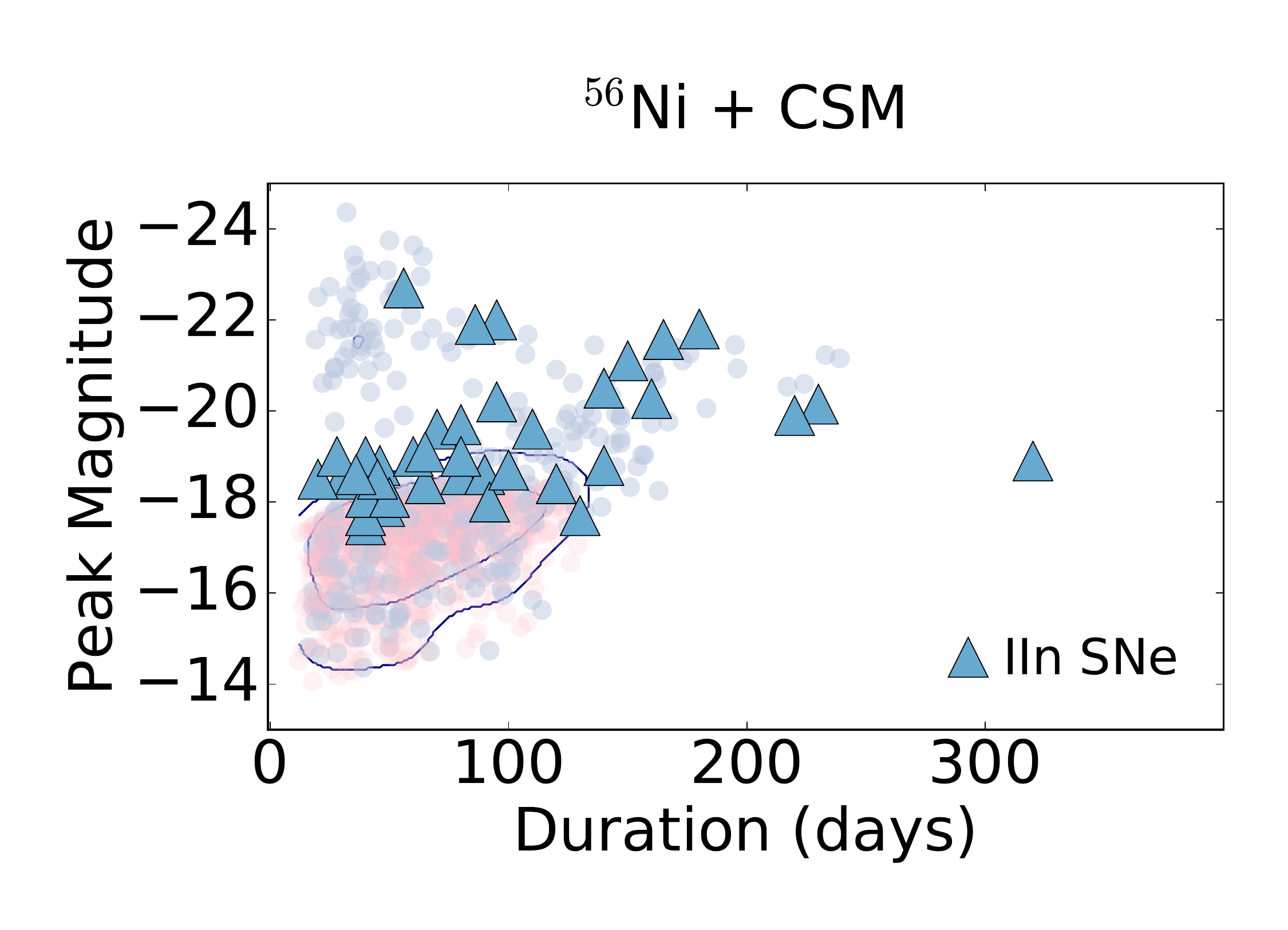}%
}
\vspace{-1.cm}
\subfloat[]{%
  \includegraphics[clip,width=\columnwidth]{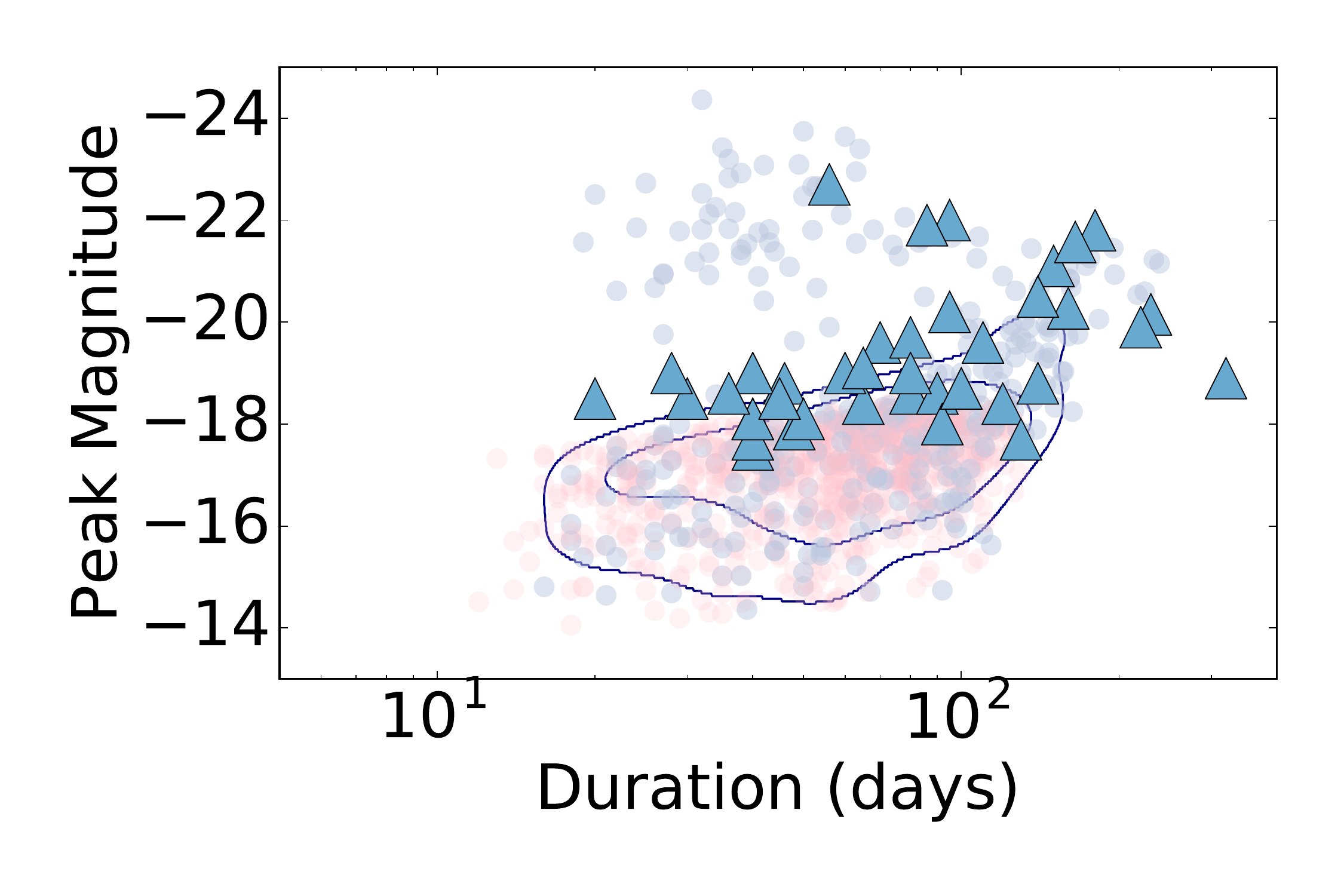}%
}
\vspace{-1.cm}
\subfloat[]{%
  \includegraphics[clip,width=\columnwidth]{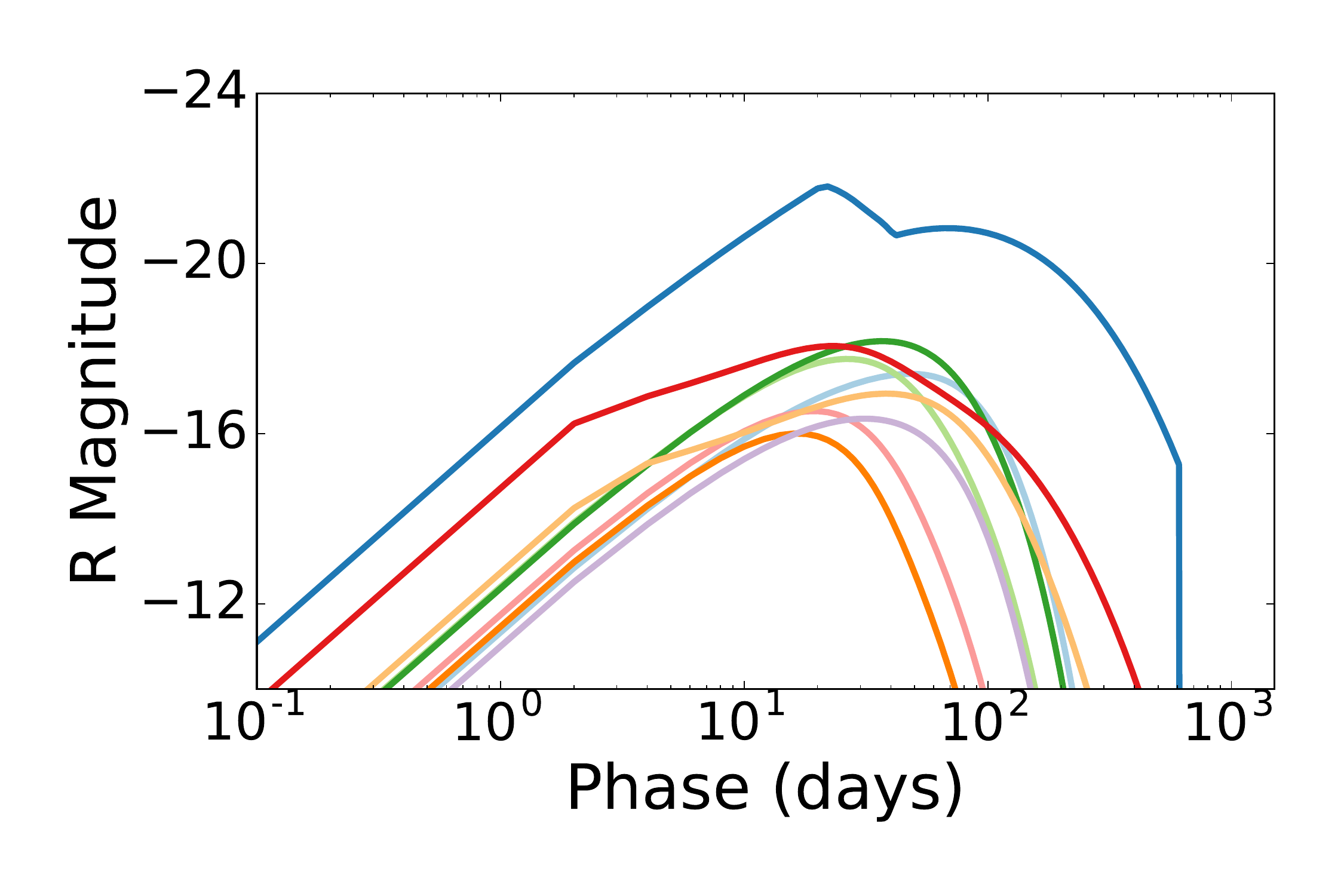}%
}
\caption{ \textit{Top \& Middle}: Simulated DLPS for explosions powered by ejecta-CSM interaction and $^{56}$Ni radioactive decay. The color indicates the dominated heating source at peak luminosity (blue for CSM interaction  and red for $^{56}$Ni). Also shown are 68th and 90th percentile contours for the realizations, estimated using a KDE. \textit{Bottom}: Representative simulated light curves }
\label{fig:csmni}
\end{figure}

Typical light curves for the case of $s=2$ (wind) and their distribution in the DLPS are shown in Figure \ref{fig:csmni}. As in the case of combined $^{56}$Ni-decay and magnetar spin-down, no transients brighter than $\sim-19.5$ mag are dominated by the $^{56}$Ni input luminosity at peak. Unlike the transients solely powered by ejecta-CSM interactions, we find no transients with durations $\lesssim10$ days, and no transients with $M_\mathrm{R}\gtrsim-14$ mag, because in such cases the timescale and luminosity are determined by radioactive heating.

The dominance of CSM interaction over $^{56}$Ni is mainly controlled by the mass loss rate, $\dot{M}$. Assuming $v_\mathrm{wind}\sim100$ km s$^{-1}$, CSM interaction dominates when $\dot{M}\gtrsim10^{-3}$ M$_\odot$ yr$^{-1}$. This is consistent with LBV mass loss rates \citep{smith2014mass}.

\section{Discussion}\label{sec:discussion}
\subsection{Specific Engine Insights}
In this section, we enumerate a number of insights which can be gained from the preceding analysis of the DLPS. We focus on the overlap of our predicted models with the observed populations and the regions of phase space that these models occupy. 

We begin with the adiabatic expansion models which lack any internal heating source. The largest progenitors (RSG-like) can produce luminous (reaching $M_\mathrm{R}\sim-18$ mag) transients on timescales similar to those of SNe ($t_\mathrm{dur}\sim20-100$ days). In fact, the RSG models span a similar range of peak magnitudes as the Type IIP/L models. The BSG-like subclass lie within the luminosity gap ($M_\mathrm{R}\sim-10$ to $-15$ mag) with durations similar to those of SNe ($t_\mathrm{dur}\sim 20$ days), and the white dwarf-like subclass have nova-like luminosities ($M_\mathrm{R}\sim-7$ to $-11$ mag) and have much shorter durations ($t_\mathrm{dur}\sim1$ day). In reality, these models will likely be paired with some radioactive heating or an additional heating source and therefore represent lower limits in both duration and luminosity. We can see from these models that for massive progenitors, we expect transients which last $\gtrsim 10$ days. In contrast, compact object (or stripped) progenitors can reach the extreme limits of this phase space and generate faster transients but only at low luminosities.

Next we discuss our simulated DLPS of $^{56}$Ni-powered transients. In the simulated and observed populations of Type Ib/c SNe, there is a dearth of short duration ($t_\mathrm{dur}\lesssim20$ days) and long duration ($t_\mathrm{dur}\gtrsim80$ days) transients. These timescales correspond to transients with small ejecta masses/high velocities and large ejecta masses/low velocities, respectively. From the literature, we find only two well-observed supernovae with durations $\gtrsim70$ days (also shown in Figure \ref{fig:ibc_objects}): iPTF15dtg \citep{taddia2016iptf15dtg} and SN 2011bm \citep{valenti2012spectroscopically}. Both objects require large $^{56}$Ni and ejecta masses to explain their extended light curves, suggesting intrinsically rare massive progenitors \citep{valenti2012spectroscopically,taddia2016iptf15dtg}. Longer duration transients are seen in the PISNe models with larger ejecta masses and kinetic energies. However, the low-metallicity progenitors of PISNe are expected to be found at high redshift, meaning that observed PISNe are likely to be even slower (time dilated by $1+z$) and redder. Shorter duration transients are seen in the Iax-like models due to their lower ejecta masses, although very few of these transients (observed or simulated) have $t_{\rm dur}\lesssim 10$ days.




When considering the radioactive decay of $^{56}$Ni as a heating source for short-duration transients, it is important to note that it is largely the ratio of the ejecta mass to the nickel mass which limits the light curve parameters. In reasonable physical models, it is unlikely that $f_{\rm Ni}\gtrsim 0.5$ (although a few Type Ia SNe with higher nickel fractions have been observed, e.g., \citealt{2015MNRAS.454.3816C}). This means that, regardless of the amount of $^{56}$Ni within the ejecta, the timescale of the transient will typically be set by $M_\mathrm{ej}$ (and other factors). This fact -- the low $^{56}$Ni fraction in physical models -- essentially eliminates luminous, short-duration transients powered by $^{56}$Ni. For example, this is why $^{56}$Ni fails as the main power source for superluminous supernovae, which have relatively short durations given their high luminosities \citep{kasen2010supernova,nicholl2013slowly}.

The kilonovae models, powered by $r$-process decay, lie in a unique area of the DLPS, with short durations ($t_\mathrm{dur}\sim$ few days) and low luminosities ($M_\mathrm{R}\sim-8$ to $-16$ mag). Their short durations couples with low luminosities follows the general trend seen in the stripped SNe models. Although there is currently large uncertainty in the opacity of Lanthanide-rich ejecta \citep{barnes2016radioactivity}, all of the models are below typical SNe luminosities and durations. Our red models in particular span to even dimmer events than those explored in \cite{barnes2013effect} and \cite{metzger2016kilonova}, consistent with the recent result by \cite{2017arXiv170507084W}. We additionally note that a brighter and longer-lived, magnetar-powered kilonova has been recently proposed \citep{yu2013bright,metzger2014optical,siegel2016electromagnetic} which was not explored in this work. Such a kilonova could peak at $\sim10^{44}-10^{45}$ erg s$^{-1}$ with a duration of several days, although it would represent a small fraction of the kilonova population.

We next examine the magnetar models explored in Section \ref{sec:magnetar}. We find that the models span a broad range in both duration ($t_\mathrm{dur}\sim20-250$ days) and luminosity ($M_\mathrm{R}\sim-16$ to $-23$ mag). Our models reproduce both the detailed theoretical predictions and observed light curves of SLSNe-I. However, the SLSN-I light curves span a narrower range of the DLPS, primarily at the bright end. This indicates that at least those magnetar-powered events have a narrower range of parameters than explored in this work, as suggested recently by \cite{2017arXiv170600825N}. Transients which have weak contributions from the magnetar's spin-down are likely dominated by $^{56}$Ni-decay (as discussed in Section \ref{sec:magni}) and are classified as normal Type Ib/c SNe. We also note that several Type I SLSNe have been accompanied with early-time bumps with several day durations and SN-like luminosities \citep{leloudas2012sn,nicholl2016seeing} which were not explored in this paper. The origin of these bumps is currently unknown, although several theoretical explanations have been posed (e.g., \citealt{kasen2016magnetar,2017arXiv170501103M}). 

Our ejecta-CSM interaction models span the widest range of the DLPS of the models presented here, due to both a large number of free parameters (which may not be independent, as assumed) and the simplifying assumptions used \citep{chatzopoulos2012generalized}. One of the most striking features is the difference between the wind-like and shell-like CSM geometries, with shell-like models producing brighter transients with somewhat shorter durations ($t_\mathrm{dur}\sim100$ days for wind-like vs $t_\mathrm{dur}\sim50$ days for shell-like). Although wind-like models can reproduce both low and high luminosity Type IIn SNe, the shell-like models with SN-like ejecta masses and kinetic energies do not extend to the luminosities of normal Type IIn SNe. Focusing on the wind-like models, we find that luminosity and duration are positively correlated at shorter ($t_\mathrm{dur}\lesssim20$ days) durations, with no models brighter than $M_\mathrm{R}\sim-14$ mag in this regime. 

We note that \cite{chatzopoulos2013analytical} find that, when fitting SLSNe with the semi-analytical model used in this work, both $s=0$ and $s=2$ can generally be used to find acceptable fits, but the models lead to substantially different explosion parameters. For the normal Type IIn SNe, \cite{moriya2014mass} estimated the CSM profile ($s$) from the post-peak light curves of 11 Type IIn SNe and found that most showed $s\sim2$. This implies that the shell model ($s=0$) is less physical for at least the Type IIn events.

 The heterogeneous group of transients discussed in Section \ref{sec:other:novae} span a broad range of the DLPS, but their phase space is not particularly unique. Many of the models with likely compact object progenitors (e.g. Ca-rich transients and sub-Chandrasekar models), are confined to a small area similar to the Iax-like models we explored in Section \ref{sec:iaxultra}. The electron-capture SN models overlap with the Type Ib/c and Type IIP SNe, and the LRNe are broadly consistent with the CSM interaction outburst-like models. Only the AIC models and classical novae extend to novel regions of the DLPS at $t_{\rm dur}\lesssim 10$ days durations but invariably with low luminosities ($M_\mathrm{R}\gtrsim -14$ mag).

 
 Finally, we focus on our combined models with radioactive decay coupled to either magnetar spin-down or ejecta-CSM interaction. In the ejecta-CSM interaction case, the addition of $^{56}$Ni decay eliminates both short-duration ($t_\mathrm{dur}\lesssim10$ days) and dim ($M_\mathrm{R}\gtrsim-14$ mag) transients that are otherwise produced by this model. The former is due to the fact that the decay of $^{56}$Ni dominates the CSM interaction light curves, eliminating the artificial cutoffs to the input luminosities. Additionally, there is a clear separation of transients which are dominated by $^{56}$Ni decay or CSM interaction/magnetar spin-down in the DLPS around $M_\mathrm{R}\sim-19.5$ mag. In the ejecta-CSM interaction case, this separation roughly coincides with where the estimated mass-loss rate of the progenitor star roughly matches typical LBV mass loss rates ($\dot{M}\sim10^{-3}$ M$_\odot$ yr$^{-1}$; \citealt{smith2014mass}), and where many identified Type IIn SNe lie. In the case of magnetar spin-down, the separation occurs at $B\sim5\times10^{13}$, about the cutoff for expected magnetar magnetic field strengths \citep{zhang2000high}.

\begin{figure*}[htbp]
\centering
  \includegraphics[clip,width=\textwidth]{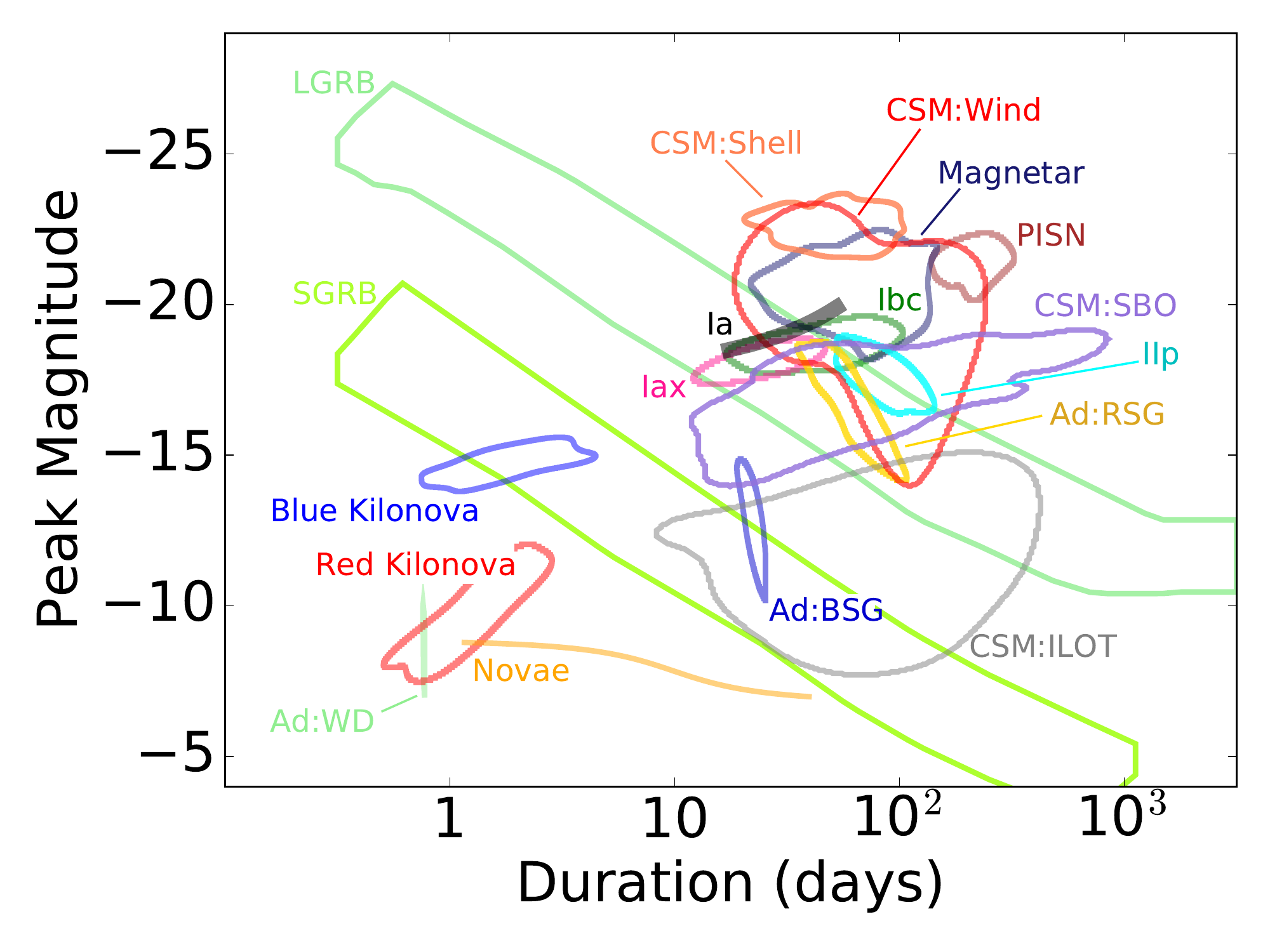}%
\caption{The full DLPS explored in this work. Each colored region represents a contour that contains roughly 68\% of the Monte Carlo realizations for each class, estimated using a KDE. An interactive version of this plot, including 90\% contours and linear-space version can be found at ashleyvillar.com/dlps.}
\label{fig:all_bokeh}
\end{figure*}


\subsection{The Optical Transient Landscape}

In this section, we summarize the overarching results from the DLPS analysis and highlight several regions of interest. To begin, we present all of the classes simulated in this paper (excluding TDEs and the ``other transients" in Section \ref{sec:other_subclasses}) in Figure \ref{fig:all_bokeh}. An interactive version of this plot, which can be used to compare user-uploaded transients to the complete DLPS, is available online\footnote{ashleyvillar.com/dlps. This applet allows a user to selectively plot transient classes and to add their own datapoints to the simulated DLPS.}. 

There is substantial overlap among models, especially between $t_\mathrm{dur}\sim10-100$ days and $M_\mathrm{R}\sim-18$ to $-20$ mag. Interestingly, this is also where most observed transients lie in the DLPS. A wide range of explosion physics and internal heating sources lead to similar optical light curve properties, in part due to similar kinetic energies and ejecta masses. This highly populated regime highlights the fact that the abundance of observed transients with $\sim$ month-long durations and SN-like luminosities is likely not due to observational biases but a reflection of the underlying physics. But what about the more extreme areas of the DLPS?

We begin by focusing on fast ($t_\mathrm{dur}\lesssim10$ days) and bright ($M_\mathrm{R}\lesssim-18$ mag) transients within our explored models. We find essentially no models that can produce transients in this regime, with the exception of on-axis GRB afterglows. Luminous and fast optical transients cannot be powered by radioactivity, magnetar spin-down or CSM interaction; however, they can be powered by relativistic outflows. Relativistic sources (with $\Gamma\gtrsim$ a few) like GRBs are rare compared to other optical transients. For example, the GRB volumetric rate at $z\lesssim0.5$ is only 0.1\% of the CCSNe rate \citep{dahlen2004high,wanderman2010luminosity}. Given this low rate and current lack of other physically-motivated models, we argue that this portion of the DLPS is, and will continue to be, sparsely populated due to intrinsically rare physics.

A number of heating sources can produce transients that are fast ($t_\mathrm{dur}\lesssim10$ days) but invariably dim ($M_\mathrm{R}\gtrsim-14$ mag), including novae, adiabatic explosions of white dwarfs, $r$-process kilonovae and CSM interaction models. However, most of these models require unique combinations of parameters, mainly very low ejecta masses, and represent a small fraction of the DLPS explored. Therefore, short-duration transients seem intrinsically rare, even at lower luminosities.

At the other extreme, we find several models that can produce exceptionally luminous transients ($M_\mathrm{R}\lesssim-22$ mag), including $^{56}$Ni decay (in the context of PISNe), magnetar spin-down, GRB afterglows and ejecta-CSM interactions. TDEs may also reach these high luminosities (as seen in the case of ASASSN-15lh; \citealt{margutti2016x}). All of these models require extreme parameters to reach such bright luminosities, implying that such events are intrinsically rare. However, these luminous transients are invariably long-duration ($t_\mathrm{dur}\gtrsim50$ days).

The dimmest transients ($M_\mathrm{R}\gtrsim-14$ mag), are generated from adiabatic explosions of white dwarfs, off-axis GRB afterglows, outburst-like ILOTs, classical novae and $r$-process kilonovae, with a broad range of durations $(t_\mathrm{dur}\sim1-300$ days). Of these, few lie in the intermediate luminosity gap between the brightest classical novae and dimmest SNe ($M_\mathrm{R}\sim-10$ to $-14$ mag). Due to the low rates of GRB afterglows and kilonovae, the most commonly discovered class in this gap will likely be powered by CSM interaction in the context of massive star eruptions (rather than explosions) as inferred for the small sample of known ILOTs (e.g., \citealt{kochanek2012unmasking}). 

To summarize, we find three sparse regimes of the DLPS: (i) bright and fast transients ($t_\mathrm{dur}\lesssim10$ days and $M_\mathrm{R}\lesssim-16$ mag); (ii) intermediate luminosity transients ($M_\mathrm{R} \approx -10$ to $-14$ mag) across all durations; and (iii) luminous transients ($M_\mathrm{R} \lesssim -21$ mag).  Of these, the most sparsely occupied by theoretical models is the first.  On the other hand, the typical parameter ranges for SNe (i.e., $t_\mathrm{dur}\sim10-100$ days and $M_\mathrm{R}\sim-18$ to $-20$ mag) contain a number of overlapping models, consistent with the fact that most observed optical transients lie within this regime.

\begin{figure*}
\includegraphics[width=18cm]{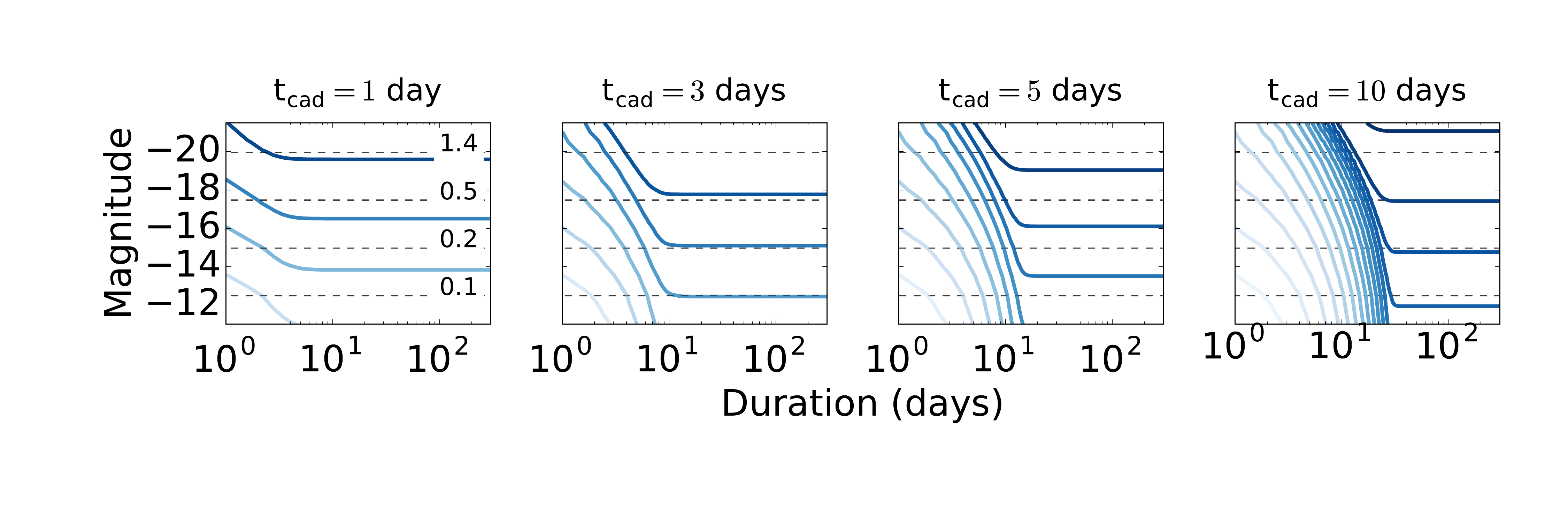}
\centering
\caption{Detection rate of optical transients given a luminosity, duration, and survey cadence ($t_\mathrm{cad}$) and depth ($24.5$ mag). The lines are exponential contours, with the darkest shade of blue being the most detectable transients. The numbers (and corresponding horizontal lines) in the left panel represent $z_\mathrm{lim}$ for a given absolute magnitude (see text for details).}
\label{fig:rate}
\end{figure*}
\subsection{Observability \& Survey Considerations}

Until now we have investigated theoretical models of transients that occupy the DLPS. The \textit{observed} DLPS of transients will be modified by each class's volumetric rate and luminosity function (which we will explore in a follow-up paper). In this section, we will consider the effects of a given survey's parameters (cadence and area) on the observed DLPS. We perform a simple calculation to explore the effect of a transient's luminosity and duration on its survey discovery potential, or its relative discovery rate assuming a constant volumetric rate ($\mathcal{R}$) for every transient, in a flat cosmology ($\Omega_\mathrm{M}=0.3$; $\Omega_\Lambda=0.7$; $H_0=70$ km s$^{-1}$ Mpc$^{-1}$). 

The number ($\mathcal{N}$) of transients of a certain luminosity and duration discovered in a given magnitude-limited survey is proportional to:

\begin{equation}
    \mathcal{N}\propto\int_{z=0}^{z=z_{\mathrm{lim}}}\epsilon\mathcal{R}dV
\end{equation}
where $z_\mathrm{lim}$ is the redshift where the apparent magnitude of the transient is equal to the limiting magnitude of the survey. The parameter $\epsilon$ represents a detection efficiency of the survey, defined by our heuristic equation:

\begin{equation}\label{eqn:eff}
    \epsilon = \frac{1}{1+e^{-\left(t_\mathrm{dur}\left(1+z\right)-N_\mathrm{D}t_\mathrm{cad}\right)}}
\end{equation}

where $t_\mathrm{dur}$ is the transient duration in its rest frame, $t_\mathrm{cad}$ is the survey cadence and $N_\mathrm{D}$ is a penalty term to simulate the need for multiple datapoints to ``detect" a transient; here we choose $N_\mathrm{D}=3$ for illustrative purposes. The chosen efficiency function goes to one when $t_\mathrm{dur}\gg t_\mathrm{cad}$ and to zero when $t_\mathrm{dur}\ll t_\mathrm{cad}$. When $t_\mathrm{dur} = N_\mathrm{D}t_\mathrm{cad}$, the efficiency is 0.5.

In Figure \ref{fig:rate}, we assume a limiting $R$-band limiting magnitude of 24.5 (matched to LSST) and calculate $\mathcal{N}$ for a given transient's absolute magnitude and duration assuming a constant volumetric rate and ignoring any k-corrections. We find that the expected detection rates of transients drop off exponentially decreasing luminosity, as well as with shorter duration as it approaches the survey cadence. Specifically, this simple example demonstrates the fact that, even with a relatively high cadence, a wide-field survey will detect 100 -- 1000 times more SN-like transients ($M_\mathrm{R}\lesssim-18$ mag) compared to ILOTs ($M_\mathrm{R}\sim-10$ to $-14$ mag). Similarly, a survey with a cadence of several ($\approx 3$) days will detect 10 -- 100 times more transients with SN-like durations ($\sim20-30$ days) compared to transients with short durations ($\lesssim10$ days). To counter these facts, one could design a survey with a faster survey cadence, but there is a trade off between a survey's cadence, depth and coverage area. A high survey cadence requires a much smaller coverage area, even with a large field of view. A more efficient approach to search for dim transients may be a targeted survey of nearby galaxies.

Bringing together the above conclusions with those of the previous section, we conclude that quickly evolving transients are invariably dim. Both characteristics lead to diminishing survey potential and therefore lower observed rates. In contrast, bright ($M_\mathrm{R}\lesssim-22$ mag) transients tend to have longer durations and are therefore easier to observe. However, their volumetric rates are low and have not typically been found in large numbers in untargeted surveys. Between these two regimes, SN-like transients with relatively bright luminosities and intermediate-durations typically have higher volumetric rates, allowing them to be some of the most commonly observed extragalactic phenomena in wide-field surveys.

\section{Conclusion}\label{sec:conclusion}

We utilized semi-analytical, one-zone models to explore a wide range of heating sources that are either known to or expected to power optical transients.  For each heating source we generated model light curves for a physically motivated set of parameters. We also investigated the effects of the parameters on the light curves and the locus of simulated light curves within the DLPS. Our main conclusions are as follows:

\begin{itemize}
    \item Most model transients lie at $t_\mathrm{dur}\sim20-100$ days and $M_\mathrm{R}\sim-18$ to $-20$ mag, consistent with the observed properties of the bulk of optical transients.
    \item Only sources with relativistic expansion can produce luminous ($M_\mathrm{R}<-18$ mag) and fast ($t_\mathrm{dur}<10$ days) transients.  However, such sources (i.e., GRBs) are known to be intrinsically rare.
    \item Luminosity and duration are positively correlated for most heating sources, implying that short-duration transients ($t_{\rm dur}\lesssim 10$ days) also have low peak luminosities ($M_R\gtrsim 15$ mag). 
    \item There is a paucity of heating sources that produce transients in the luminosity gap between classical novae and SNe ($M_\mathrm{R}\sim-10$ to $-14$ mag) with most models in this regime powered by CSM interaction with low ejecta masses relevant to stellar eruptions rather than explosions.  
    \item Transients with short duration and/or low luminosity are exponentially more difficult to detect in a wide field time-domain survey as the survey cadence approaches the transient duration.  Since fast transients have low luminosity this implies an even more significant reduction in the survey potential for fast transients.
\end{itemize}

The rarity of fast and luminous transients seems unavoidable given our understanding of basic physical processes in optical transients; therefore fast transients will mostly be dim. In addition to this fact, the relative difficulty of detecting short and dim transients (compared to luminous and long duration) is inherent to any time-domain survey. 
We argue that our approach is essential for the survey designs of future missions (e.g., LSST, WFIRST, etc).  Also, we argue that rapid cadence may be more relevant for capturing early phases in the evolution of ``slow" transients, rather than for the discovery of intrinsically fast transients.

\acknowledgments
We thank Matt Nicholl and Peter Blanchard for valuable discussion and feedback on this work. The Berger Time-Domain Group at Harvard is supported in
part by the NSF under grant AST-1411763 and by NASA under
grant NNX15AE50G. VAV acknowledges support by the National Science Foundation through a Graduate Research Fellowship. This paper
greatly benefited from the Open Supernova Catalog \citep{guillochon2017open}.

\newpage
\clearpage
\appendix

\section{Detailed CSM Models}
In this appendix, we outline the CSM models used in this work in greater detail. Broadly speaking, both the \cite{chatzopoulos2012generalized} and \cite{ofek2014sn} models follow the interaction of a supernova's ejecta (whose ejecta density distribution is spherically symmetric and a described by broken power-law) and a pre-existing CSM (whose density distribution is spherically symmetric and described by a single power-law). This interaction produces forward and reverse shocks which independently power the optical transient as their kinetic energy is converted into radiation \citep{chatzopoulos2012generalized}:

\begin{equation}\label{eqn:csm_lum}
    L = \epsilon\frac{dE_\mathrm{KE}}{dt} = \epsilon\frac{d}{dt}\left(\frac{1}{2}M_\mathrm{sw}v_\mathrm{sh}^2\right)
    = \epsilon M_\mathrm{sw}v_\mathrm{sh}\frac{dv_\mathrm{sh}}{dt}+\epsilon\frac{1}{2}\frac{dM_\mathrm{sw}}{dt}v_\mathrm{sh}^2
\end{equation}
where $\epsilon=0.5$ is an efficiency factor, $r_\mathrm{sh}$ is the shock (forward or reverse) radius at time $t$, $v_\mathrm{sh}=dr_\mathrm{sh}/dt$ is the shock's velocity at time $t$, $M_\mathrm{sw}=4\pi\int_{R_0}^{r_\mathrm{fs}}\rho_\mathrm{CSM}(r)r^2dr$ is the swept-up CSM mass at time $t$ and $\rho_\mathrm{CSM}(r)$ is the CSM density at radius $r$. \cite{dessart2015numerical} found that the conversion efficiency ($\epsilon$) depends on the ratio of the CSM and ejecta masses, reaching as low as $\epsilon\sim0.3$  and as high as $\epsilon\sim0.7$. 

The progenitor star is embedded in spherically symmetric CSM shell described by a power law ($\rho_\mathrm{CSM}(r)=qr^{-s}$; \citealt{chevalier1994emission}), where $q=\rho_{\mathrm{CSM}}R_0^s$. Note that $\rho_\mathrm{CSM}$ is a constant while $\rho_\mathrm{CSM}(r)$ is a function of $r$ with $\rho_\mathrm{CSM}(R_0)=\rho_\mathrm{CSM}$ at $r=R_0$. The index of the CSM profile can vary from $s=0$ (roughly corresponding to shell-like, eruptive mass-loss histories) to $s=2$ (a wind mass-loss history). The SN ejecta's density profile profile $\rho_{SN}=g^nt^{n-3}r^{-n}$ is described by a broken power law (as described in \citealt{chevalier1994emission} and seen observationally in SN 1987A): 

\begin{equation}
    g^n = \frac{1}{4\pi(\delta-n)}\frac{\Big(2(5-\delta)(n-5)E_{KE}\Big)^{\frac{n-3}{2}}}{\Big(\left(3-\delta\right)\left(n-3\right)M_\mathrm{ej}\Big)^{\frac{n-5}{2}}}
\end{equation}
where $\delta$ is the index of the inner profile and $n$ is the index of the outer profile. The light curves are fairly insensitive to $\delta$, so we set its value to 0. The value of $n$ depends on the polytropic index of the progenitor star, varying from 7-- 12 for convective to degenerate cores. 

The self-similar solutions for the forward and reverse shocks are, respectively \citep{chatzopoulos2012generalized,chevalier1994emission}:

\begin{equation}
    r_\mathrm{fs}=R_0+\beta_F\left(\frac{Ag^n}{q}\right)t^\frac{n-3}{n-s}
\end{equation}
and
\begin{equation}
    r_\mathrm{rs}=R_0+\beta_R\left(\frac{Ag^n}{q}\right)t^\frac{n-3}{n-s}
\end{equation}
$\beta_F$, $\beta_R$ and $A$ are constants which depend on $n$ and $s$ and are order unity in most cases. We use interpolated values from those listed in \cite{chevalier1994emission}. 

\subsection{Recovering the Cacoon SBO Solution}
If we assume that the shock deceleration is small $dv_\mathrm{sh}/dt=0$ and the geometric factors $\beta_F=\beta_R=1$, we can recover the bolometric luminosity solution of \cite{ofek2014sn} from Equation \ref{eqn:csm_lum}:

\begin{equation}
    L = \epsilon\frac{dE_\mathrm{KE}}{dt} = \epsilon\frac{\epsilon}{2}\frac{dM_\mathrm{sw}}{dt}v_\mathrm{sh}=2\pi\epsilon\rho_\mathrm{CSM}(r_\mathrm{sh})r_\mathrm{sh}^2v_\mathrm{sh}^3
\end{equation}

Furthermore, we assume that this shock efficiently diffuses through the CSM and has an effective diffusion time $t_\mathrm{d}=0$ (i.e., the input luminosity is equal to the observed luminosity; \citealt{ofek2014sn}). 

The temperature can then be estimated as \citep{chevalier2011shock, ofek2014sn}:

\begin{equation}
    T(t) = \left(\frac{18}{7a}\rho_\mathrm{CSM} v_\mathrm{fs}^2\right)^{1/4}
\end{equation}
where a is the radiation constant. 

\subsubsection{Full CSM interaction Solution}
If we loosen the assumptions made to reproduce the light curve solution from \cite{ofek2014sn}, we will reproduce the generalized solution presented by \cite{chatzopoulos2012generalized}. To do this, we calculate the contributions to the total luminosity from both the forward and reverse shocks and diffuse this input luminosity through the CSM. 

We explore both shell-like and wind-like CSM profiles ($s=0$ and $s=2$, respectively) and leave the inner radius of the CSM as a free parameter ($R_0$). By also allowing the total mass of the CSM to be a free parameter ($M_{\mathrm{CSM}}$), we can define the total radius of the CSM as:

\begin{equation}
    R_\mathrm{CSM} = \left(\frac{3M_\mathrm{CSM}}{4\pi q}+R_0^{3}\right)^{\frac{1}{3}}
\end{equation}

We can further define the photospheric radius as:

\begin{equation}
    R_\mathrm{ph} = \frac{2\kappa q}{3}+R_\mathrm{CSM}
\end{equation}
where $\kappa=0.34$ g cm$^{-3}$. $R_\mathrm{CSM}$ and $R_\mathrm{ph}$ will be important for setting physical constraints on our generated models.

The input luminosity arises from the conversion of the forward and reverse shocks' kinetic energy into radiation, which can be described as:
\begin{multline}
    L_\mathrm{in}(t)=\frac{2\pi\epsilon}{(n-s)^3}g^{n\frac{5-s}{n-s}}q^{\frac{n-5}{n-s}}(n-3)^2(n-5)\beta_F^{5-s} A^{\frac{5-s}{n-s}} t^{\frac{2n+6s-ns-15}{n-s}}\theta(t_\mathrm{FS}-t) + 2\pi\epsilon \left(\frac{Ag^n}{q}\right)^{\frac{5-n}{n-s}}\beta_R^{5-n}g^n\\\times\left(\frac{3-s}{n-s}\right)^3 t^{\frac{2n+6s-ns-15}{n-s}}\theta(t_\mathrm{RS}-t)
\end{multline}
$\theta$ is the Heaviside step function, which designates which components (the forward or reverse shocks) are contributing to the total luminosity based on the shock termination times:

\begin{align}
    t_\mathrm{FS} &= \left|\frac{(3-s)q^{\frac{3-n}{n-s}}(Ag^n)^{\frac{s-3}{n-s}}}{4\pi\beta_F^{3-s}}\right|^\frac{n-s}{(n-3)(3-s)}M_\mathrm{CSM}^\frac{n-s}{(n-3)(3-s)} \\
    t_\mathrm{RS} &= \left(\left(\frac{v_\mathrm{ph}}{\beta_R}\right)\left(\frac{q}{Ag^n}^\frac{1}{n-s}\right)\left(1-\frac{(3-n)M_\mathrm{ej}}{4\pi v_\mathrm{ph}^{3-n}g^n}\right)^{\frac{1}{3-n}}\right)^\frac{n-s}{s-3}
\end{align}

\newpage

\setlength{\extrarowheight}{10pt}
\renewcommand{\arraystretch}{0.6}
\begin{table}
\centering
\caption{Engines and Parameters}
\begin{tabularx}{\textwidth}{X X X X X X X}
\hline\hline
\multicolumn{7}{l}{\textbf{Adiabatic Expansion (White Dwarf)}} \\
$M$\textsubscript{ej}/M\textsubscript{$\odot$} & $E$\textsubscript{KE}/$10^{51}$ erg & $R$\textsubscript{0}/R\textsubscript{$\odot$} &  &  & &\\
 $0.1 - 1$ & $0.01 - 1^{*}$ & 0.01 &  &  & &$\vspace{0.1cm}$\\
\hline
\multicolumn{7}{l}{\textbf{Adiabatic Expansion (BSG)}} \\
$M$\textsubscript{ej}/M\textsubscript{$\odot$} & $E$\textsubscript{KE}/$10^{51}$ erg & $R$\textsubscript{0}/R\textsubscript{$\odot$} &  &  & &\\ 
 $1 - 10$ & $0.01 - 1^*$ & 10 &  &  & & $\vspace{0.1cm}$\\
\hline
\multicolumn{7}{l}{\textbf{Adiabatic Expansion (RSG)}} \\
$M$\textsubscript{ej}/M\textsubscript{$\odot$} & $E$\textsubscript{KE}/$10^{51}$ erg & $R$\textsubscript{0}/R\textsubscript{$\odot$} & & &  &\\ 
$ 1 - 10$ & $0.01 - 1^*$ & 500 &  &  &  & $\vspace{0.1cm}$\\
\hline
\multicolumn{7}{l}{\textbf{\textsuperscript{56}Ni (Ib/c Supernovae)}} \\
$M$\textsubscript{ej}/M\textsubscript{$\odot$}  & $E$\textsubscript{KE}/$10^{51}$ erg & $F$\textsubscript{Ni} &  &  & &  \\
 $1 - 10$ & $1 - 10$ & $0.01 - 0.15$ &  &  & & $\vspace{0.1cm}$\\
\hline
\multicolumn{7}{l}{\textbf{\textsuperscript{56}Ni (WD Eruptions, Iax Supernovae)}} \\
$M$\textsubscript{ej}/M\textsubscript{$\odot$} & $E$\textsubscript{KE}/$10^{51}$ erg & $F$\textsubscript{Ni} &  &  & & \\
 $0.01 - 1.0^*$ & $0.01 - 1^*$  & $0.1 - 0.3$ &  &  & &$\vspace{0.1cm}$\\
\hline
\multicolumn{7}{l}{\textbf{\textsuperscript{56}Ni (Pair Instability Supernovae)}} \\
$M$\textsubscript{ej}/M\textsubscript{$\odot$} & $E$\textsubscript{KE}/$10^{51}$ erg & $F$\textsubscript{Ni} &  &  & & \\
 $50 - 250$ & $10 - 100$ & $0.01 - 0.3$ & & & & $\vspace{0.1cm}$\\
 \hline
\multicolumn{7}{l}{\textbf{CSM Interaction (Supernova, Shell)}} \\
$M$\textsubscript{ej}/M\textsubscript{$\odot$} & $E$\textsubscript{KE}/$10^{51}$ erg & $M$\textsubscript{CSM}/M\textsubscript{$\odot$} & $\rho$/cm$^{-3}$ & $n$ & $R$\textsubscript{0}/AU \\ 
 $1 - 10$ & $1 - 10$ & $0.1 - 10^*$ & $10^{-17}-10^{-13*}$ & $7-12$ & $1 - 100^*$ $\vspace{0.1cm}$\\
 \hline
\multicolumn{7}{l}{\textbf{CSM Interaction (Supernova, Wind)}}\\
$M$\textsubscript{ej}/M\textsubscript{$\odot$} & $E$\textsubscript{KE}/$10^{51}$ erg & $M$\textsubscript{CSM}/M\textsubscript{$\odot$} & $\rho$/cm$^{-3}$ & $n$ & $\delta$ & $R$\textsubscript{0}/R\textsubscript{$\odot$} \\
 $1 - 10$ & $1 - 10$ & $0.1 - 10^*$ & $10^{-17}-10^{-13*}$ & $7-12$ & 0 &$1 - 100^*$ $\vspace{0.1cm}$\\
 \hline
\multicolumn{7}{l}{\textbf{CSM Interaction (ILOT/outburst-like)}} \\
$M$\textsubscript{ej}/M\textsubscript{$\odot$} & $v$/km s$^{-1}$ & $M$\textsubscript{CSM}/M\textsubscript{$\odot$} & $\rho$/cm$^{-3}$ & n & $R$\textsubscript{0}/R\textsubscript{$\odot$} \\
 $0.1 - 10^*$ & $100 - 1000$ & $0.1 - 10^*$ & $10^{-17}-10^{-13*}$ & $7-12$ &  $1 - 100^*$ $\vspace{0.1cm}$\\
\hline
\multicolumn{7}{l}{\textbf{Hydrogen Recombination (IIP/L Supernovae)}} \\
$M$\textsubscript{ej}/M\textsubscript{$\odot$} & $E$\textsubscript{KE}/$10^{51}$ erg & $R_0$/R$_\odot$ &  &  & & \\ 
 $5 - 18$ & $0.1 - 5$ & $100 - 1000$ &  &  & &$\vspace{0.1cm}$\\
 \hline
 \multicolumn{7}{l}{\textbf{Magnetar Spin-down}} \\
$M$\textsubscript{ej}/M\textsubscript{$\odot$} & $E$\textsubscript{KE}/$10^{51}$ erg & $P$/ms & $B$/$10^{14}$G &  & & \\ 
 $1 - 10$ & $1 - 10$ & $1 - 10$ & $0.1 - 10^{*}$ & &  &$\vspace{0.1cm}$\\
\hline
\multicolumn{7}{l}{\textbf{$r$-process Decay (Red Kilonovae)}} \\
$M$\textsubscript{ej}/M\textsubscript{$\odot$} & $v$/1000 km s$^{-1}$ & $\kappa$/cm$^2$g$^{-1}$ &  &  & & \\
 $0.001 - 0.1$ & $30 - 90$ & $10 - 100$ &  &  &  & $\vspace{0.1cm}$\\
\hline
\multicolumn{7}{l}{\textbf{$r$-process Decay (Blue Kilonovae)}} \\
$M$\textsubscript{ej}/M\textsubscript{$\odot$} & $v$/1000 km s$^{-1}$ & $\kappa$/cm$^2$g$^{-1}$ &  &  & & \\ 
 $0.001 - 0.1$ & $30 - 90$ & 0.2 &  &  &  & $\vspace{0.1cm}$\\
\hline\hline
\multicolumn{7}{r}{{Note: $^{*}$ Sampled log-uniformly. }}\\

\label{tab:models}
\end{tabularx}
\end{table}

\newpage
\begin{table}
\centering
\caption{\texttt{MOSFiT} Modules}
\begin{tabularx}{\textwidth}{X X X X X X}
Heating Source & Engine & Diffusion & Photosphere & SED & Constraints\\
\hline
None (Adiabatic)$^*$ & - & - & - & - & - \\
$^{56}$Ni Decay & \texttt{nickelcolbalt} & \texttt{diffusion} & \texttt{temperature\_floor} & \texttt{blackbody} & -\\
CSM Interaction & \texttt{csm} & \texttt{diffusion\_csm} & \texttt{temperature\_floor} & \texttt{blackbody} & \texttt{csmconstraints}\\
Hydrogen Recombination$^*$  & - & - & - & - & -\\
Magnetar Spin-down & \texttt{magnetar} & \texttt{diffusion} & \texttt{temperature\_floor} & \texttt{blackbody} & -\\
$r$-process Decay & \texttt{rprocess} & \texttt{diffusion} & \texttt{temperature\_floor } & \texttt{blackbody} & -\\
\hline\hline
\multicolumn{6}{r}{{Note: $^{*}$ Not implemented in \texttt{MOSFiT} }}\\

\label{tab:mosfit}
\end{tabularx}
\end{table}

\bibliographystyle{apj}
\bibliography{mybib}

\begin{thebibliography}{}
\expandafter\ifx\csname natexlab\endcsname\relax\def\natexlab#1{#1}\fi

\bibitem[{Alexander {et~al.}(2016)Alexander, Berger, Guillochon, Zauderer, \&
  Williams}]{alexander2016discovery}
Alexander, K.~D., Berger, E., Guillochon, J., Zauderer, B.~A., \& Williams, P.
  K.~G. 2016, The Astrophysical Journal Letters, 819, L25

\bibitem[{Arcavi {et~al.}(2014)Arcavi, Gal-Yam, Sullivan, Pan, Cenko, Horesh,
  Ofek, De~Cia, Yan, Yang, {et~al.}}]{arcavi2014continuum}
Arcavi, I., Gal-Yam, A., Sullivan, M., {et~al.} 2014, The Astrophysical
  Journal, 793, 38

\bibitem[{Arcavi {et~al.}(2016)Arcavi, Wolf, Howell, Bildsten, Leloudas,
  Hardin, Prajs, Perley, Svirski, Gal-Yam, {et~al.}}]{arcavi2016rapidly}
Arcavi, I., Wolf, W.~M., Howell, D.~A., {et~al.} 2016, The Astrophysical
  Journal, 819, 35

\bibitem[{Arnett \& Livne(1994)}]{arnett1994delayed}
Arnett, D., \& Livne, E. 1994, The Astrophysical Journal, 427, 315

\bibitem[{Arnett(1979)}]{arnett1979theory}
Arnett, W.~D. 1979, The Astrophysical Journal, 230, L37

\bibitem[{Arnett(1980)}]{arnett1980analytic}
---. 1980, The Astrophysical Journal, 237, 541

\bibitem[{Arnett(1982)}]{arnett1982type}
---. 1982, The Astrophysical Journal, 253, 785

\bibitem[{Baade(1938)}]{baade1938absolute}
Baade, W. 1938, The Astrophysical Journal, 88, 285

\bibitem[{Bailey {et~al.}(2007)Bailey, Aragon, Romano, Thomas, Weaver, \&
  Wong}]{bailey2007find}
Bailey, S., Aragon, C., Romano, R., {et~al.} 2007, The Astrophysical Journal,
  665, 1246

\bibitem[{Bailyn \& Grindlay(1990)}]{bailyn1990neutron}
Bailyn, C.~D., \& Grindlay, J.~E. 1990, The Astrophysical Journal, 353, 159

\bibitem[{Barkat {et~al.}(1967)Barkat, Rakavy, \& Sack}]{barkat1967dynamics}
Barkat, Z., Rakavy, G., \& Sack, N. 1967, Physical Review Letters, 18, 379

\bibitem[{Barnes \& Kasen(2013)}]{barnes2013effect}
Barnes, J., \& Kasen, D. 2013, The Astrophysical Journal, 775, 18

\bibitem[{Barnes {et~al.}(2016)Barnes, Kasen, Wu, \&
  Mart{\'\i}nez-Pinedo}]{barnes2016radioactivity}
Barnes, J., Kasen, D., Wu, M.-R., \& Mart{\'\i}nez-Pinedo, G. 2016, The
  Astrophysical Journal, 829, 110

\bibitem[{{Beloborodov}(2014)}]{2014MNRAS.438..169B}
{Beloborodov}, A.~M. 2014, \mnras, 438, 169

\bibitem[{Berger(2014)}]{berger2014short}
Berger, E. 2014, Annual review of Astronomy and Astrophysics, 52, 43

\bibitem[{Berger {et~al.}(2013{\natexlab{a}})Berger, Fong, \&
  Chornock}]{berger2013r}
Berger, E., Fong, W., \& Chornock, R. 2013{\natexlab{a}}, The Astrophysical
  Journal Letters, 774, L23

\bibitem[{Berger {et~al.}(2013{\natexlab{b}})Berger, Leibler, Chornock, Rest,
  Foley, Soderberg, Price, Burgett, Chambers, Flewelling,
  {et~al.}}]{berger2013search}
Berger, E., Leibler, C., Chornock, R., {et~al.} 2013{\natexlab{b}}, The
  Astrophysical Journal, 779, 18

\bibitem[{Bersten \& Hamuy(2009)}]{bersten2009bolometric}
Bersten, M.~C., \& Hamuy, M. 2009, The Astrophysical Journal, 701, 200

\bibitem[{Betoule {et~al.}(2014)Betoule, Kessler, Guy, Mosher, Hardin, Biswas,
  Astier, El-Hage, Konig, Kuhlmann, {et~al.}}]{betoule2014improved}
Betoule, M., Kessler, R., Guy, J., {et~al.} 2014, Astronomy \& Astrophysics,
  568, A22

\bibitem[{Bildsten {et~al.}(2007)Bildsten, Shen, Weinberg, \&
  Nelemans}]{bildsten2007faint}
Bildsten, L., Shen, K.~J., Weinberg, N.~N., \& Nelemans, G. 2007, The
  Astrophysical Journal Letters, 662, L95

\bibitem[{{Blagorodnova} {et~al.}(2017){Blagorodnova}, {Kotak}, {Polshaw},
  {Kasliwal}, {Cao}, {Cody}, {Doran}, {Elias-Rosa}, {Fraser}, {Fremling},
  {Gonzalez-Fernandez}, {Harmanen}, {Jencson}, {Kankare}, {Kudritzki},
  {Kulkarni}, {Magnier}, {Manulis}, {Masci}, {Mattila}, {Nugent}, {Ochner},
  {Pastorello}, {Reynolds}, {Smith}, {Sollerman}, {Taddia}, {Terreran},
  {Tomasella}, {Turatto}, {Vreeswijk}, {Wozniak}, \&
  {Zaggia}}]{2017ApJ...834..107B}
{Blagorodnova}, N., {Kotak}, R., {Polshaw}, J., {et~al.} 2017, \apj, 834, 107

\bibitem[{Brown {et~al.}(2014)Brown, Breeveld, Holland, Kuin, \&
  Pritchard}]{brown2014sousa}
Brown, P.~J., Breeveld, A.~A., Holland, S., Kuin, P., \& Pritchard, T. 2014,
  Astrophysics and Space Science, 354, 89

\bibitem[{Cao {et~al.}(2012)Cao, Kasliwal, Neill, Kulkarni, Lou, Ben-Ami,
  Bloom, Cenko, Law, Nugent, {et~al.}}]{cao2012classical}
Cao, Y., Kasliwal, M.~M., Neill, J.~D., {et~al.} 2012, The Astrophysical
  Journal, 752, 133

\bibitem[{{Capaccioli} {et~al.}(1990){Capaccioli}, {della Valle}, {D'Onofrio},
  \& {Rosino}}]{1990ApJ...360...63C}
{Capaccioli}, M., {della Valle}, M., {D'Onofrio}, M., \& {Rosino}, L. 1990,
  \apj, 360, 63

\bibitem[{{Charnock} \& {Moss}(2017)}]{2017ApJ...837L..28C}
{Charnock}, T., \& {Moss}, A. 2017, \apjl, 837, L28

\bibitem[{Chatzopoulos {et~al.}(2012)Chatzopoulos, Wheeler, \&
  Vinko}]{chatzopoulos2012generalized}
Chatzopoulos, E., Wheeler, J.~C., \& Vinko, J. 2012, The Astrophysical Journal,
  746, 121

\bibitem[{Chatzopoulos {et~al.}(2013)Chatzopoulos, Wheeler, Vink{\'o}, Horvath,
  \& Nagy}]{chatzopoulos2013analytical}
Chatzopoulos, E., Wheeler, J.~C., Vink{\'o}, J., Horvath, Z., \& Nagy, A. 2013,
  The Astrophysical Journal, 773, 76

\bibitem[{Chevalier(1982)}]{chevalier1982self}
Chevalier, R.~A. 1982, The Astrophysical Journal, 258, 790

\bibitem[{Chevalier \& Fransson(1994)}]{chevalier1994emission}
Chevalier, R.~A., \& Fransson, C. 1994, The Astrophysical Journal, 420, 268

\bibitem[{Chevalier \& Irwin(2011)}]{chevalier2011shock}
Chevalier, R.~A., \& Irwin, C.~M. 2011, The Astrophysical Journal Letters, 729,
  L6

\bibitem[{{Childress} {et~al.}(2015){Childress}, {Hillier}, {Seitenzahl},
  {Sullivan}, {Maguire}, {Taubenberger}, {Scalzo}, {Ruiter}, {Blagorodnova},
  {Camacho}, {Castillo}, {Elias-Rosa}, {Fraser}, {Gal-Yam}, {Graham}, {Howell},
  {Inserra}, {Jha}, {Kumar}, {Mazzali}, {McCully}, {Morales-Garoffolo},
  {Pandya}, {Polshaw}, {Schmidt}, {Smartt}, {Smith}, {Sollerman}, {Spyromilio},
  {Tucker}, {Valenti}, {Walton}, {Wolf}, {Yaron}, {Young}, {Yuan}, \&
  {Zhang}}]{2015MNRAS.454.3816C}
{Childress}, M.~J., {Hillier}, D.~J., {Seitenzahl}, I., {et~al.} 2015, \mnras,
  454, 3816

\bibitem[{Chornock {et~al.}(2013)Chornock, Berger, Gezari, Zauderer, Rest,
  Chomiuk, Kamble, Soderberg, Czekala, Dittmann,
  {et~al.}}]{chornock2013ultraviolet}
Chornock, R., Berger, E., Gezari, S., {et~al.} 2013, The Astrophysical Journal,
  780, 44

\bibitem[{Colgate \& McKee(1969)}]{colgate1969early}
Colgate, S.~A., \& McKee, C. 1969, The Astrophysical Journal, 157, 623

\bibitem[{{Contreras} {et~al.}(2010){Contreras}, {Hamuy}, {Phillips},
  {Folatelli}, {Suntzeff}, {Persson}, {Stritzinger}, {Boldt}, {Gonz{\'a}lez},
  {Krzeminski}, {Morrell}, {Roth}, {Salgado}, {Jos{\'e} Maureira}, {Burns},
  {Freedman}, {Madore}, {Murphy}, {Wyatt}, {Li}, \&
  {Filippenko}}]{2010AJ....139..519C}
{Contreras}, C., {Hamuy}, M., {Phillips}, M.~M., {et~al.} 2010, \aj, 139, 519

\bibitem[{Dahlen {et~al.}(2004)Dahlen, Strolger, Riess, Mobasher, Chary,
  Conselice, Ferguson, Fruchter, Giavalisco, Livio, {et~al.}}]{dahlen2004high}
Dahlen, T., Strolger, L.-G., Riess, A.~G., {et~al.} 2004, The Astrophysical
  Journal, 613, 189

\bibitem[{Darbha {et~al.}(2010)Darbha, Metzger, Quataert, Kasen, Nugent, \&
  Thomas}]{darbha2010nickel}
Darbha, S., Metzger, B., Quataert, E., {et~al.} 2010, Monthly Notices of the
  Royal Astronomical Society, 409, 846

\bibitem[{{Darbha} {et~al.}(2010){Darbha}, {Metzger}, {Quataert}, {Kasen},
  {Nugent}, \& {Thomas}}]{2010MNRAS.409..846D}
{Darbha}, S., {Metzger}, B.~D., {Quataert}, E., {et~al.} 2010, \mnras, 409, 846

\bibitem[{Della~Valle \& Livio(1995)}]{della1995calibration}
Della~Valle, M., \& Livio, M. 1995, The Astrophysical Journal, 452, 704

\bibitem[{{Demircan} \& {Kahraman}(1991)}]{1991Ap&SS.181..313D}
{Demircan}, O., \& {Kahraman}, G. 1991, \apss, 181, 313

\bibitem[{Dessart {et~al.}(2015)Dessart, Audit, \&
  Hillier}]{dessart2015numerical}
Dessart, L., Audit, E., \& Hillier, D.~J. 2015, Monthly Notices of the Royal
  Astronomical Society, 449, 4304

\bibitem[{Dessart {et~al.}(2006)Dessart, Burrows, Ott, Livne, Yoon, \&
  Langer}]{dessart2006multidimensional}
Dessart, L., Burrows, A., Ott, C., {et~al.} 2006, The Astrophysical Journal,
  644, 1063

\bibitem[{Dessart {et~al.}(2012)Dessart, Hillier, Waldman, Livne, \&
  Blondin}]{dessart2012superluminous}
Dessart, L., Hillier, D.~J., Waldman, R., Livne, E., \& Blondin, S. 2012,
  Monthly Notices of the Royal Astronomical Society: Letters, 426, L76

\bibitem[{Dessart {et~al.}(2013)Dessart, Waldman, Livne, Hillier, \&
  Blondin}]{dessart2013radiative}
Dessart, L., Waldman, R., Livne, E., Hillier, D.~J., \& Blondin, S. 2013,
  Monthly Notices of the Royal Astronomical Society, 428, 3227

\bibitem[{Dong {et~al.}(2016)Dong, Shappee, Prieto, Jha, Stanek, Holoien,
  Kochanek, Thompson, Morrell, Thompson, {et~al.}}]{dong2016asassn}
Dong, S., Shappee, B., Prieto, J., {et~al.} 2016, Science, 351, 257

\bibitem[{Drout {et~al.}(2014)Drout, Chornock, Soderberg, Sanders, McKinnon,
  Rest, Foley, Milisavljevic, Margutti, Berger, {et~al.}}]{drout2014rapidly}
Drout, M., Chornock, R., Soderberg, A.~M., {et~al.} 2014, The Astrophysical
  Journal, 794, 23

\bibitem[{Drout {et~al.}(2011)Drout, Soderberg, Gal-Yam, Cenko, Fox, Leonard,
  Sand, Moon, Arcavi, \& Green}]{drout2011first}
Drout, M.~R., Soderberg, A.~M., Gal-Yam, A., {et~al.} 2011, The Astrophysical
  Journal, 741, 97

\bibitem[{{Ferretti} {et~al.}(2016){Ferretti}, {Amanullah}, {Goobar},
  {Johansson}, {Vreeswijk}, {Butler}, {Cao}, {Cenko}, {Doran}, {Filippenko},
  {Freeland}, {Hosseinzadeh}, {Howell}, {Lundqvist}, {Mattila}, {Nordin},
  {Nugent}, {Petrushevska}, {Valenti}, {Vogt}, \&
  {Wozniak}}]{2016yCat..35920040F}
{Ferretti}, R., {Amanullah}, R., {Goobar}, A., {et~al.} 2016, VizieR Online
  Data Catalog, 359

\bibitem[{Firth {et~al.}(2015)Firth, Sullivan, Gal-Yam, Howell, Maguire,
  Nugent, Piro, Baltay, Feindt, Hadjiyksta, {et~al.}}]{firth2015rising}
Firth, R., Sullivan, M., Gal-Yam, A., {et~al.} 2015, Monthly Notices of the
  Royal Astronomical Society, 446, 3895

\bibitem[{Foley {et~al.}(2013)Foley, Challis, Chornock, Ganeshalingam, Li,
  Marion, Morrell, Pignata, Stritzinger, Silverman, {et~al.}}]{foley2013type}
Foley, R.~J., Challis, P.~J., Chornock, R., {et~al.} 2013, The Astrophysical
  Journal, 767, 57

\bibitem[{Fong {et~al.}(2015)Fong, Berger, Margutti, \&
  Zauderer}]{fong2015decade}
Fong, W.-f., Berger, E., Margutti, R., \& Zauderer, B.~A. 2015, The
  Astrophysical Journal, 815, 102

\bibitem[{Frank \& Rees(1976)}]{frank1976effects}
Frank, J., \& Rees, M.~J. 1976, Monthly Notices of the Royal Astronomical
  Society, 176, 633

\bibitem[{Fryer {et~al.}(1999)Fryer, Benz, Herant, \& Colgate}]{fryer1999can}
Fryer, C., Benz, W., Herant, M., \& Colgate, S.~A. 1999, The Astrophysical
  Journal, 516, 892

\bibitem[{Gal-Yam(2012)}]{gal2012luminous}
Gal-Yam, A. 2012, Science, 337, 927

\bibitem[{Gallagher \& Starrfield(1978)}]{gallagher1978theory}
Gallagher, J., \& Starrfield, S. 1978, Annual review of astronomy and
  astrophysics, 16, 171

\bibitem[{{Ganeshalingam} {et~al.}(2010){Ganeshalingam}, {Li}, {Filippenko},
  {Anderson}, {Foster}, {Gates}, {Griffith}, {Grigsby}, {Joubert}, {Leja},
  {Lowe}, {Macomber}, {Pritchard}, {Thrasher}, \&
  {Winslow}}]{2010ApJS..190..418G}
{Ganeshalingam}, M., {Li}, W., {Filippenko}, A.~V., {et~al.} 2010, \apjs, 190,
  418

\bibitem[{Gezari {et~al.}(2006)Gezari, Martin, Milliard, Basa, Halpern,
  Forster, Friedman, Morrissey, Neff, Schiminovich,
  {et~al.}}]{gezari2006ultraviolet}
Gezari, S., Martin, D., Milliard, B., {et~al.} 2006, The Astrophysical Journal
  Letters, 653, L25

\bibitem[{Gezari {et~al.}(2012)Gezari, Chornock, Rest, Huber, Forster, Berger,
  Challis, Neill, Martin, Heckman, {et~al.}}]{gezari2012ultraviolet}
Gezari, S., Chornock, R., Rest, A., {et~al.} 2012, Nature, 485, 217

\bibitem[{Goranskij {et~al.}(2016)Goranskij, Barsukova, Spiridonova, Valeev,
  Fatkhullin, Moskvitin, Vozyakova, Cheryasov, Safonov, Zharova,
  {et~al.}}]{goranskij2016photometry}
Goranskij, V., Barsukova, E., Spiridonova, O., {et~al.} 2016, Astrophysical
  Bulletin, 71, 82

\bibitem[{Guillochon {et~al.}(2014)Guillochon, Manukian, \&
  Ramirez-Ruiz}]{guillochon2014ps1}
Guillochon, J., Manukian, H., \& Ramirez-Ruiz, E. 2014, The Astrophysical
  Journal, 783, 23

\bibitem[{Guillochon {et~al.}(2017)Guillochon, Parrent, Kelley, \&
  Margutti}]{guillochon2017open}
Guillochon, J., Parrent, J., Kelley, L.~Z., \& Margutti, R. 2017, The
  Astrophysical Journal, 835, 64

\bibitem[{Guillochon \& Ramirez-Ruiz(2013)}]{guillochon2013hydrodynamical}
Guillochon, J., \& Ramirez-Ruiz, E. 2013, The Astrophysical Journal, 767, 25

\bibitem[{Guillochon {et~al.}(2009)Guillochon, Ramirez-Ruiz, Rosswog, \&
  Kasen}]{guillochon2009three}
Guillochon, J., Ramirez-Ruiz, E., Rosswog, S., \& Kasen, D. 2009, The
  Astrophysical Journal, 705, 844

\bibitem[{Guy {et~al.}(2005)Guy, Astier, Nobili, Regnault, \&
  Pain}]{guy2005salt}
Guy, J., Astier, P., Nobili, S., Regnault, N., \& Pain, R. 2005, Astronomy \&
  Astrophysics, 443, 781

\bibitem[{Hamuy(2003)}]{hamuy2003observed}
Hamuy, M. 2003, The Astrophysical Journal, 582, 905

\bibitem[{{Hicken} {et~al.}(2009){Hicken}, {Challis}, {Jha}, {Kirshner},
  {Matheson}, {Modjaz}, {Rest}, {Wood-Vasey}, {Bakos}, {Barton}, {Berlind},
  {Bragg}, {Brice{\~n}o}, {Brown}, {Caldwell}, {Calkins}, {Cho}, {Ciupik},
  {Contreras}, {Dendy}, {Dosaj}, {Durham}, {Eriksen}, {Esquerdo}, {Everett},
  {Falco}, {Fernandez}, {Gaba}, {Garnavich}, {Graves}, {Green}, {Groner},
  {Hergenrother}, {Holman}, {Hradecky}, {Huchra}, {Hutchison}, {Jerius},
  {Jordan}, {Kilgard}, {Krauss}, {Luhman}, {Macri}, {Marrone}, {McDowell},
  {McIntosh}, {McNamara}, {Megeath}, {Mochejska}, {Munoz}, {Muzerolle},
  {Naranjo}, {Narayan}, {Pahre}, {Peters}, {Peterson}, {Rines}, {Ripman},
  {Roussanova}, {Schild}, {Sicilia-Aguilar}, {Sokoloski}, {Smalley}, {Smith},
  {Spahr}, {Stanek}, {Barmby}, {Blondin}, {Stubbs}, {Szentgyorgyi}, {Torres},
  {Vaz}, {Vikhlinin}, {Wang}, {Westover}, {Woods}, \&
  {Zhao}}]{2009ApJ...700..331H}
{Hicken}, M., {Challis}, P., {Jha}, S., {et~al.} 2009, \apj, 700, 331

\bibitem[{{Hicken} {et~al.}(2012){Hicken}, {Challis}, {Kirshner}, {Rest},
  {Cramer}, {Wood-Vasey}, {Bakos}, {Berlind}, {Brown}, {Caldwell}, {Calkins},
  {Currie}, {de Kleer}, {Esquerdo}, {Everett}, {Falco}, {Fernandez},
  {Friedman}, {Groner}, {Hartman}, {Holman}, {Hutchins}, {Keys}, {Kipping},
  {Latham}, {Marion}, {Narayan}, {Pahre}, {Pal}, {Peters}, {Perumpilly},
  {Ripman}, {Sipocz}, {Szentgyorgyi}, {Tang}, {Torres}, {Vaz}, {Wolk}, \&
  {Zezas}}]{2012ApJS..200...12H}
{Hicken}, M., {Challis}, P., {Kirshner}, R.~P., {et~al.} 2012, \apjs, 200, 12

\bibitem[{Hills(1988)}]{hills1988hyper}
Hills, J.~G. 1988, Nature, 331, 687

\bibitem[{Holoien {et~al.}(2014)Holoien, Prieto, Bersier, Kochanek, Stanek,
  Shappee, Grupe, Basu, Beacom, Brimacombe, {et~al.}}]{holoien2014asassn}
Holoien, T.-S., Prieto, J., Bersier, D., {et~al.} 2014, Monthly Notices of the
  Royal Astronomical Society, 445, 3263

\bibitem[{{Howell} {et~al.}(2013){Howell}, {Kasen}, {Lidman}, {Sullivan},
  {Conley}, {Astier}, {Balland}, {Carlberg}, {Fouchez}, {Guy}, {Hardin},
  {Pain}, {Palanque-Delabrouille}, {Perrett}, {Pritchet}, {Regnault}, {Rich},
  \& {Ruhlmann-Kleider}}]{2013ApJ...779...98H}
{Howell}, D.~A., {Kasen}, D., {Lidman}, C., {et~al.} 2013, \apj, 779, 98

\bibitem[{Hoyle \& Fowler(1960)}]{hoyle1960nucleosynthesis}
Hoyle, F., \& Fowler, W.~A. 1960, The Astrophysical Journal, 132, 565

\bibitem[{Humphreys \& Davidson(1994)}]{humphreys1994luminous}
Humphreys, R.~M., \& Davidson, K. 1994, Publications of the Astronomical
  Society of the Pacific, 106, 1025

\bibitem[{{Inserra} {et~al.}(2013){Inserra}, {Smartt}, {Jerkstrand}, {Valenti},
  {Fraser}, {Wright}, {Smith}, {Chen}, {Kotak}, {Pastorello}, {Nicholl},
  {Bresolin}, {Kudritzki}, {Benetti}, {Botticella}, {Burgett}, {Chambers},
  {Ergon}, {Flewelling}, {Fynbo}, {Geier}, {Hodapp}, {Howell}, {Huber},
  {Kaiser}, {Leloudas}, {Magill}, {Magnier}, {McCrum}, {Metcalfe}, {Price},
  {Rest}, {Sollerman}, {Sweeney}, {Taddia}, {Taubenberger}, {Tonry},
  {Wainscoat}, {Waters}, \& {Young}}]{2013ApJ...770..128I}
{Inserra}, C., {Smartt}, S.~J., {Jerkstrand}, A., {et~al.} 2013, \apj, 770, 128

\bibitem[{Inserra {et~al.}(2013)Inserra, Pastorello, Turatto, Pumo, Benetti,
  Cappellaro, Botticella, Bufano, Elias-Rosa, Harutyunyan,
  {et~al.}}]{inserra2013vizier}
Inserra, C., Pastorello, A., Turatto, M., {et~al.} 2013, VizieR Online Data
  Catalog, 355, 50142

\bibitem[{{Ivanova} {et~al.}(2013){Ivanova}, {Justham}, {Avendano Nandez}, \&
  {Lombardi}}]{2013Sci...339..433I}
{Ivanova}, N., {Justham}, S., {Avendano Nandez}, J.~L., \& {Lombardi}, J.~C.
  2013, Science, 339, 433

\bibitem[{{Jha} {et~al.}(2006){Jha}, {Kirshner}, {Challis}, {Garnavich},
  {Matheson}, {Soderberg}, {Graves}, {Hicken}, {Alves}, {Arce}, {Balog},
  {Barmby}, {Barton}, {Berlind}, {Bragg}, {Brice{\~n}o}, {Brown}, {Buckley},
  {Caldwell}, {Calkins}, {Carter}, {Concannon}, {Donnelly}, {Eriksen},
  {Fabricant}, {Falco}, {Fiore}, {Garcia}, {G{\'o}mez}, {Grogin}, {Groner},
  {Groot}, {Haisch}, {Hartmann}, {Hergenrother}, {Holman}, {Huchra},
  {Jayawardhana}, {Jerius}, {Kannappan}, {Kim}, {Kleyna}, {Kochanek},
  {Koranyi}, {Krockenberger}, {Lada}, {Luhman}, {Luu}, {Macri}, {Mader},
  {Mahdavi}, {Marengo}, {Marsden}, {McLeod}, {McNamara}, {Megeath}, {Moraru},
  {Mossman}, {Muench}, {Mu{\~n}oz}, {Muzerolle}, {Naranjo}, {Nelson-Patel},
  {Pahre}, {Patten}, {Peters}, {Peters}, {Raymond}, {Rines}, {Schild},
  {Sobczak}, {Spahr}, {Stauffer}, {Stefanik}, {Szentgyorgyi}, {Tollestrup},
  {V{\"a}is{\"a}nen}, {Vikhlinin}, {Wang}, {Willner}, {Wolk}, {Zajac}, {Zhao},
  \& {Stanek}}]{2006AJ....131..527J}
{Jha}, S., {Kirshner}, R.~P., {Challis}, P., {et~al.} 2006, \aj, 131, 527

\bibitem[{Karpenka {et~al.}(2012)Karpenka, Feroz, \&
  Hobson}]{karpenka2012simple}
Karpenka, N.~V., Feroz, F., \& Hobson, M. 2012, Monthly Notices of the Royal
  Astronomical Society, sts412

\bibitem[{Kasen \& Bildsten(2010)}]{kasen2010supernova}
Kasen, D., \& Bildsten, L. 2010, The Astrophysical Journal, 717, 245

\bibitem[{Kasen {et~al.}(2015)Kasen, Fern{\'a}ndez, \&
  Metzger}]{kasen2015kilonova}
Kasen, D., Fern{\'a}ndez, R., \& Metzger, B.~D. 2015, Monthly Notices of the
  Royal Astronomical Society, 450, 1777

\bibitem[{Kasen {et~al.}(2016)Kasen, Metzger, \& Bildsten}]{kasen2016magnetar}
Kasen, D., Metzger, B.~D., \& Bildsten, L. 2016, The Astrophysical Journal,
  821, 36

\bibitem[{Kasen \& Woosley(2009)}]{kasen2009type}
Kasen, D., \& Woosley, S. 2009, The Astrophysical Journal, 703, 2205

\bibitem[{Kasen {et~al.}(2011)Kasen, Woosley, \& Heger}]{kasen2011pair}
Kasen, D., Woosley, S., \& Heger, A. 2011, The Astrophysical Journal, 734, 102

\bibitem[{Kashi \& Soker(2010)}]{kashi2010common}
Kashi, A., \& Soker, N. 2010, arXiv preprint arXiv:1011.1222

\bibitem[{Kasliwal(2012)}]{kasliwal2012bridging}
Kasliwal, M.~M. 2012, Bridging the gap: elusive explosions in the local
  universe (Universal-Publishers)

\bibitem[{Kasliwal {et~al.}(2011)Kasliwal, Kulkarni, Arcavi, Quimby, Ofek,
  Nugent, Jacobsen, Gal-Yam, Green, Yaron, {et~al.}}]{kasliwal2011ptf}
Kasliwal, M.~M., Kulkarni, S.~R., Arcavi, I., {et~al.} 2011, The Astrophysical
  Journal, 730, 134

\bibitem[{Kasliwal {et~al.}(2012)Kasliwal, Kulkarni, Gal-Yam, Nugent, Sullivan,
  Bildsten, Yaron, Perets, Arcavi, Ben-Ami, {et~al.}}]{kasliwal2012calcium}
Kasliwal, M.~M., Kulkarni, S., Gal-Yam, A., {et~al.} 2012, The Astrophysical
  Journal, 755, 161

\bibitem[{Khokhlov(1991)}]{khokhlov1991delayed}
Khokhlov, A. 1991, Astronomy and Astrophysics, 245, 114

\bibitem[{Kiewe {et~al.}(2011)Kiewe, Gal-Yam, Arcavi, Leonard, Enriquez, Cenko,
  Fox, Moon, Sand, Soderberg, {et~al.}}]{kiewe2011caltech}
Kiewe, M., Gal-Yam, A., Arcavi, I., {et~al.} 2011, The Astrophysical Journal,
  744, 10

\bibitem[{Kochanek {et~al.}(2012)Kochanek, Szczygie{\l}, \&
  Stanek}]{kochanek2012unmasking}
Kochanek, C., Szczygie{\l}, D., \& Stanek, K. 2012, The Astrophysical Journal,
  758, 142

\bibitem[{Korobkin {et~al.}(2012)Korobkin, Rosswog, Arcones, \&
  Winteler}]{korobkin2012astrophysical}
Korobkin, O., Rosswog, S., Arcones, A., \& Winteler, C. 2012, Monthly Notices
  of the Royal Astronomical Society, 426, 1940

\bibitem[{Kulkarni {et~al.}(2007)Kulkarni, Ofek, Rau, Cenko, Soderberg, Fox,
  Gal-Yam, Capak, Moon, Li, {et~al.}}]{kulkarni2007unusually}
Kulkarni, S., Ofek, E., Rau, A., {et~al.} 2007, Nature, 447, 458

\bibitem[{{Kuncarayakti} {et~al.}(2015){Kuncarayakti}, {Harmanen}, {Kangas},
  {Mattila}, {Fraser}, {Botticella}, {Inserra}, {Kankare}, {Maguire}, {Smartt},
  {Smith}, {Sullivan}, {Valenti}, {Yaron}, {Young}, {Manulis}, {Baltay},
  {Ellman}, {Hadjiyska}, {McKinnon}, {Rabinowitz}, {Rostami}, {Feindt},
  {Kowalski}, {Nugent}, {Wright}, {Chambers}, {Flewelling}, {Huber}, {Magnier},
  {Tonry}, {Waters}, \& {Wainscoat}}]{2015ATel.8325....1K}
{Kuncarayakti}, H., {Harmanen}, J., {Kangas}, T., {et~al.} 2015, The
  Astronomer's Telegram, 8325

\bibitem[{Leloudas {et~al.}(2012)Leloudas, Chatzopoulos, Dilday, Gorosabel,
  Vinko, Gallazzi, Wheeler, Bassett, Fischer, Frieman,
  {et~al.}}]{leloudas2012sn}
Leloudas, G., Chatzopoulos, E., Dilday, B., {et~al.} 2012, Astronomy \&
  Astrophysics, 541, A129

\bibitem[{Leloudas {et~al.}(2016)Leloudas, Fraser, Stone, van Velzen, Jonker,
  Arcavi, Fremling, Maund, Smartt, Kruhler,
  {et~al.}}]{leloudas2016superluminous}
Leloudas, G., Fraser, M., Stone, N., {et~al.} 2016, arXiv preprint
  arXiv:1609.02927

\bibitem[{Li \& Paczy{\'n}ski(1998)}]{li1998transient}
Li, L.-X., \& Paczy{\'n}ski, B. 1998, The Astrophysical Journal Letters, 507,
  L59

\bibitem[{{Lochner} {et~al.}(2016){Lochner}, {McEwen}, {Peiris}, {Lahav}, \&
  {Winter}}]{2016ApJS..225...31L}
{Lochner}, M., {McEwen}, J.~D., {Peiris}, H.~V., {Lahav}, O., \& {Winter},
  M.~K. 2016, \apjs, 225, 31

\bibitem[{Lodato(2012)}]{lodato2012challenges}
Lodato, G. 2012, in EPJ Web of Conferences, Vol.~39, EDP Sciences, 01001

\bibitem[{{Lunnan} {et~al.}(2013){Lunnan}, {Chornock}, {Berger},
  {Milisavljevic}, {Drout}, {Sanders}, {Challis}, {Czekala}, {Foley}, {Fong},
  {Huber}, {Kirshner}, {Leibler}, {Marion}, {McCrum}, {Narayan}, {Rest},
  {Roth}, {Scolnic}, {Smartt}, {Smith}, {Soderberg}, {Stubbs}, {Tonry},
  {Burgett}, {Chambers}, {Kudritzki}, {Magnier}, \&
  {Price}}]{2013ApJ...771...97L}
{Lunnan}, R., {Chornock}, R., {Berger}, E., {et~al.} 2013, \apj, 771, 97

\bibitem[{{Lunnan} {et~al.}(2016){Lunnan}, {Chornock}, {Berger},
  {Milisavljevic}, {Jones}, {Rest}, {Fong}, {Fransson}, {Margutti}, {Drout},
  {Blanchard}, {Challis}, {Cowperthwaite}, {Foley}, {Kirshner}, {Morrell},
  {Riess}, {Roth}, {Scolnic}, {Smartt}, {Smith}, {Villar}, {Chambers},
  {Draper}, {Huber}, {Kaiser}, {Kudritzki}, {Magnier}, {Metcalfe}, \&
  {Waters}}]{2016ApJ...831..144L}
---. 2016, \apj, 831, 144

\bibitem[{{Lunnan} {et~al.}(2017){Lunnan}, {Kasliwal}, {Cao}, {Hangard},
  {Yaron}, {Parrent}, {McCully}, {Gal-Yam}, {Mulchaey}, {Ben-Ami},
  {Filippenko}, {Fremling}, {Fruchter}, {Howell}, {Koda}, {Kupfer}, {Kulkarni},
  {Laher}, {Masci}, {Nugent}, {Ofek}, {Yagi}, \& {Yan}}]{2017ApJ...836...60L}
{Lunnan}, R., {Kasliwal}, M.~M., {Cao}, Y., {et~al.} 2017, \apj, 836, 60

\bibitem[{Lyman {et~al.}(2014)Lyman, Levan, Church, Davies, \&
  Tanvir}]{lyman2014progenitors}
Lyman, J., Levan, A., Church, R., Davies, M.~B., \& Tanvir, N. 2014, Monthly
  Notices of the Royal Astronomical Society, 444, 2157

\bibitem[{{Margalit} \& {Metzger}(2016)}]{2016MNRAS.461.1154M}
{Margalit}, B., \& {Metzger}, B.~D. 2016, \mnras, 461, 1154

\bibitem[{{Margalit} {et~al.}(2017){Margalit}, {Metzger}, {Thompson},
  {Nicholl}, \& {Sukhbold}}]{2017arXiv170501103M}
{Margalit}, B., {Metzger}, B.~D., {Thompson}, T.~A., {Nicholl}, M., \&
  {Sukhbold}, T. 2017, ArXiv e-prints, arXiv:1705.01103

\bibitem[{Margutti {et~al.}(2013)Margutti, Milisavljevic, Soderberg, Chornock,
  Zauderer, Murase, Guidorzi, Sanders, Kuin, Fransson,
  {et~al.}}]{margutti2013panchromatic}
Margutti, R., Milisavljevic, D., Soderberg, A.~M., {et~al.} 2013, The
  Astrophysical Journal, 780, 21

\bibitem[{Margutti {et~al.}(2016)Margutti, Metzger, Chornock, Milisavljevic,
  Berger, Blanchard, Guidorzi, Migliori, Kamble, Lunnan,
  {et~al.}}]{margutti2016x}
Margutti, R., Metzger, B., Chornock, R., {et~al.} 2016, arXiv preprint
  arXiv:1610.01632

\bibitem[{Martini {et~al.}(1999)Martini, Wagner, Tomaney, Rich, Della~Valle, \&
  Hauschildt}]{martini1999nova}
Martini, P., Wagner, R.~M., Tomaney, A., {et~al.} 1999, The Astronomical
  Journal, 118, 1034

\bibitem[{Matzner \& McKee(1999)}]{matzner1999expulsion}
Matzner, C.~D., \& McKee, C.~F. 1999, The Astrophysical Journal, 510, 379

\bibitem[{{McCrum} {et~al.}(2014){McCrum}, {Smartt}, {Kotak}, {Rest},
  {Jerkstrand}, {Inserra}, {Rodney}, {Chen}, {Howell}, {Huber}, {Pastorello},
  {Tonry}, {Bresolin}, {Kudritzki}, {Chornock}, {Berger}, {Smith},
  {Botticella}, {Foley}, {Fraser}, {Milisavljevic}, {Nicholl}, {Riess},
  {Stubbs}, {Valenti}, {Wood-Vasey}, {Wright}, {Young}, {Drout}, {Czekala},
  {Burgett}, {Chambers}, {Draper}, {Flewelling}, {Hodapp}, {Kaiser}, {Magnier},
  {Metcalfe}, {Price}, {Sweeney}, \& {Wainscoat}}]{2014MNRAS.437..656M}
{McCrum}, M., {Smartt}, S.~J., {Kotak}, R., {et~al.} 2014, \mnras, 437, 656

\bibitem[{Melandri {et~al.}(2014)Melandri, Covino, Rogantini, Salvaterra,
  Sbarufatti, Bernardini, Campana, D’avanzo, D’elia, Fugazza,
  {et~al.}}]{melandri2014optical}
Melandri, A., Covino, S., Rogantini, D., {et~al.} 2014, Astronomy \&
  Astrophysics, 565, A72

\bibitem[{Metzger {et~al.}(2009)Metzger, Piro, \& Quataert}]{metzger2009nickel}
Metzger, B., Piro, A., \& Quataert, E. 2009, Monthly Notices of the Royal
  Astronomical Society, 396, 1659

\bibitem[{Metzger {et~al.}(2010)Metzger, Mart{\'\i}nez-Pinedo, Darbha,
  Quataert, Arcones, Kasen, Thomas, Nugent, Panov, \&
  Zinner}]{metzger2010electromagnetic}
Metzger, B., Mart{\'\i}nez-Pinedo, G., Darbha, S., {et~al.} 2010, Monthly
  Notices of the Royal Astronomical Society, 406, 2650

\bibitem[{Metzger(2012)}]{metzger2012nuclear}
Metzger, B.~D. 2012, Monthly Notices of the Royal Astronomical Society, 419,
  827

\bibitem[{Metzger(2016)}]{metzger2016kilonova}
---. 2016, arXiv preprint arXiv:1610.09381

\bibitem[{Metzger {et~al.}(2015)Metzger, Margalit, Kasen, \&
  Quataert}]{metzger2015diversity}
Metzger, B.~D., Margalit, B., Kasen, D., \& Quataert, E. 2015, Monthly Notices
  of the Royal Astronomical Society, 454, 3311

\bibitem[{{Metzger} \& {Pejcha}(2017)}]{2017arXiv170503895M}
{Metzger}, B.~D., \& {Pejcha}, O. 2017, ArXiv e-prints, arXiv:1705.03895

\bibitem[{Metzger \& Piro(2014)}]{metzger2014optical}
Metzger, B.~D., \& Piro, A.~L. 2014, Monthly Notices of the Royal Astronomical
  Society, 439, 3916

\bibitem[{Miyaji {et~al.}(1980)Miyaji, Nomoto, Yokoi, \&
  Sugimoto}]{miyaji1980supernova}
Miyaji, S., Nomoto, K., Yokoi, K., \& Sugimoto, D. 1980, Publications of the
  Astronomical Society of Japan, 32, 303

\bibitem[{Moriya {et~al.}(2013{\natexlab{a}})Moriya, Blinnikov, Baklanov,
  Sorokina, \& Dolgov}]{moriya2013synthetic}
Moriya, T.~J., Blinnikov, S.~I., Baklanov, P.~V., Sorokina, E.~I., \& Dolgov,
  A.~D. 2013{\natexlab{a}}, Monthly Notices of the Royal Astronomical Society,
  stt011

\bibitem[{Moriya {et~al.}(2013{\natexlab{b}})Moriya, Maeda, Taddia, Sollerman,
  Blinnikov, \& Sorokina}]{moriya2013analytic}
Moriya, T.~J., Maeda, K., Taddia, F., {et~al.} 2013{\natexlab{b}}, Monthly
  Notices of the Royal Astronomical Society, 435, 1520

\bibitem[{Moriya {et~al.}(2014)Moriya, Maeda, Taddia, Sollerman, Blinnikov, \&
  Sorokina}]{moriya2014mass}
---. 2014, Monthly Notices of the Royal Astronomical Society, stu163

\bibitem[{Moriya {et~al.}(2016)Moriya, Mazzali, Tominaga, Hachinger, Blinnikov,
  Tauris, Takahashi, Tanaka, Langer, \& Podsiadlowski}]{moriya2016light}
Moriya, T.~J., Mazzali, P.~A., Tominaga, N., {et~al.} 2016, Monthly Notices of
  the Royal Astronomical Society, stw3225

\bibitem[{{M{\"u}ller} {et~al.}(2017){M{\"u}ller}, {Prieto}, {Pejcha}, \&
  {Clocchiatti}}]{2017arXiv170200416M}
{M{\"u}ller}, T., {Prieto}, J.~L., {Pejcha}, O., \& {Clocchiatti}, A. 2017,
  ArXiv e-prints, arXiv:1702.00416

\bibitem[{Munari {et~al.}(2002)Munari, Henden, Kiyota, Laney, Marang, Zwitter,
  Corradi, Desidera, Marrese, Giro, {et~al.}}]{munari2002mysterious}
Munari, U., Henden, A., Kiyota, S., {et~al.} 2002, Astronomy \& Astrophysics,
  389, L51

\bibitem[{{Nicholl} {et~al.}(2017{\natexlab{a}}){Nicholl}, {Berger},
  {Margutti}, {Blanchard}, {Milisavljevic}, {Challis}, {Metzger}, \&
  {Chornock}}]{2017ApJ...835L...8N}
{Nicholl}, M., {Berger}, E., {Margutti}, R., {et~al.} 2017{\natexlab{a}},
  \apjl, 835, L8

\bibitem[{{Nicholl} {et~al.}(2017{\natexlab{b}}){Nicholl}, {Guillochon}, \&
  {Berger}}]{2017arXiv170600825N}
{Nicholl}, M., {Guillochon}, J., \& {Berger}, E. 2017{\natexlab{b}}, ArXiv
  e-prints, arXiv:1706.00825

\bibitem[{Nicholl \& Smartt(2016)}]{nicholl2016seeing}
Nicholl, M., \& Smartt, S. 2016, Monthly Notices of the Royal Astronomical
  Society: Letters, 457, L79

\bibitem[{{Nicholl} {et~al.}(2013){Nicholl}, {Smartt}, {Jerkstrand}, {Inserra},
  {McCrum}, {Kotak}, {Fraser}, {Wright}, {Chen}, {Smith}, {Young}, {Sim},
  {Valenti}, {Howell}, {Bresolin}, {Kudritzki}, {Tonry}, {Huber}, {Rest},
  {Pastorello}, {Tomasella}, {Cappellaro}, {Benetti}, {Mattila}, {Kankare},
  {Kangas}, {Leloudas}, {Sollerman}, {Taddia}, {Berger}, {Chornock}, {Narayan},
  {Stubbs}, {Foley}, {Lunnan}, {Soderberg}, {Sanders}, {Milisavljevic},
  {Margutti}, {Kirshner}, {Elias-Rosa}, {Morales-Garoffolo}, {Taubenberger},
  {Botticella}, {Gezari}, {Urata}, {Rodney}, {Riess}, {Scolnic}, {Wood-Vasey},
  {Burgett}, {Chambers}, {Flewelling}, {Magnier}, {Kaiser}, {Metcalfe},
  {Morgan}, {Price}, {Sweeney}, \& {Waters}}]{2013Natur.502..346N}
{Nicholl}, M., {Smartt}, S.~J., {Jerkstrand}, A., {et~al.} 2013, \nat, 502, 346

\bibitem[{Nicholl {et~al.}(2013)Nicholl, Smartt, Jerkstrand, Inserra, McCrum,
  Kotak, Fraser, Wright, Chen, Smith, {et~al.}}]{nicholl2013slowly}
Nicholl, M., Smartt, S., Jerkstrand, A., {et~al.} 2013, Nature, 502, 346

\bibitem[{{Nicholl} {et~al.}(2014){Nicholl}, {Smartt}, {Jerkstrand}, {Inserra},
  {Anderson}, {Baltay}, {Benetti}, {Chen}, {Elias-Rosa}, {Feindt}, {Fraser},
  {Gal-Yam}, {Hadjiyska}, {Howell}, {Kotak}, {Lawrence}, {Leloudas},
  {Margheim}, {Mattila}, {McCrum}, {McKinnon}, {Mead}, {Nugent}, {Rabinowitz},
  {Rest}, {Smith}, {Sollerman}, {Sullivan}, {Taddia}, {Valenti}, {Walker}, \&
  {Young}}]{2014MNRAS.444.2096N}
{Nicholl}, M., {Smartt}, S.~J., {Jerkstrand}, A., {et~al.} 2014, \mnras, 444,
  2096

\bibitem[{{Nicholl} {et~al.}(2015){Nicholl}, {Smartt}, {Jerkstrand}, {Sim},
  {Inserra}, {Anderson}, {Baltay}, {Benetti}, {Chambers}, {Chen}, {Elias-Rosa},
  {Feindt}, {Flewelling}, {Fraser}, {Gal-Yam}, {Galbany}, {Huber}, {Kangas},
  {Kankare}, {Kotak}, {Kr{\"u}hler}, {Maguire}, {McKinnon}, {Rabinowitz},
  {Rostami}, {Schulze}, {Smith}, {Sullivan}, {Tonry}, {Valenti}, \&
  {Young}}]{2015ApJ...807L..18N}
---. 2015, \apjl, 807, L18

\bibitem[{{Nicholl} {et~al.}(2016){Nicholl}, {Berger}, {Smartt}, {Margutti},
  {Kamble}, {Alexander}, {Chen}, {Inserra}, {Arcavi}, {Blanchard}, {Cartier},
  {Chambers}, {Childress}, {Chornock}, {Cowperthwaite}, {Drout}, {Flewelling},
  {Fraser}, {Gal-Yam}, {Galbany}, {Harmanen}, {Holoien}, {Hosseinzadeh},
  {Howell}, {Huber}, {Jerkstrand}, {Kankare}, {Kochanek}, {Lin}, {Lunnan},
  {Magnier}, {Maguire}, {McCully}, {McDonald}, {Metzger}, {Milisavljevic},
  {Mitra}, {Reynolds}, {Saario}, {Shappee}, {Smith}, {Valenti}, {Villar},
  {Waters}, \& {Young}}]{2016ApJ...826...39N}
{Nicholl}, M., {Berger}, E., {Smartt}, S.~J., {et~al.} 2016, \apj, 826, 39

\bibitem[{Nomoto \& Kondo(1991)}]{nomoto1991conditions}
Nomoto, K., \& Kondo, Y. 1991, The Astrophysical Journal, 367, L19

\bibitem[{Nugent {et~al.}(2002)Nugent, Kim, \& Perlmutter}]{nugent2002k}
Nugent, P., Kim, A., \& Perlmutter, S. 2002, Publications of the Astronomical
  Society of the Pacific, 114, 803

\bibitem[{Ofek {et~al.}(2014{\natexlab{a}})Ofek, Sullivan, Shaviv, Steinbok,
  Arcavi, Gal-Yam, Tal, Kulkarni, Nugent, Ben-Ami,
  {et~al.}}]{ofek2014precursors}
Ofek, E.~O., Sullivan, M., Shaviv, N.~J., {et~al.} 2014{\natexlab{a}}, The
  Astrophysical Journal, 789, 104

\bibitem[{Ofek {et~al.}(2014{\natexlab{b}})Ofek, Zoglauer, Boggs, Barri{\'e}re,
  Reynolds, Fryer, Harrison, Cenko, Kulkarni, Gal-Yam, {et~al.}}]{ofek2014sn}
Ofek, E.~O., Zoglauer, A., Boggs, S.~E., {et~al.} 2014{\natexlab{b}}, The
  Astrophysical Journal, 781, 42

\bibitem[{{Papadopoulos} {et~al.}(2015){Papadopoulos}, {D'Andrea}, {Sullivan},
  {Nichol}, {Barbary}, {Biswas}, {Brown}, {Covarrubias}, {Finley}, {Fischer},
  {Foley}, {Goldstein}, {Gupta}, {Kessler}, {Kovacs}, {Kuhlmann}, {Lidman},
  {March}, {Nugent}, {Sako}, {Smith}, {Spinka}, {Wester}, {Abbott}, {Abdalla},
  {Allam}, {Banerji}, {Bernstein}, {Bernstein}, {Carnero}, {da Costa}, {DePoy},
  {Desai}, {Diehl}, {Eifler}, {Evrard}, {Flaugher}, {Frieman}, {Gerdes},
  {Gruen}, {Honscheid}, {James}, {Kuehn}, {Kuropatkin}, {Lahav}, {Maia},
  {Makler}, {Marshall}, {Merritt}, {Miller}, {Miquel}, {Ogando}, {Plazas},
  {Roe}, {Romer}, {Rykoff}, {Sanchez}, {Santiago}, {Scarpine}, {Schubnell},
  {Sevilla}, {Soares-Santos}, {Suchyta}, {Swanson}, {Tarle}, {Thaler},
  {Tucker}, {Wechsler}, \& {Zuntz}}]{2015MNRAS.449.1215P}
{Papadopoulos}, A., {D'Andrea}, C.~B., {Sullivan}, M., {et~al.} 2015, \mnras,
  449, 1215

\bibitem[{{Pastorello} {et~al.}(2007){Pastorello}, {Mazzali}, {Pignata},
  {Benetti}, {Cappellaro}, {Filippenko}, {Li}, {Meikle}, {Arkharov}, {Blanc},
  {Bufano}, {Derekas}, {Dolci}, {Elias-Rosa}, {Foley}, {Ganeshalingam},
  {Harutyunyan}, {Kiss}, {Kotak}, {Larionov}, {Lucey}, {Napoleone},
  {Navasardyan}, {Patat}, {Rich}, {Ryder}, {Salvo}, {Schmidt}, {Stanishev},
  {Sz{\'e}kely}, {Taubenberger}, {Temporin}, {Turatto}, \&
  {Hillebrandt}}]{2007MNRAS.377.1531P}
{Pastorello}, A., {Mazzali}, P.~A., {Pignata}, G., {et~al.} 2007, \mnras, 377,
  1531

\bibitem[{Patat {et~al.}(1994)Patat, Barbon, Cappellaro, \&
  Turatto}]{patat1994light}
Patat, F., Barbon, R., Cappellaro, E., \& Turatto, M. 1994, Astronomy and
  Astrophysics, 282, 731

\bibitem[{Phillips {et~al.}(2007)Phillips, Li, Frieman, Blinnikov, DePoy,
  Prieto, Milne, Contreras, Folatelli, Morrell,
  {et~al.}}]{phillips2007peculiar}
Phillips, M.~M., Li, W., Frieman, J.~A., {et~al.} 2007, Publications of the
  Astronomical Society of the Pacific, 119, 360

\bibitem[{Popov(1993)}]{popov1993analytical}
Popov, D. 1993, The Astrophysical Journal, 414, 712

\bibitem[{{Prieto} {et~al.}(2006){Prieto}, {Rest}, \&
  {Suntzeff}}]{2006ApJ...647..501P}
{Prieto}, J.~L., {Rest}, A., \& {Suntzeff}, N.~B. 2006, \apj, 647, 501

\bibitem[{Quimby {et~al.}(2011)Quimby, Kulkarni, Kasliwal, Gal-Yam, Arcavi,
  Sullivan, Nugent, Thomas, Howell, Nakar, {et~al.}}]{quimby2011hydrogen}
Quimby, R., Kulkarni, S., Kasliwal, M., {et~al.} 2011, Nature, 474, 487

\bibitem[{{Quimby} {et~al.}(2011){Quimby}, {Kulkarni}, {Kasliwal}, {Gal-Yam},
  {Arcavi}, {Sullivan}, {Nugent}, {Thomas}, {Howell}, {Nakar}, {Bildsten},
  {Theissen}, {Law}, {Dekany}, {Rahmer}, {Hale}, {Smith}, {Ofek}, {Zolkower},
  {Velur}, {Walters}, {Henning}, {Bui}, {McKenna}, {Poznanski}, {Cenko}, \&
  {Levitan}}]{2011Natur.474..487Q}
{Quimby}, R.~M., {Kulkarni}, S.~R., {Kasliwal}, M.~M., {et~al.} 2011, \nat,
  474, 487

\bibitem[{Rau {et~al.}(2007)Rau, Kulkarni, Ofek, \& Yan}]{rau2007spitzer}
Rau, A., Kulkarni, S., Ofek, E., \& Yan, L. 2007, The Astrophysical Journal,
  659, 1536

\bibitem[{{Rest} {et~al.}(2011){Rest}, {Foley}, {Gezari}, {Narayan}, {Draine},
  {Olsen}, {Huber}, {Matheson}, {Garg}, {Welch}, {Becker}, {Challis},
  {Clocchiatti}, {Cook}, {Damke}, {Meixner}, {Miknaitis}, {Minniti}, {Morelli},
  {Nikolaev}, {Pignata}, {Prieto}, {Smith}, {Stubbs}, {Suntzeff}, {Walker},
  {Wood-Vasey}, {Zenteno}, {Wyrzykowski}, {Udalski}, {Szyma{\'n}ski}, {Kubiak},
  {Pietrzy{\'n}ski}, {Soszy{\'n}ski}, {Szewczyk}, {Ulaczyk}, \&
  {Poleski}}]{2011ApJ...729...88R}
{Rest}, A., {Foley}, R.~J., {Gezari}, S., {et~al.} 2011, \apj, 729, 88

\bibitem[{Retter \& Marom(2003)}]{retter2003model}
Retter, A., \& Marom, A. 2003, Monthly Notices of the Royal Astronomical
  Society, 345, L25

\bibitem[{Rhoads(1997)}]{rhoads1997tell}
Rhoads, J.~E. 1997, The Astrophysical Journal Letters, 487, L1

\bibitem[{Richards {et~al.}(2011)Richards, Starr, Butler, Bloom, Brewer,
  Crellin-Quick, Higgins, Kennedy, \& Rischard}]{richards2011machine}
Richards, J.~W., Starr, D.~L., Butler, N.~R., {et~al.} 2011, The Astrophysical
  Journal, 733, 10

\bibitem[{{Riess} {et~al.}(1999){Riess}, {Kirshner}, {Schmidt}, {Jha},
  {Challis}, {Garnavich}, {Esin}, {Carpenter}, {Grashius}, {Schild}, {Berlind},
  {Huchra}, {Prosser}, {Falco}, {Benson}, {Brice{\~n}o}, {Brown}, {Caldwell},
  {dell'Antonio}, {Filippenko}, {Goodman}, {Grogin}, {Groner}, {Hughes},
  {Green}, {Jansen}, {Kleyna}, {Luu}, {Macri}, {McLeod}, {McLeod}, {McNamara},
  {McLean}, {Milone}, {Mohr}, {Moraru}, {Peng}, {Peters}, {Prestwich},
  {Stanek}, {Szentgyorgyi}, \& {Zhao}}]{1999AJ....117..707R}
{Riess}, A.~G., {Kirshner}, R.~P., {Schmidt}, B.~P., {et~al.} 1999, \aj, 117,
  707

\bibitem[{Rossi {et~al.}(2002)Rossi, Lazzati, \& Rees}]{rossi2002afterglow}
Rossi, E., Lazzati, D., \& Rees, M.~J. 2002, Monthly Notices of the Royal
  Astronomical Society, 332, 945

\bibitem[{Rubin {et~al.}(2016)Rubin, Gal-Yam, De~Cia, Horesh, Khazov, Ofek,
  Kulkarni, Arcavi, Manulis, Yaron, {et~al.}}]{rubin2016type}
Rubin, A., Gal-Yam, A., De~Cia, A., {et~al.} 2016, The Astrophysical Journal,
  820, 33

\bibitem[{Sanders {et~al.}(2015)Sanders, Soderberg, Gezari, Betancourt,
  Chornock, Berger, Foley, Challis, Drout, Kirshner,
  {et~al.}}]{sanders2015toward}
Sanders, N., Soderberg, A., Gezari, S., {et~al.} 2015, The Astrophysical
  Journal, 799, 208

\bibitem[{{Sari} {et~al.}(1998){Sari}, {Piran}, \&
  {Narayan}}]{1998ApJ...497L..17S}
{Sari}, R., {Piran}, T., \& {Narayan}, R. 1998, \apjl, 497, L17

\bibitem[{Shen {et~al.}(2010)Shen, Kasen, Weinberg, Bildsten, \&
  Scannapieco}]{shen2010thermonuclear}
Shen, K.~J., Kasen, D., Weinberg, N.~N., Bildsten, L., \& Scannapieco, E. 2010,
  The Astrophysical Journal, 715, 767

\bibitem[{{Shen} {et~al.}(2010){Shen}, {Kasen}, {Weinberg}, {Bildsten}, \&
  {Scannapieco}}]{2010ApJ...715..767S}
{Shen}, K.~J., {Kasen}, D., {Weinberg}, N.~N., {Bildsten}, L., \&
  {Scannapieco}, E. 2010, \apj, 715, 767

\bibitem[{Siegel \& Ciolfi(2016)}]{siegel2016electromagnetic}
Siegel, D.~M., \& Ciolfi, R. 2016, The Astrophysical Journal, 819, 14

\bibitem[{Silverman {et~al.}(2011)Silverman, Ganeshalingam, Li, Filippenko,
  Miller, \& Poznanski}]{silverman2011fourteen}
Silverman, J.~M., Ganeshalingam, M., Li, W., {et~al.} 2011, Monthly Notices of
  the Royal Astronomical Society, 410, 585

\bibitem[{{Silverman} {et~al.}(2012){Silverman}, {Foley}, {Filippenko},
  {Ganeshalingam}, {Barth}, {Chornock}, {Griffith}, {Kong}, {Lee}, {Leonard},
  {Matheson}, {Miller}, {Steele}, {Barris}, {Bloom}, {Cobb}, {Coil},
  {Desroches}, {Gates}, {Ho}, {Jha}, {Kandrashoff}, {Li}, {Mandel}, {Modjaz},
  {Moore}, {Mostardi}, {Papenkova}, {Park}, {Perley}, {Poznanski}, {Reuter},
  {Scala}, {Serduke}, {Shields}, {Swift}, {Tonry}, {Van Dyk}, {Wang}, \&
  {Wong}}]{2012MNRAS.425.1789S}
{Silverman}, J.~M., {Foley}, R.~J., {Filippenko}, A.~V., {et~al.} 2012, \mnras,
  425, 1789

\bibitem[{Sim {et~al.}(2010)Sim, R{\"o}pke, Hillebrandt, Kromer, Pakmor, Fink,
  Ruiter, \& Seitenzahl}]{sim2010detonations}
Sim, S., R{\"o}pke, F., Hillebrandt, W., {et~al.} 2010, The Astrophysical
  Journal Letters, 714, L52

\bibitem[{{Sim} {et~al.}(2012){Sim}, {Fink}, {Kromer}, {R{\"o}pke}, {Ruiter},
  \& {Hillebrandt}}]{2012MNRAS.420.3003S}
{Sim}, S.~A., {Fink}, M., {Kromer}, M., {et~al.} 2012, \mnras, 420, 3003

\bibitem[{Smartt {et~al.}(2009)Smartt, Eldridge, Crockett, \&
  Maund}]{smartt2009death}
Smartt, S., Eldridge, J., Crockett, R., \& Maund, J.~R. 2009, Monthly Notices
  of the Royal Astronomical Society, 395, 1409

\bibitem[{Smartt {et~al.}(2015)Smartt, Valenti, Fraser, Inserra, Young,
  Sullivan, Pastorello, Benetti, Gal-Yam, Knapic, {et~al.}}]{smartt2015pessto}
Smartt, S., Valenti, S., Fraser, M., {et~al.} 2015, Astronomy \& Astrophysics,
  579, A40

\bibitem[{{Smartt} {et~al.}(2015){Smartt}, {Valenti}, {Fraser}, {Inserra},
  {Young}, {Sullivan}, {Pastorello}, {Benetti}, {Gal-Yam}, {Knapic},
  {Molinaro}, {Smareglia}, {Smith}, {Taubenberger}, {Yaron}, {Anderson},
  {Ashall}, {Balland}, {Baltay}, {Barbarino}, {Bauer}, {Baumont}, {Bersier},
  {Blagorodnova}, {Bongard}, {Botticella}, {Bufano}, {Bulla}, {Cappellaro},
  {Campbell}, {Cellier-Holzem}, {Chen}, {Childress}, {Clocchiatti},
  {Contreras}, {Dall'Ora}, {Danziger}, {de Jaeger}, {De Cia}, {Della Valle},
  {Dennefeld}, {Elias-Rosa}, {Elman}, {Feindt}, {Fleury}, {Gall},
  {Gonzalez-Gaitan}, {Galbany}, {Morales Garoffolo}, {Greggio}, {Guillou},
  {Hachinger}, {Hadjiyska}, {Hage}, {Hillebrandt}, {Hodgkin}, {Hsiao}, {James},
  {Jerkstrand}, {Kangas}, {Kankare}, {Kotak}, {Kromer}, {Kuncarayakti},
  {Leloudas}, {Lundqvist}, {Lyman}, {Hook}, {Maguire}, {Manulis}, {Margheim},
  {Mattila}, {Maund}, {Mazzali}, {McCrum}, {McKinnon}, {Moreno-Raya},
  {Nicholl}, {Nugent}, {Pain}, {Pignata}, {Phillips}, {Polshaw}, {Pumo},
  {Rabinowitz}, {Reilly}, {Romero-Ca{\~n}izales}, {Scalzo}, {Schmidt},
  {Schulze}, {Sim}, {Sollerman}, {Taddia}, {Tartaglia}, {Terreran},
  {Tomasella}, {Turatto}, {Walker}, {Walton}, {Wyrzykowski}, {Yuan}, \&
  {Zampieri}}]{2015A&A...579A..40S}
{Smartt}, S.~J., {Valenti}, S., {Fraser}, M., {et~al.} 2015, \aap, 579, A40

\bibitem[{Smith(2013)}]{smith2013model}
Smith, N. 2013, Monthly Notices of the Royal Astronomical Society, 429, 2366

\bibitem[{Smith(2014)}]{smith2014mass}
---. 2014, Annual Review of Astronomy and Astrophysics, 52, 487

\bibitem[{Smith(2016)}]{smith2016interacting}
---. 2016, arXiv preprint arXiv:1612.02006

\bibitem[{Smith \& Frew(2011)}]{smith2011revised}
Smith, N., \& Frew, D.~J. 2011, Monthly Notices of the Royal Astronomical
  Society, 415, 2009

\bibitem[{Smith {et~al.}(2011)Smith, Li, Silverman, Ganeshalingam, \&
  Filippenko}]{smith2011luminous}
Smith, N., Li, W., Silverman, J.~M., Ganeshalingam, M., \& Filippenko, A.~V.
  2011, Monthly Notices of the Royal Astronomical Society, 415, 773

\bibitem[{Smith {et~al.}(2012)Smith, Mauerhan, Silverman, Ganeshalingam,
  Filippenko, Cenko, Clubb, \& Kandrashoff}]{smith2012sn}
Smith, N., Mauerhan, J.~C., Silverman, J.~M., {et~al.} 2012, Monthly Notices of
  the Royal Astronomical Society, 426, 1905

\bibitem[{Smith \& McCray(2007)}]{smith2007shell}
Smith, N., \& McCray, R. 2007, The Astrophysical Journal Letters, 671, L17

\bibitem[{{Smith} {et~al.}(2007){Smith}, {Li}, {Foley}, {Wheeler}, {Pooley},
  {Chornock}, {Filippenko}, {Silverman}, {Quimby}, {Bloom}, \&
  {Hansen}}]{2007ApJ...666.1116S}
{Smith}, N., {Li}, W., {Foley}, R.~J., {et~al.} 2007, \apj, 666, 1116

\bibitem[{Soker \& Tylenda(2006)}]{soker2006violent}
Soker, N., \& Tylenda, R. 2006, Monthly Notices of the Royal Astronomical
  Society, 373, 733

\bibitem[{Stone(2013)}]{stone2013tidal}
Stone, N.~C. 2013, PhD thesis

\bibitem[{{Stritzinger} {et~al.}(2002){Stritzinger}, {Hamuy}, {Suntzeff},
  {Smith}, {Phillips}, {Maza}, {Strolger}, {Antezana}, {Gonz{\'a}lez},
  {Wischnjewsky}, {Candia}, {Espinoza}, {Gonz{\'a}lez}, {Stubbs}, {Becker},
  {Rubenstein}, \& {Galaz}}]{2002AJ....124.2100S}
{Stritzinger}, M., {Hamuy}, M., {Suntzeff}, N.~B., {et~al.} 2002, \aj, 124,
  2100

\bibitem[{Taddia {et~al.}(2013)Taddia, Stritzinger, Sollerman, Phillips,
  Anderson, Boldt, Campillay, Castellon, Contreras, Folatelli,
  {et~al.}}]{taddia2013vizier}
Taddia, F., Stritzinger, M., Sollerman, J., {et~al.} 2013, VizieR Online Data
  Catalog, 355, 50010

\bibitem[{Taddia {et~al.}(2015)Taddia, Sollerman, Leloudas, Stritzinger,
  Valenti, Galbany, Kessler, Schneider, \& Wheeler}]{taddia2015early}
Taddia, F., Sollerman, J., Leloudas, G., {et~al.} 2015, Astronomy \&
  Astrophysics, 574, A60

\bibitem[{Taddia {et~al.}(2016)Taddia, Fremling, Sollerman, Corsi, Gal-Yam,
  Karamehmetoglu, Lunnan, Bue, Ergon, Kasliwal, {et~al.}}]{taddia2016iptf15dtg}
Taddia, F., Fremling, C., Sollerman, J., {et~al.} 2016, Astronomy \&
  Astrophysics, 592, A89

\bibitem[{Tanvir {et~al.}(2013)Tanvir, Levan, Fruchter, Hjorth, Hounsell,
  Wiersema, \& Tunnicliffe}]{tanvir2013kilonova}
Tanvir, N., Levan, A., Fruchter, A., {et~al.} 2013, Nature, 500, 547

\bibitem[{Tartaglia {et~al.}(2016)Tartaglia, Pastorello, Sullivan, Baltay,
  Rabinowitz, Nugent, Drake, Djorgovski, Gal-Yam, Fabrika,
  {et~al.}}]{tartaglia2016interacting}
Tartaglia, L., Pastorello, A., Sullivan, M., {et~al.} 2016, Monthly Notices of
  the Royal Astronomical Society, 459, 1039

\bibitem[{Tauris {et~al.}(2015)Tauris, Langer, \&
  Podsiadlowski}]{tauris2015ultra}
Tauris, T.~M., Langer, N., \& Podsiadlowski, P. 2015, Monthly Notices of the
  Royal Astronomical Society, 451, 2123

\bibitem[{Tominaga {et~al.}(2013)Tominaga, Blinnikov, \&
  Nomoto}]{tominaga2013supernova}
Tominaga, N., Blinnikov, S.~I., \& Nomoto, K. 2013, The Astrophysical Journal
  Letters, 771, L12

\bibitem[{{Tominaga} {et~al.}(2013){Tominaga}, {Blinnikov}, \&
  {Nomoto}}]{2013ApJ...771L..12T}
{Tominaga}, N., {Blinnikov}, S.~I., \& {Nomoto}, K. 2013, \apjl, 771, L12

\bibitem[{Tripp \& Branch(1999)}]{tripp1999determination}
Tripp, R., \& Branch, D. 1999, The Astrophysical Journal, 525, 209

\bibitem[{Tsvetkov \& Bartunov(1993)}]{tsvetkov1993sternberg}
Tsvetkov, D.~Y., \& Bartunov, O. 1993, Bulletin d'Information du Centre de
  Donnees Stellaires, 42, 17

\bibitem[{Valenti {et~al.}(2012)Valenti, Taubenberger, Pastorello, Aramyan,
  Botticella, Fraser, Benetti, Smartt, Cappellaro, Elias-Rosa,
  {et~al.}}]{valenti2012spectroscopically}
Valenti, S., Taubenberger, S., Pastorello, A., {et~al.} 2012, The Astrophysical
  Journal Letters, 749, L28

\bibitem[{{Van Dyk} {et~al.}(2006){Van Dyk}, {Li}, {Filippenko}, {Humphreys},
  {Chornock}, {Foley}, \& {Challis}}]{2006astro.ph..3025V}
{Van Dyk}, S.~D., {Li}, W., {Filippenko}, A.~V., {et~al.} 2006, ArXiv
  Astrophysics e-prints, astro-ph/0603025

\bibitem[{{van Eerten} {et~al.}(2010){van Eerten}, {Zhang}, \&
  {MacFadyen}}]{2010ApJ...722..235V}
{van Eerten}, H., {Zhang}, W., \& {MacFadyen}, A. 2010, \apj, 722, 235

\bibitem[{Van~Eerten \& MacFadyen(2012)}]{van2012gamma}
Van~Eerten, H.~J., \& MacFadyen, A.~I. 2012, The Astrophysical Journal Letters,
  747, L30

\bibitem[{{van Velzen}(2017)}]{2017arXiv170703458V}
{van Velzen}, S. 2017, ArXiv e-prints, arXiv:1707.03458

\bibitem[{Villar {et~al.}(2016)Villar, Berger, Chornock, Margutti, Laskar,
  Brown, Blanchard, Czekala, Lunnan, \& Reynolds}]{villar2016intermediate}
Villar, V.~A., Berger, E., Chornock, R., {et~al.} 2016, The Astrophysical
  Journal, 830, 11

\bibitem[{{Vink{\'o}} {et~al.}(2003){Vink{\'o}}, {B{\'{\i}}r{\'o}}, {Cs{\'a}k},
  {Csizmadia}, {Derekas}, {Fur{\'e}sz}, {Heiner}, {S{\'a}rneczky}, {Sipocz},
  {Szab{\'o}}, {Szab{\'o}}, {Szil{\'a}di}, \&
  {Szatm{\'a}ry}}]{2003A&A...397..115V}
{Vink{\'o}}, J., {B{\'{\i}}r{\'o}}, I.~B., {Cs{\'a}k}, B., {et~al.} 2003, \aap,
  397, 115

\bibitem[{Vink{\'o} {et~al.}(2014)Vink{\'o}, Yuan, Quimby, Wheeler,
  Ramirez-Ruiz, Guillochon, Chatzopoulos, Marion, \&
  Akerlof}]{vinko2014luminous}
Vink{\'o}, J., Yuan, F., Quimby, R.~M., {et~al.} 2014, The Astrophysical
  Journal, 798, 12

\bibitem[{{Vreeswijk} {et~al.}(2014){Vreeswijk}, {Savaglio}, {Gal-Yam}, {De
  Cia}, {Quimby}, {Sullivan}, {Cenko}, {Perley}, {Filippenko}, {Clubb},
  {Taddia}, {Sollerman}, {Leloudas}, {Arcavi}, {Rubin}, {Kasliwal}, {Cao},
  {Yaron}, {Tal}, {Ofek}, {Capone}, {Kutyrev}, {Toy}, {Nugent}, {Laher},
  {Surace}, \& {Kulkarni}}]{2014ApJ...797...24V}
{Vreeswijk}, P.~M., {Savaglio}, S., {Gal-Yam}, A., {et~al.} 2014, \apj, 797, 24

\bibitem[{{Walker} {et~al.}(2015){Walker}, {Baltay}, {Campillay}, {Citrenbaum},
  {Contreras}, {Ellman}, {Feindt}, {Gonz{\'a}lez}, {Graham}, {Hadjiyska},
  {Hsiao}, {Krisciunas}, {McKinnon}, {Ment}, {Morrell}, {Nugent}, {Phillips},
  {Rabinowitz}, {Rostami}, {Ser{\'o}n}, {Stritzinger}, {Sullivan}, \&
  {Tucker}}]{2015ApJS..219...13W}
{Walker}, E.~S., {Baltay}, C., {Campillay}, A., {et~al.} 2015, \apjs, 219, 13

\bibitem[{Wanderman \& Piran(2010)}]{wanderman2010luminosity}
Wanderman, D., \& Piran, T. 2010, Monthly Notices of the Royal Astronomical
  Society, 406, 1944

\bibitem[{Weiler(2003)}]{weiler2003supernovae}
Weiler, K. 2003, Supernovae and Gamma-Ray Bursters, Lecture Notes in Physics
  (Springer Berlin Heidelberg)

\bibitem[{Wheeler {et~al.}(2015)Wheeler, Johnson, \&
  Clocchiatti}]{wheeler2015analysis}
Wheeler, J.~C., Johnson, V., \& Clocchiatti, A. 2015, Monthly Notices of the
  Royal Astronomical Society, 450, 1295

\bibitem[{Williams {et~al.}(2016)Williams, Darnley, Bode, \&
  Shafter}]{williams2016progenitors}
Williams, S., Darnley, M., Bode, M., \& Shafter, A. 2016, The Astrophysical
  Journal, 817, 143

\bibitem[{{Wollaeger} {et~al.}(2017){Wollaeger}, {Korobkin}, {Fontes},
  {Rosswog}, {Even}, {Fryer}, {Sollerman}, {Hungerford}, {van Rossum}, \&
  {Wollaber}}]{2017arXiv170507084W}
{Wollaeger}, R.~T., {Korobkin}, O., {Fontes}, C.~J., {et~al.} 2017, ArXiv
  e-prints, arXiv:1705.07084

\bibitem[{Woosley(1993)}]{woosley1993gamma}
Woosley, S. 1993, The Astrophysical Journal, 405, 273

\bibitem[{Woosley(2010)}]{woosley2010bright}
---. 2010, The Astrophysical Journal Letters, 719, L204

\bibitem[{Woosley \& Kasen(2011)}]{woosley2011sub}
Woosley, S., \& Kasen, D. 2011, The Astrophysical Journal, 734, 38

\bibitem[{{Woosley} \& {Kasen}(2011)}]{2011ApJ...734...38W}
{Woosley}, S.~E., \& {Kasen}, D. 2011, \apj, 734, 38

\bibitem[{{Wyrzykowski} {et~al.}(2015){Wyrzykowski}, {Kostrzewa-Rutkowska},
  {Udalski}, {Kozlowski}, {Pawlak}, {Szymanski}, {Soszynski}, \&
  {Sitek}}]{2015ATel.7207....1W}
{Wyrzykowski}, L., {Kostrzewa-Rutkowska}, Z., {Udalski}, A., {et~al.} 2015, The
  Astronomer's Telegram, 7207

\bibitem[{{Wyrzykowski} {et~al.}(2014{\natexlab{a}}){Wyrzykowski},
  {Kostrzewa-Rutkowska}, {Udalski}, {Kozlowski}, {Poleski}, \&
  {Sitek}}]{2014ATel.6861....1W}
---. 2014{\natexlab{a}}, The Astronomer's Telegram, 6861

\bibitem[{{Wyrzykowski} {et~al.}(2014{\natexlab{b}}){Wyrzykowski},
  {Kostrzewa-Rutkowska}, {Udalski}, {Kozlowski}, {Ulaczyk}, \&
  {Pietrzynski}}]{2014ATel.6700....1W}
---. 2014{\natexlab{b}}, The Astronomer's Telegram, 6700

\bibitem[{{Wyrzykowski} {et~al.}(2014{\natexlab{c}}){Wyrzykowski}, {Udalski},
  {Kozlowski}, {Kostrzewa-Rutkowska}, {Mroz}, \&
  {Ulaczyk}}]{2014ATel.5975....1W}
{Wyrzykowski}, L., {Udalski}, A., {Kozlowski}, S., {et~al.} 2014{\natexlab{c}},
  The Astronomer's Telegram, 5975

\bibitem[{{Wyrzykowski} {et~al.}(2014{\natexlab{d}}){Wyrzykowski}, {Udalski},
  {Kozlowski}, {Kostrzewa-Rutkowska}, {Szymanski}, {Mroz}, \&
  {Soszynski}}]{2014ATel.6494....1W}
---. 2014{\natexlab{d}}, The Astronomer's Telegram, 6494

\bibitem[{{Wyrzykowski} {et~al.}(2012){Wyrzykowski}, {Udalski}, {Kozlowski}, \&
  {Kubiak}}]{2012ATel.4689....1W}
{Wyrzykowski}, L., {Udalski}, A., {Kozlowski}, S., \& {Kubiak}, M. 2012, The
  Astronomer's Telegram, 4689

\bibitem[{{Wyrzykowski} {et~al.}(2014{\natexlab{e}}){Wyrzykowski}, {Udalski},
  {Kozlowski}, \& {Pietrukowicz}}]{2014ATel.5875....1W}
{Wyrzykowski}, L., {Udalski}, A., {Kozlowski}, S., \& {Pietrukowicz}, P.
  2014{\natexlab{e}}, The Astronomer's Telegram, 5875

\bibitem[{{Wyrzykowski} {et~al.}(2013){Wyrzykowski}, {Udalski}, {Kozlowski}, \&
  {Soszynski}}]{2013ATel.5663....1W}
{Wyrzykowski}, L., {Udalski}, A., {Kozlowski}, S., \& {Soszynski}, I. 2013, The
  Astronomer's Telegram, 5663

\bibitem[{{Wyrzykowski} {et~al.}(2014{\natexlab{f}}){Wyrzykowski},
  {Kostrzewa-Rutkowska}, {Koz{\l}owski}, {Udalski}, {Poleski}, {Skowron},
  {Blagorodnova}, {Kubiak}, {Szyma{\'n}ski}, {Pietrzy{\'n}ski},
  {Soszy{\'n}ski}, {Ulaczyk}, {Pietrukowicz}, \&
  {Mr{\'o}z}}]{2014AcA....64..197W}
{Wyrzykowski}, {\L}., {Kostrzewa-Rutkowska}, Z., {Koz{\l}owski}, S., {et~al.}
  2014{\natexlab{f}}, ACTA ASTRONOMICA, 64, 197

\bibitem[{Wyrzykowski {et~al.}(2014)Wyrzykowski, Kostrzewa-Rutkowska,
  Kozlowski, Udalski, Poleski, Skowron, Blagorodnova, Kubiak, Szymanski,
  Pietrzynski, {et~al.}}]{wyrzykowski2014ogle}
Wyrzykowski, L., Kostrzewa-Rutkowska, Z., Kozlowski, S., {et~al.} 2014, arXiv
  preprint arXiv:1409.1095

\bibitem[{{Young} {et~al.}(2010){Young}, {Smartt}, {Valenti}, {Pastorello},
  {Benetti}, {Benn}, {Bersier}, {Botticella}, {Corradi}, {Harutyunyan},
  {Hrudkova}, {Hunter}, {Mattila}, {de Mooij}, {Navasardyan}, {Snellen},
  {Tanvir}, \& {Zampieri}}]{2010A&A...512A..70Y}
{Young}, D.~R., {Smartt}, S.~J., {Valenti}, S., {et~al.} 2010, \aap, 512, A70

\bibitem[{Yu {et~al.}(2013)Yu, Zhang, \& Gao}]{yu2013bright}
Yu, Y.-W., Zhang, B., \& Gao, H. 2013, The Astrophysical Journal Letters, 776,
  L40

\bibitem[{Zhang \& Harding(2000)}]{zhang2000high}
Zhang, B., \& Harding, A.~K. 2000, The Astrophysical Journal Letters, 535, L51

\end{thebibliography}

\end{document}